\documentclass[useAMS,usenatbib]{mn2e}

\usepackage{graphicx}

\title[MIDI observations of terrestrial planet formation]{Resolving
  the terrestrial planet forming regions of HD113766 and HD172555 with
MIDI}
\author[R. Smith et al.]{R. Smith$^{1,2}$\thanks{E-mail:
rs@astro.keele.ac.uk}, M. C. Wyatt$^{2}$ and C. A. Haniff$^3$ \\
$^{1}$Astrophysics Group, Lennard-Jones Laboratories, Keele University,
  Keele, Staffordshire, ST5 5BG, UK \\
$^{2}$Institute of Astronomy, University of Cambridge, Madingley Road,
Cambridge, CB3 0HA, UK\\
$^{3}$Cavendish Laboratory, University of Cambridge, JJ Thomson
Avenue, Cambridge, CB3 0HE }
\begin{document}

\date{In prep.}

\pagerange{\pageref{firstpage}--\pageref{lastpage}} \pubyear{2011}

\maketitle

\label{firstpage}

\begin{abstract}
We present new MIDI interferometric and VISIR spectroscopic
observations of HD113766 and HD172555.  Additionally
we present VISIR 11$\mu$m and 18$\mu$m imaging observations of
HD113766.  These sources represent the youngest  (16Myr and 12Myr old
respectively) debris disc hosts with emission on $\ll$10AU scales.  We
find that the disc of HD113766 is partially resolved on baselines of
42--102m, with variations in resolution with baseline length
consistent with a Gaussian model for the disc with FWHM of
1.2--1.6AU (9--12mas).  This is consistent with the VISIR observations which
place an upper limit of 0\farcs14 (17AU) on the emission, with no evidence
for extended emission at larger distances.   For HD172555 the MIDI
observations are consistent with complete resolution of the disc
emission on all baselines of lengths 56--93m, putting the dust at a
distance of $>$1AU ($>$35mas).  When combined with limits from
TReCS imaging the dust at $\sim$10$\mu$m is constrained to
lie somewhere in the region 1--8AU.  Observations at
  $\sim$18$\mu$m reveal extended disc emission which could originate
  from the outer edge of a broad disc, the inner parts of which are
  also detected but not resolved at 10$\mu$m, or from a spatially
  distinct component. 
These observations provide the most accurate direct
measurements of the location of dust at 1--8AU that might 
originate from the collisions expected during terrestrial planet
formation.  These observations provide valuable constraints for models
of the composition of discs at this epoch and provide a foundation for
future studies to examine in more detail the morphology of debris
discs.    
\end{abstract}


\section{Introduction}

Since first being identified using data from the IRAS satellite, the
debris disc phenomenon has been the subject of intense study.  Spitzer
surveys have confirmed that such discs, thought to be the debris
material left over at the end of the planet formation process, are
present around $\sim$15\% of nearby stars (see
e.g. \citealt{wyattreview} and references therein).  The
spectral energy distribution (SED) of the dust emission in most debris
discs peaks at $>$60$\mu$m, implying that the dust is cool ($<$80K),
and resides in Kuiper belt-like regions ($\gg$10AU) in the systems.
However, some systems have hot dust on scales $\ll$10AU.  The current
tally of known hot 
debris discs is $\sim$20 across spectral types A--M (13 catalogued in
\citealt{wyattsmith06, wyattsmith07}, and 7 recently discovered with
\emph{AKARI}; \citealt{fujiwara}).  This tally includes systems with
multiple components where the hottest component lies at $\ll$10AU,
such as $\eta$ Tel \citep{smitheta} and $\beta$ Leo \citep{stock}.
For some multiple dust component systems (e.g. $\eta$ Corvi,
\citealt{smithmidi} and references therein) it is suggested that the
hot dust component may be fed by a parent planetesimal belt coincident
with the cooler dust belt.  For systems without a known cold dust
component, alternative models for the origin of the hot dust must be
sought.   

\begin{table*}
\caption{\label{tab:sources} Characteristics of science and
  calibration targets} 
\begin{tabular}{lclllcccc}\hline \multicolumn{8}{c}{Science targets} \\
  \hline Source & Spectral type & Age & RA & Dec & $F_\star$ at 10$\mu$m
  & $F_{\rm{disc}}$ at 10$\mu$m & Stellar angular size &
  Predicted disc size \\ HD & & Gyr & & & mJy & mJy & mas &
  mas \\ \hline 113766 & F3V & 16$^a$ & 13 06 35.8 & -46
  02 02.01 & 94 & 2359 & 0.048$\pm$0.003 & 13, 31--69$^c$ \\ 172555 &
  A5V & 12$^b$ & 18 45 26.9 & -64 52 16.53 & 721 & 973 &
  0.27$\pm$0.01 & 198$^d$ \\ \hline  
\multicolumn{8}{c}{Standard stars} \\ \hline
Source & \multicolumn{2}{c}{Spectral type} & RA & Dec &
\multicolumn{2}{c}{$F_\star$ at 10$\mu$m} & Angular size &
Instrument \\ HD & & & & &
\multicolumn{2}{c}{mJy} & mas & 
\\ \hline 111915 & \multicolumn{2}{c}{K3.5III} & 12 53 06.91 & -48 56 35.93 &
\multicolumn{2}{c}{13797} & & VISIR 
\\ 110253 & \multicolumn{2}{c}{K2III} & 12 41 09.67 & -44 06 04.27 &
\multicolumn{2}{c}{1065} & 0.87$\pm$0.02 & MIDI 
\\ 112213 & \multicolumn{2}{c}{M0III} & 12 55 19.43 & -42 54 56.50 &
\multicolumn{2}{c}{14542} & 3.16$\pm$0.02 & MIDI 
\\ 116870 & \multicolumn{2}{c}{K5III} & 13 26 43.17 & -12 42 27.60 &
\multicolumn{2}{c}{10416} & 2.58$\pm$0.01 & MIDI 
\\ 152186 & \multicolumn{2}{c}{K1III} & 16 55 34.43 & -60 40 38.77 &
\multicolumn{2}{c}{785} & 1.00$\pm$0.02 & MIDI 
\\ 156277 & \multicolumn{2}{c}{K2III} & 17 21 59.48 & -67 46 14.30 &
\multicolumn{2}{c}{7004} & 2.00$\pm$0.01 & VISIR, MIDI 
\\ 169767 & \multicolumn{2}{c}{G9III} & 18 28 49.86 & -49 04 14.10 &
\multicolumn{2}{c}{9006} & 2.15$\pm$0.01 & MIDI 
\\ 171212 & \multicolumn{2}{c}{K1III} & 18 36 41.43 & -56 13 37.12 &
\multicolumn{2}{c}{754} & 1.54$\pm$0.02 & MIDI 
\\ 171759 & \multicolumn{2}{c}{K0III} & 18 43 02.14 & -71 25 21.20 &
\multicolumn{2}{c}{12640} & 2.68$\pm$0.01 & TReCS, MIDI 
\\ \hline 
\end{tabular}
\begin{flushleft}
\small{$^a$ Age taken from Sco-Cen association membership. $^b$ Age
  from membership of $\beta$ Pic moving group. $^c$ From fit to IRS
  spectra by \citet{lisse08}. $^d$ From fit to IRS spectra by
  \citet{lisse09}. Estimated values of $F_\star$ arise from fitting a
  Kurucz model photosphere of appropriate spectral type to the 2MASS
  K band measured photometry.  For the science targets $F_{\rm{disc}}$
  is determined from the Spitzer IRS spectra of the target after
  subtraction of the photospheric model.  For the standard targets used
in MIDI observations angular size was taken from the CalVin tool
available at http://www.eso.org/intruments/midi/tools where
available (HDs 112213, 116870, 156277, 169767, and  171759).  For
the remaining standard targets and the stellar components of
  the science targets the angular size was estimated by assuming
that the stars have a diameter typical for their spectral type (taken
from \citealt{allen}) and using the Hipparcos parallax to determine their
distance. Standard stars are used as calibrators for the
  instruments listed (see text for details).}
\end{flushleft}
\end{table*}

Amongst these sources, HD113766 ($\sim$16Myr old) and HD172555
($\sim$12Myr old, \citealt{zuckerman}) are the youngest.  They also
have some of the brightest levels of excess (\citealt{wyattsmith06,
  wyattsmith07}; see Table \ref{tab:sources} for further source
  details).  The favoured interpretation for the emission 
observed around these sources is that we are witnessing ongoing
terrestrial planet formation \citep{kenyon04II}, since a detailed
analysis of the Spitzer IRS spectra of both sources indicates that the dust
composition is similar to that expected from the catastrophic
disruption of terrestrial planet embryos \citep{lisse08,
  lisse09}. However, alternatives to the planet formation origin model
for such hot emission do exist, including the scattering of comets from
tens of AU in the system \citep{gomes}, the sublimation of one
supercomet \citep{beichman05}, a recent collision between two massive
asteroids \citep{song05}, a disc of planetesimals on highly
eccentric orbits \citep{wyatt10}, or that it is in fact a steady-state
phenomenon.  This has recently been suggested for the HD69830 system
\citep{heng11}, which had previously been identified as a host of
transient debris emission, \citep{wyattsmith06}.  

A correct interpretation of the hot dust depends critically on its
radial location.  We expect different dust distributions from the
different theories for the origin of the dust, e.g., a population of 
comets scattering inwards would be expected to be observed at multiple
distances from the star (the parent belt location and the dust
sublimation radius) whereas dust from terrestrial planet formation
would be expected to be confined to a narrow ring. Modelling
the SED provides poor constraints on the radial distribution of the
dust, as one may either: (i) underestimate the size of the dusty
region, because the emitting grains are small and hotter than
blackbody (observed discs can be a factor of 3 larger than predicted
due to this effect, \citealt{schneider}); or (ii) overestimate the
size of the disc, because the dust is in an extended distribution (for
example the predicted size of the $\zeta$ Lep disc was more than
double the observed size as multiple disc components over a range of
distances from the star had been fit by a single disc temperature, see
\citealt{moerchen, smithastar}).
The most direct way to resolve these ambiguities is with very high
spatial resolution observations.  In this paper we present VISIR
imaging of HD113766 together with VISIR spectroscopy and MIDI
observations of both HD113766 and HD172555. We also present a
re-analysis of archival TReCS imaging of HD 172555. These observations
are compared with models for the distribution of the dust and
constraints placed on the location of the emitting material.

\section{Observations with 8m instruments}

\subsection{VISIR imaging and photometry of HD113766}

\begin{figure*}
\begin{minipage}{1cm}
\hspace{1cm} 
\end{minipage}
\begin{minipage}{4.0cm}
\center{Standard}
\end{minipage}
\begin{minipage}{4.0cm}
\center{HD113766A}
\end{minipage}
\begin{minipage}{4.0cm}
\center{Residual} 
\end{minipage}
\begin{minipage}{4.0cm}
\center{HD113766B} 
\end{minipage} \\
\begin{minipage}{1cm}
N band
\end{minipage}
\begin{minipage}{4.0cm}
\includegraphics[width=4.0cm]{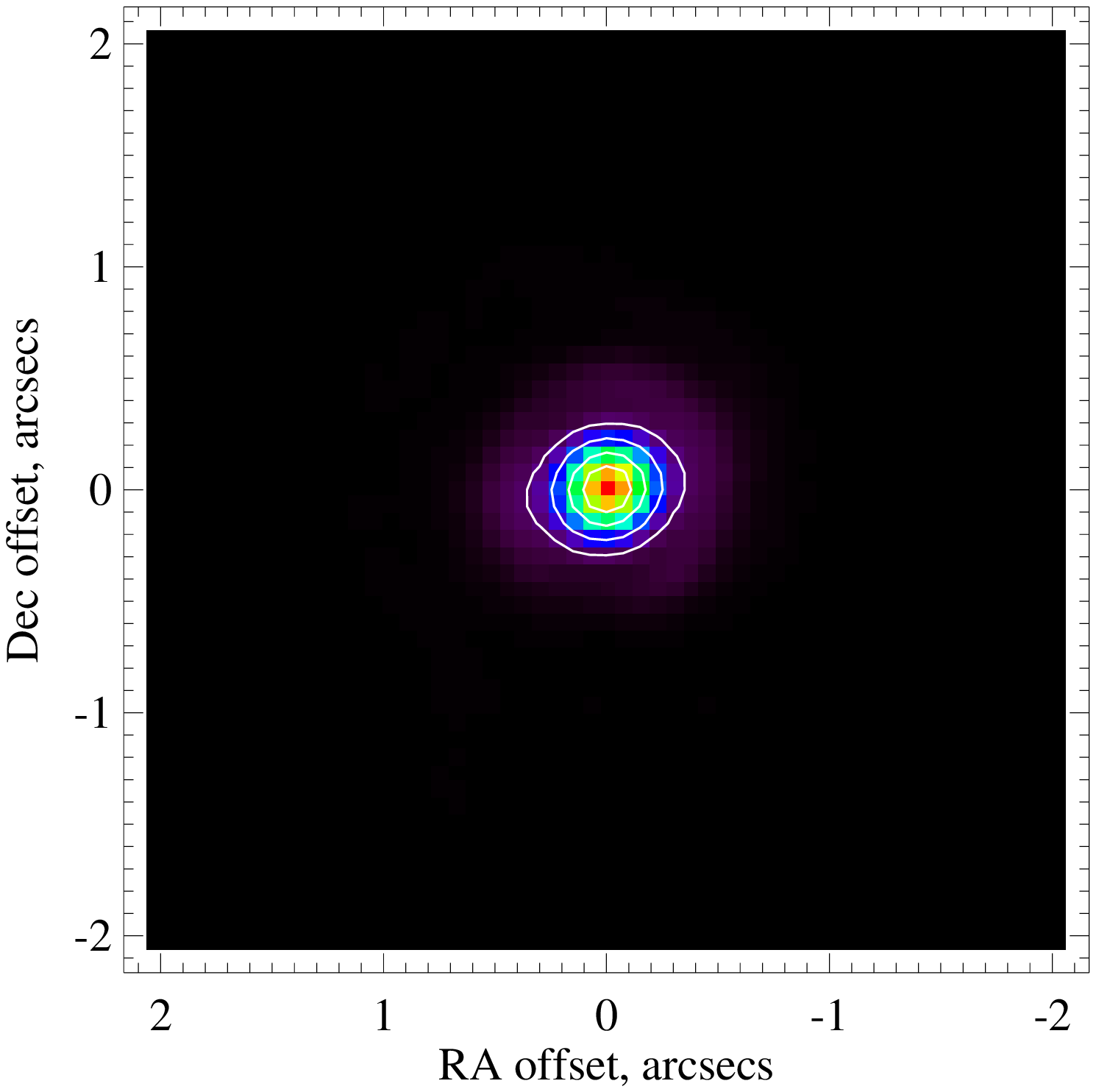}
\end{minipage}
\begin{minipage}{4.0cm}
\includegraphics[width=4.0cm]{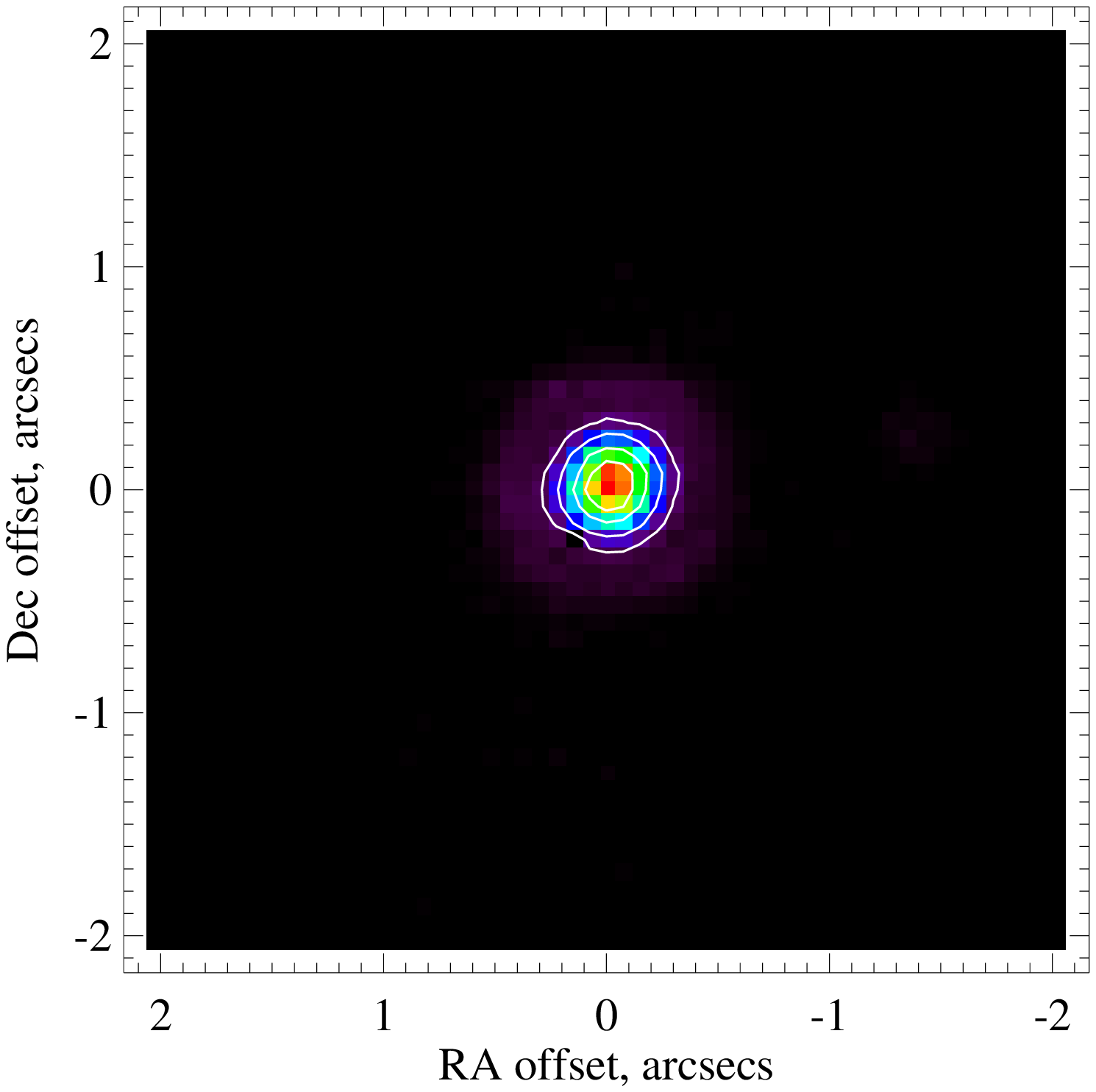}
\end{minipage}
\begin{minipage}{4.0cm}
\includegraphics[width=4.0cm]{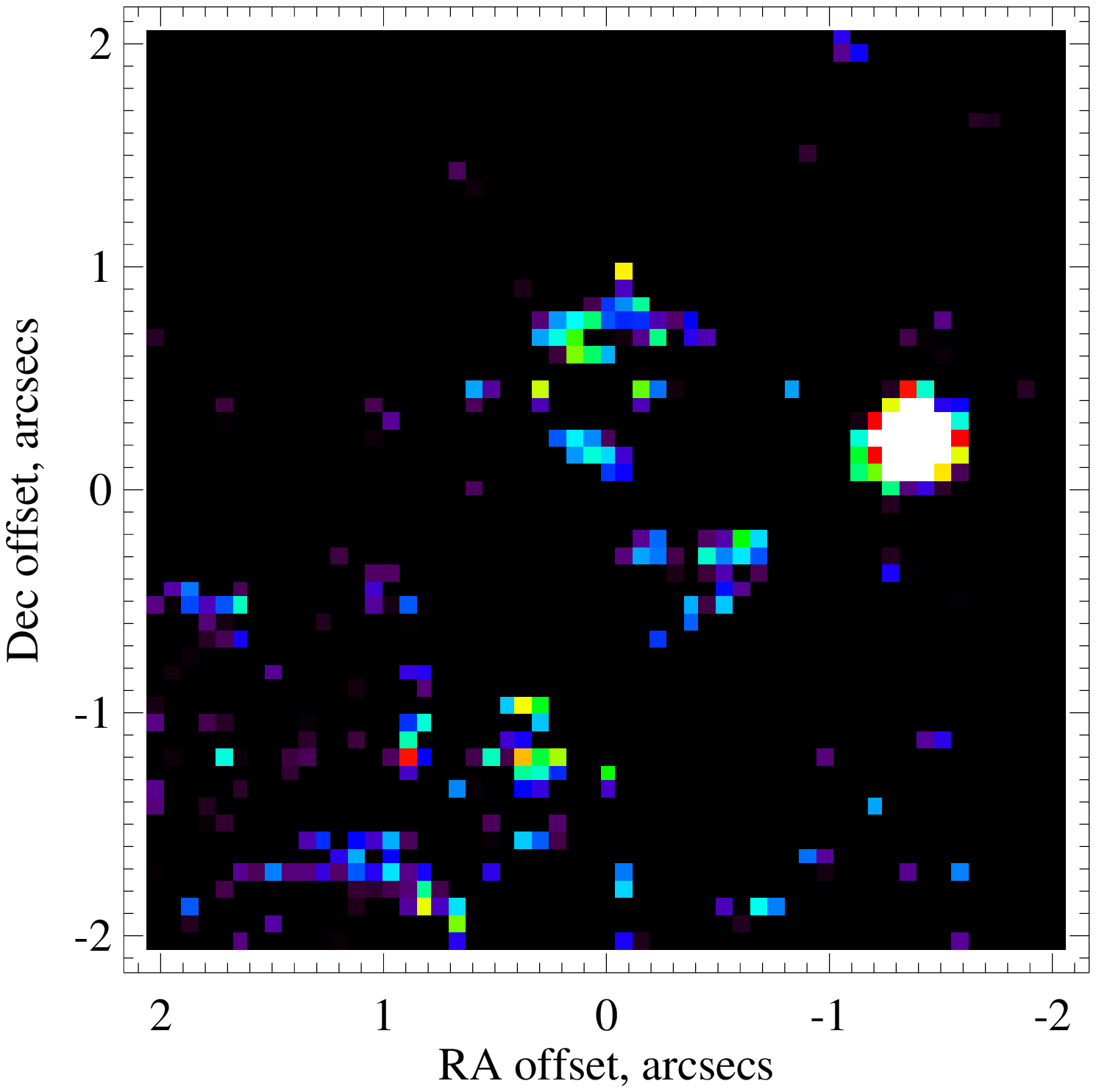}
\end{minipage}
\begin{minipage}{4.0cm}
\includegraphics[width=4.0cm]{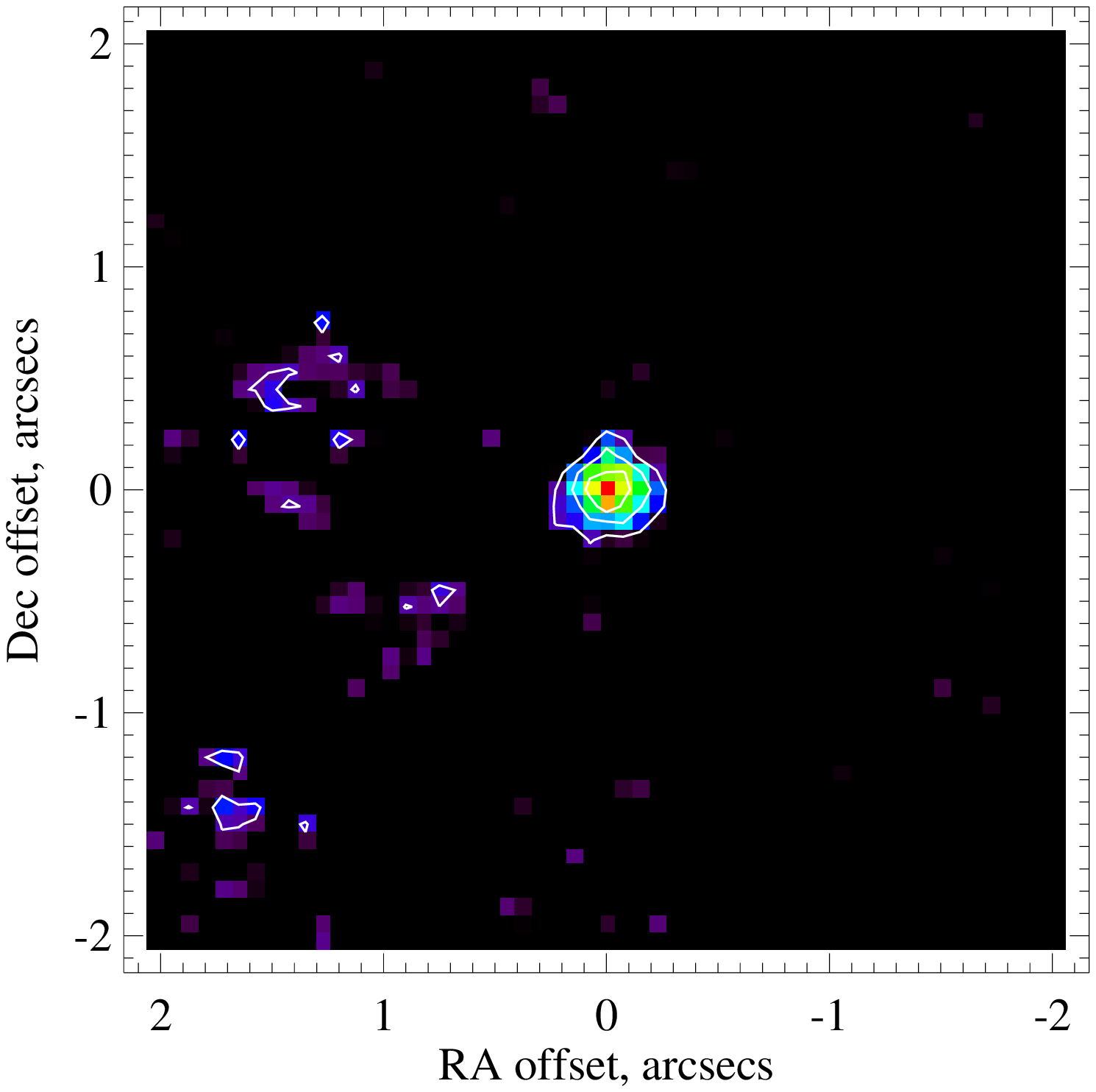}
\end{minipage} \\
\begin{minipage}{1cm}
Q band
\end{minipage}
\begin{minipage}{4.0cm}
\includegraphics[width=4.0cm]{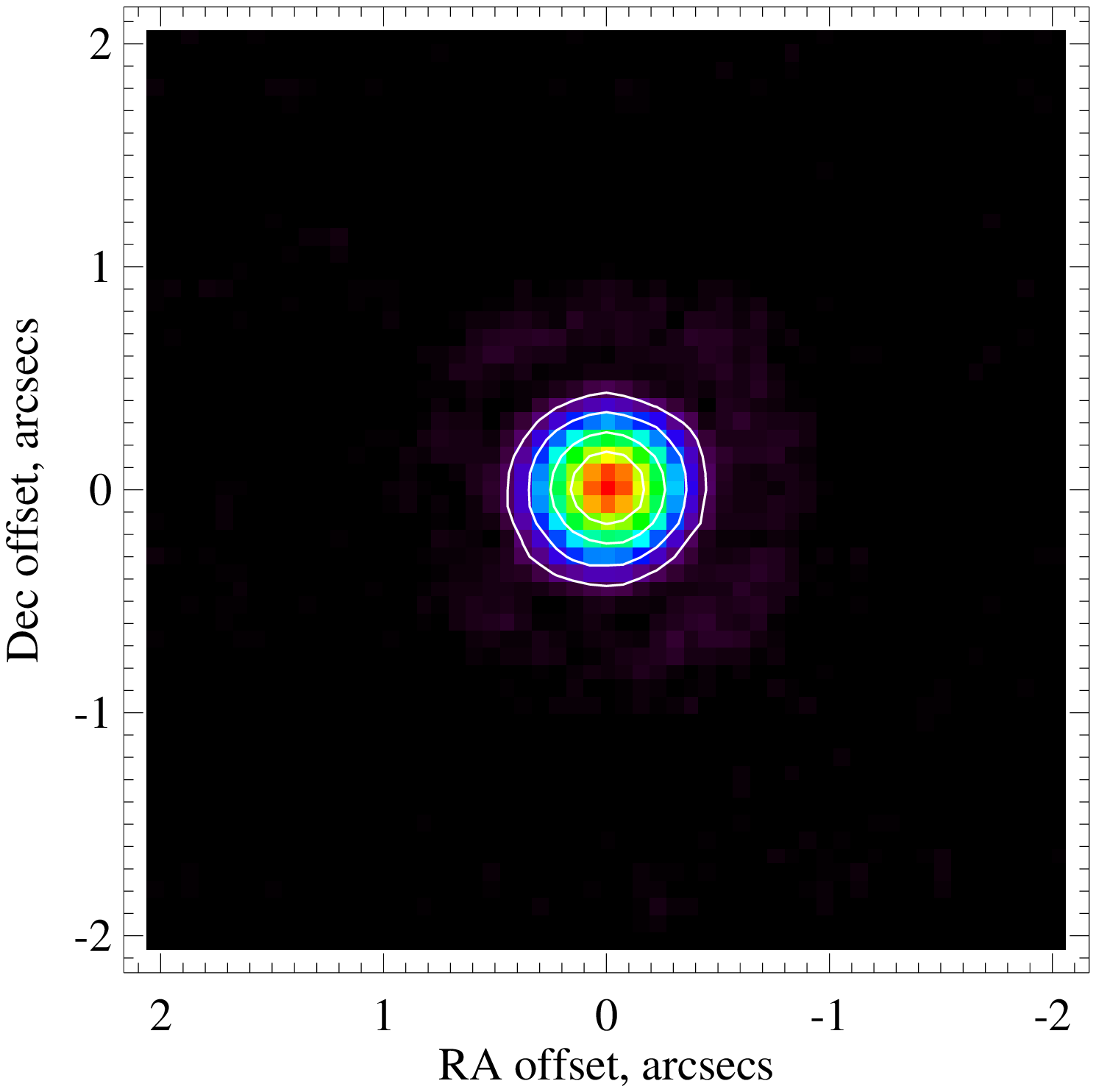}
\end{minipage}
\begin{minipage}{4.0cm}
\includegraphics[width=4.0cm]{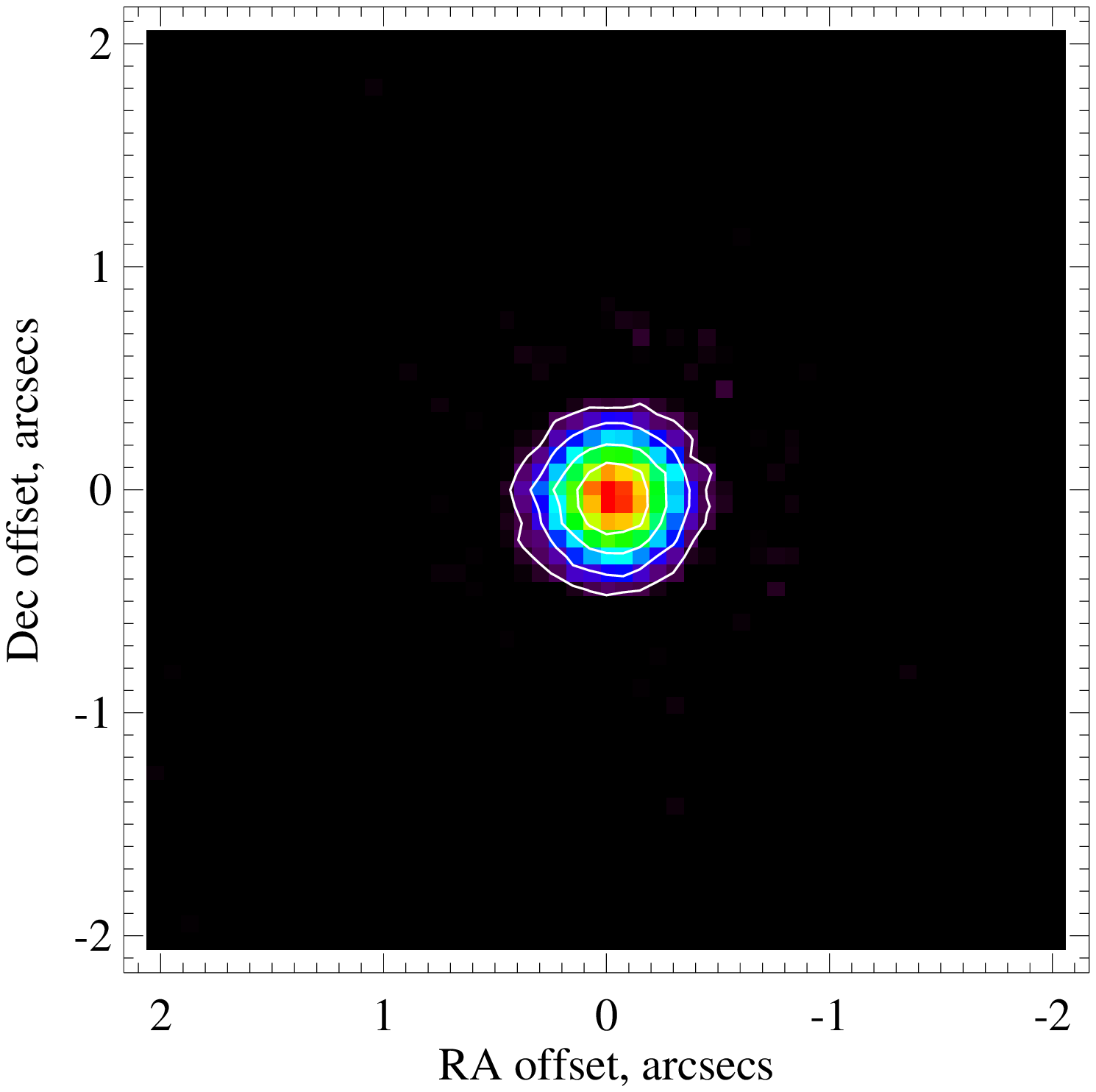}
\end{minipage}
\begin{minipage}{4.0cm}
\includegraphics[width=4.0cm]{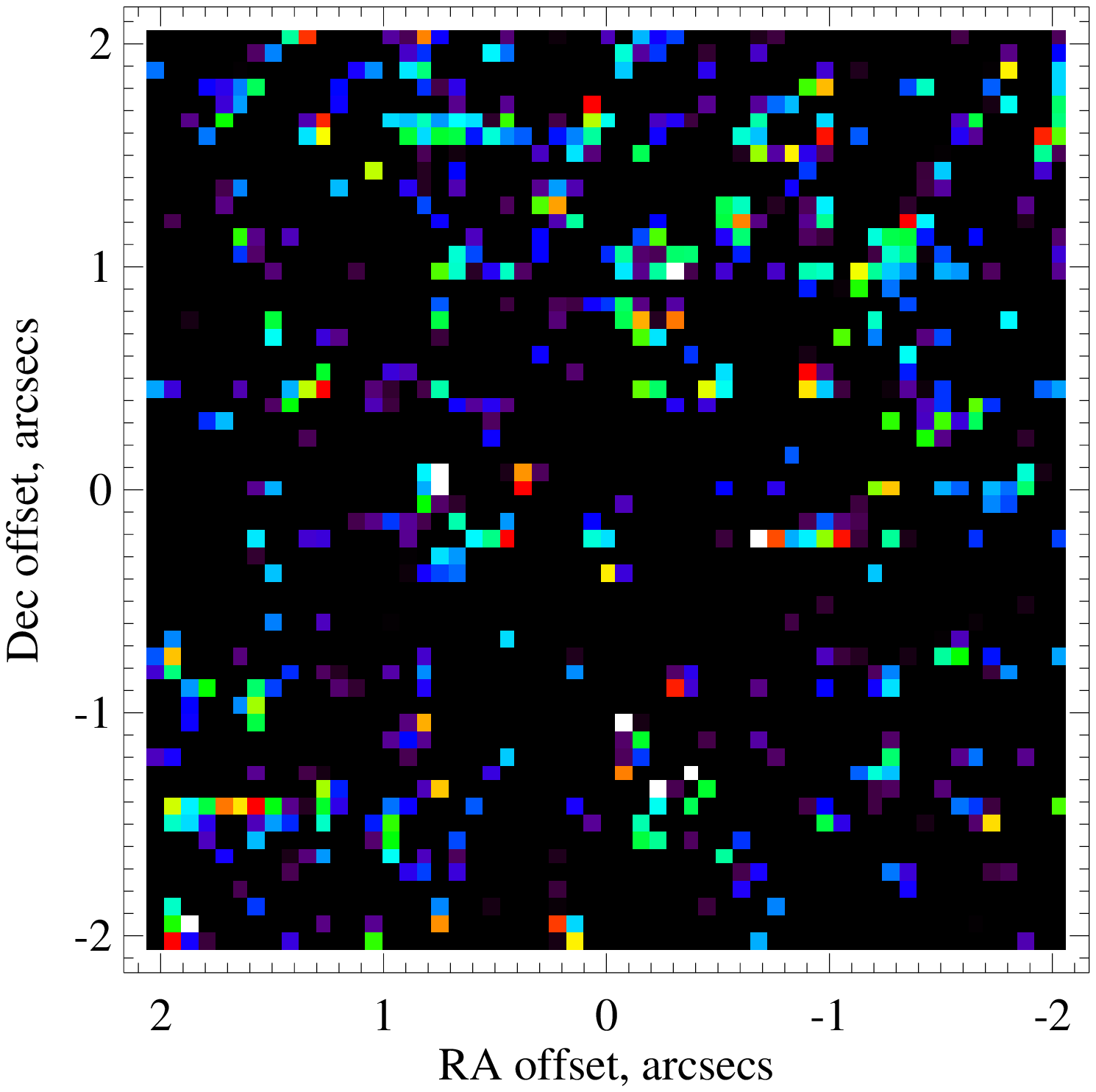}
\end{minipage}
\begin{minipage}{4.0cm}
\includegraphics[width=4.0cm]{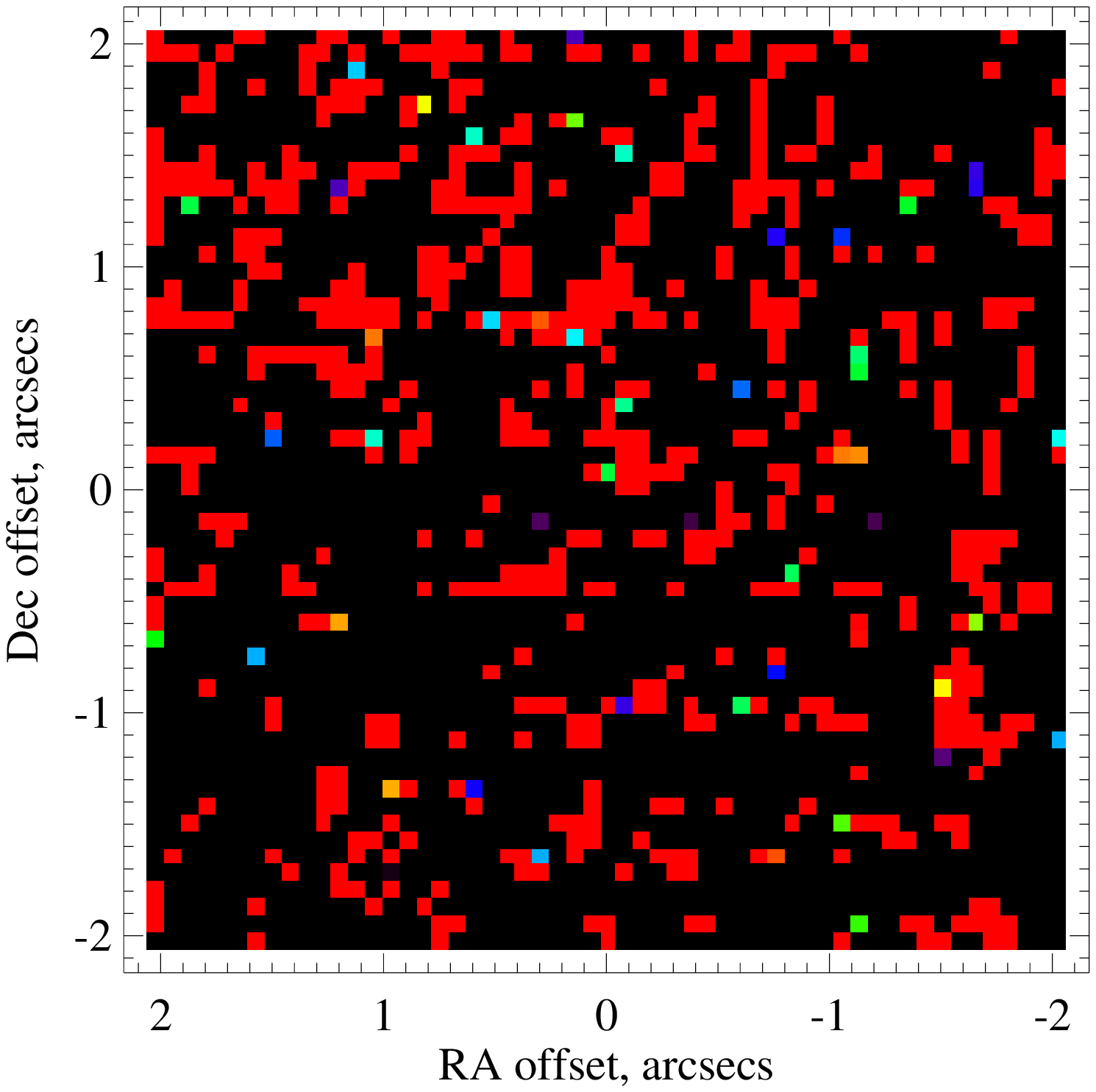}
\end{minipage} \\
\caption{\label{fig:visimg} The final co-added images for HD111915
  (the standard star) and HD113766 observed with VISIR (two leftmost
  columns).  The contours 
  are at 10, 25, 50 and 75\% of the peak and show that there is no
  evidence for extended emission beyond the size of the PSF.  The
  known binary companion HD113766B is difficult to see in the final
  images of HD1136766A.  In the
  residual image, created by subtracting the final standard star image
  from the HD113766A image after scaling to the peak, there is no
  evidence for significant residual emission which might be a resolved
  disc.  The signal seen in the N band residual image is from the
  binary companion.  In these images black pixels are $<$1$\sigma$ per
  pixel.  For the residual image the maximum pixel value (white) is
  3$\sigma$ per pixel.  The HD113766B image for the Q band contains no
  signal as there is no detection of the binary companion at this
  wavelength.} 
\end{figure*}

High resolution imaging with 8m-class telescopes can reveal debris
disc structure on $>$0\farcs5 scales which corresponds to tens of AU
for nearby stars (see e.g. \citealt{smitheta}).  We
used VISIR on the VLT to search for emission around HD113766 on such
scales. Observations were performed in filters SiC ($\lambda_c =
11.85\mu$m, $\Delta\lambda = 2.34\mu$m, hereafter referred to as N
band) and Q2 ($\lambda_c = 18.72\mu$m, $\Delta\lambda = 0.88\mu$m,
hereafter referred to as Q band) under
observing program 079.C-0259(A) in April 2007.  Observations were
performed in a perpendicular chop-nod pattern.  The science
observations were calibrated using observations of HD111915 taken from
the Cohen catalogue of mid-infrared standards \citep{cohen}.  Standard
star observations were performed immediately before and after the
science observations to allow measurements of variations in photometry
and the PSF, crucial in the search for extended emission.  A summary
of the observations is given in Table \ref{tab:visimg}.  Data
reduction was performed with custom routines, the details of which are
given in  \citet{smithhot}.  Data reduction involved determination of
a gain map using the mean values of each frame to determine pixel
responsivity (masking off pixels on which source emission 
could fall, equivalent to a sky flat).  In addition a dc-offset was
determined by calculating the mean pixel values in columns and rows
(excluding pixels on which source emission was detected) and this was
subtracted from the final image to ensure a flat background.  Pixels
showing high or low gain, or those that showed great variation
throughout the observation, were masked off. Finally, the four images
(two negative, two positive) resulting from the perpendicular chop-nod
pattern were co-added to give a final image for HD113766 and
observations of the standard. The center of each image was determined
through fitting with a two-dimensional Gaussian, with the center of
each image aligned in the co-addition step.   Aperture photometry
centered on the Gaussian peaks of the final images was performed using
1\arcsec radius apertures (noise levels were determined from an
annulus with inner radius 2\arcsec and outer radius 4\arcsec).
The calibrated flux found in the final images was 1673$\pm$42mJy for
HD113766A at N and 1895$\pm$34mJy at Q.  These uncertainties include
calibration errors of 2\% and 6\% respectively determined from
variation in calibration factors between the two standard star
observations in each filter.  These fluxes are compatible within the
errors with the IRS photometry, which taken over the filters used here
are 1599mJy at N and 1867mJy at Q. 

\begin{table}
\caption{\label{tab:visimg} Log of the VISIR imaging observations of
  HD113766 and standard star HD 111915. }
\begin{tabular}{ccccc} \hline OB ID & Filter & Int. time (s) & Object
  type & Star \\ \hline 265452 & Q2 & 400 & Cal & HD111915 \\ 265453 &
  Q2 & 1800 & Sci & HD113766 \\ 265450 & Q2 & 400 & Cal & HD111915
  \\ 265455 & SiC & 105 & Cal & HD111915 \\ 265457 & SiC & 900 & Sci &
  HD113766 \\ 265456 & SiC & 105 & Cal & HD111915 \\ \hline
\end{tabular}
\end{table}

The final VISIR images of the PSF reference (standard star image) and
HD113766 are shown in Figure \ref{fig:visimg} (first and second
columns respectively; top row is N band images, Q band images are
shown in the bottom row).  The final images are shown  with contours
at 10, 25, 50 and 75\% of the peak.  There is no evidence that there is
extended emission around HD113766 in either the N or Q band images
from the contour lines (as compared to those of the standard star; low
level ellipticity in the N band standard star image is below a
3$\sigma$ significance level).  To
search for low-level extended emission the standard star image was
scaled to the peak of the HD113766 image (in each band) and subtracted
from the science image. The result is shown in the third column of
Figure \ref{fig:visimg} (labelled residual).   The source appearing to
the North-West in the N band is a known binary companion discussed in
the following paragraph.  Excluding this source the residual images
are compatible with the noise levels of the image.  These images
reveal no evidence for residual emission extended beyond the PSF in
either band.  

\begin{figure}
\includegraphics[width=7cm]{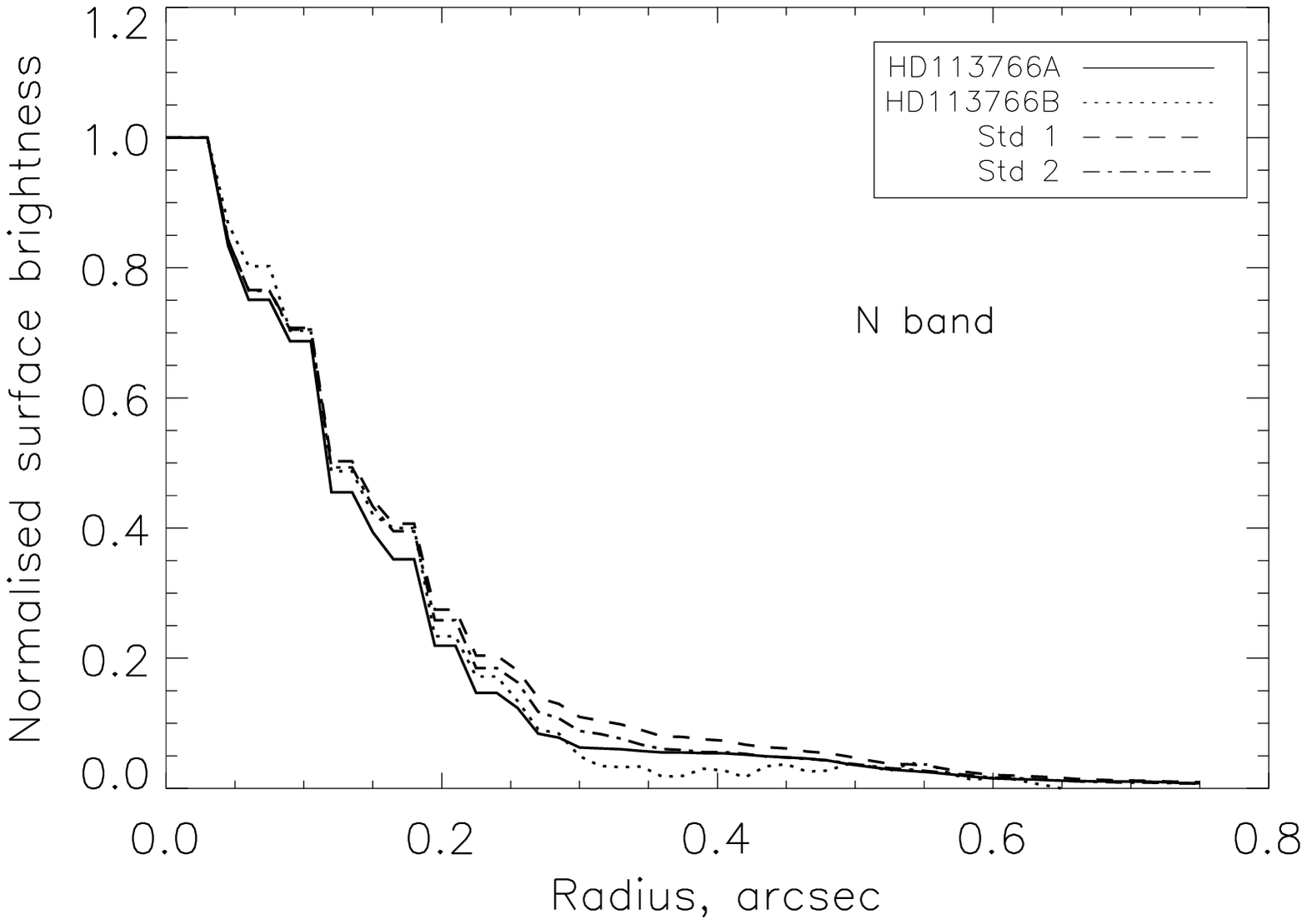} \\
\includegraphics[width=7cm]{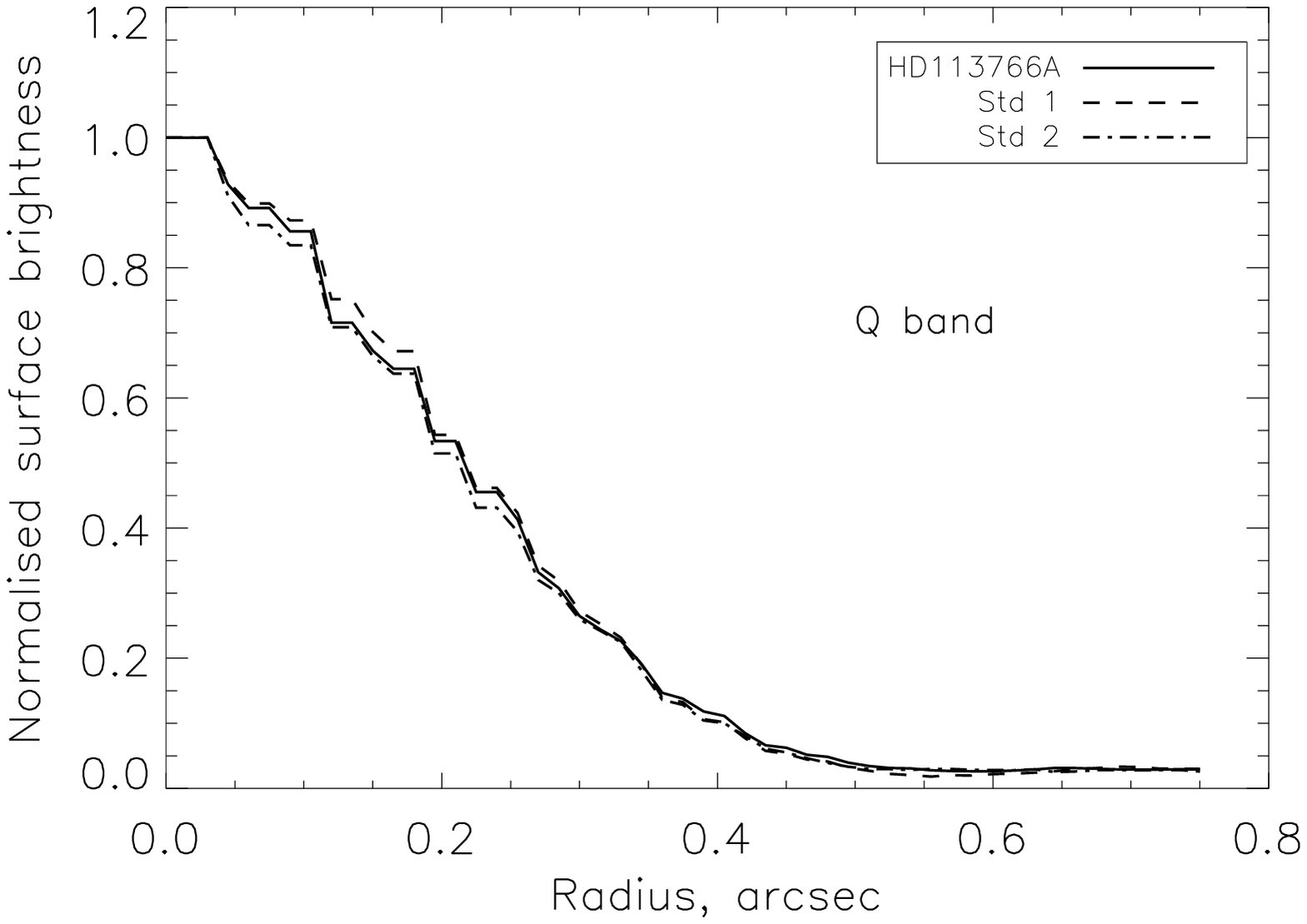}
\caption{\label{fig:sbprof} The azimuthally averaged surface
  brightness profiles of the observations of HD111915 (standard) and
  HD113766.  The profiles show that the HD113766 images are very
  similar in shape to the standard star, again showing we have no
  evidence for extended emission beyond the PSF.  The binary companion
  surface brightness profile is not shown at Q as there is no
  significant emission detected at its location in this band. }
\end{figure} 

HD113766 is a known binary star with F3/F5 components.  In the
Hipparcos catalogue the B component is listed at an offset of
1\farcs335 at a position angle of 281$^\circ$ East of North from the
primary.   A source is clearly seen in the residual image taken with
filter SiC at an offset of 1\farcs37 at a PA of 279$^\circ$ East of
North from the primary, based on Gaussian fits to determine the
centers of the source images. At a parallax distance of 131pc this
translates to an on-sky separation of 157AU.  Aperture photometry at
this location gives a flux of 49$\pm$5mJy for the binary in the N
band.  We see no significant 
flux at this location in the Q band image. This is consistent with a
SED fit to the binary companion (a Kurucz profile of spectral type F5
scaled to the Hipparcos V band flux of the binary companion) which
predicts a flux of 14mJy.  Such emission is below the 1$\sigma$
detection threshold on the Q band image (1$\sigma$ in a 1\arcsec
radius aperture is 15mJy). The contours on the HD113766B image in the
N band (Figure \ref{fig:visimg} top right) are more uneven than the
PSF as measured on the standard star, but there is no evidence of
extended emission from the contours.  As a further test the PSF image
scaled to the peak of the binary companion was subtracted from the
HD113766B image and the residual emission was found to be consistent
with the noise on the image.   

As a final test for extended emission around HD113766 the surface
brightness profiles 
of the standard star images and HD113766 were examined.  There is no
evidence in these profiles for extended emission around HD113766
beyond the shape seen for the point-like standard star targets (Figure
\ref{fig:sbprof}).   If we take sub-integrations of the science
observations and the standard star observations to consider the
variation in the profile we have 2 sub-integrations of the standard
star observation and 5 for HD113766 at N.  The mean and standard
deviations for the FWHM measurements for these sub-integrations are
0\farcs324$\pm$0\farcs002 for the standard star and
0\farcs322$\pm$0\farcs004 for HD113766.  For the Q band images we have
4 sub-integrations of the standard star observation and 11 for
HD113766.  The mean and standard deviations of the FWHM measurements
for these sub-integrations are 0\farcs498$\pm$ 0\farcs006 for the
standard star and 0\farcs498$\pm$0\farcs002 for HD113766. These
results also show no evidence for extended emission around HD113766 in
either band.  The increase in FWHM in the Q band compared to the N band
is exactly what would be expected from the increased diffraction limit
at the longer wavelength of observation.  

\begin{figure}
\includegraphics[width=7cm]{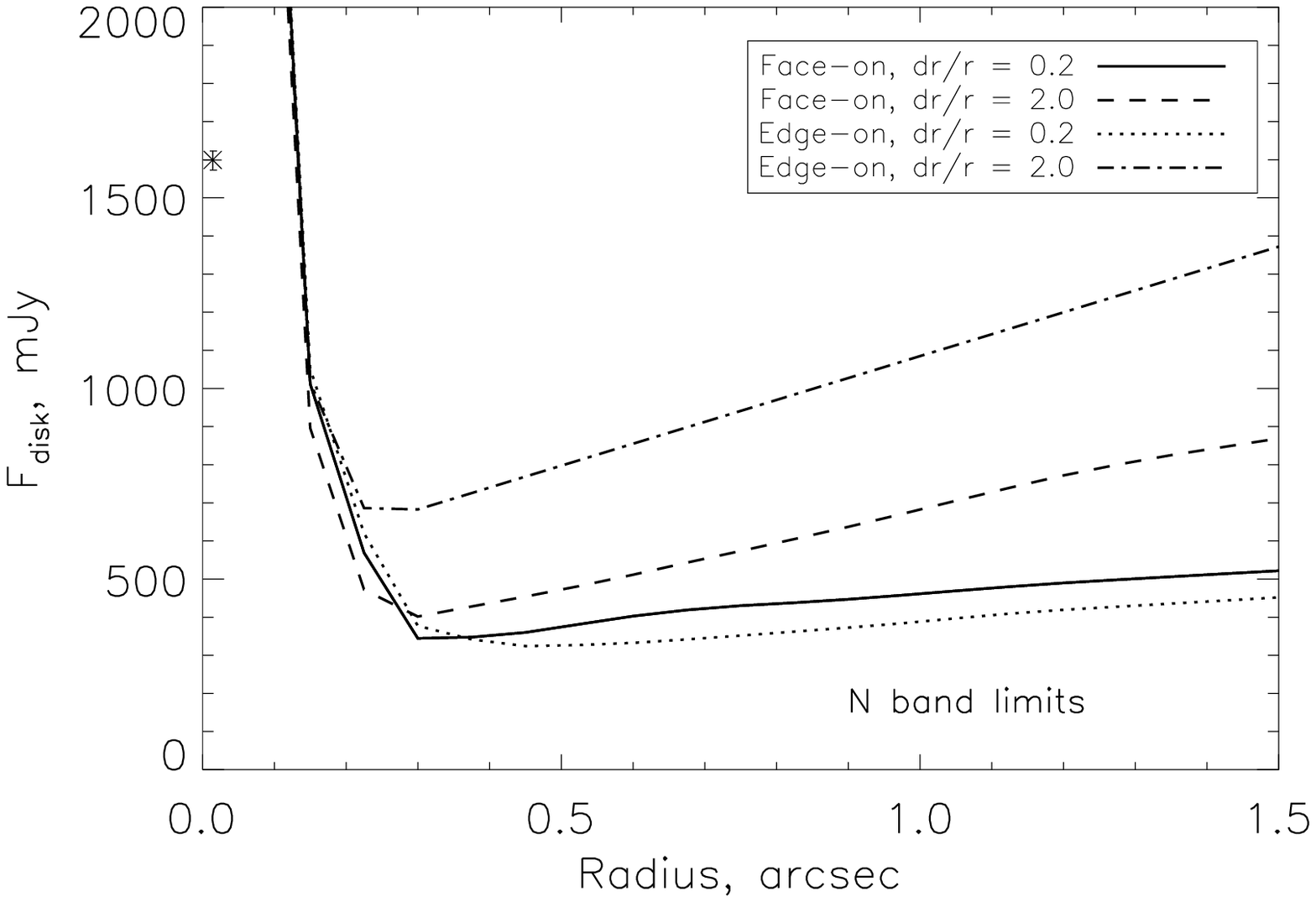}
\includegraphics[width=7cm]{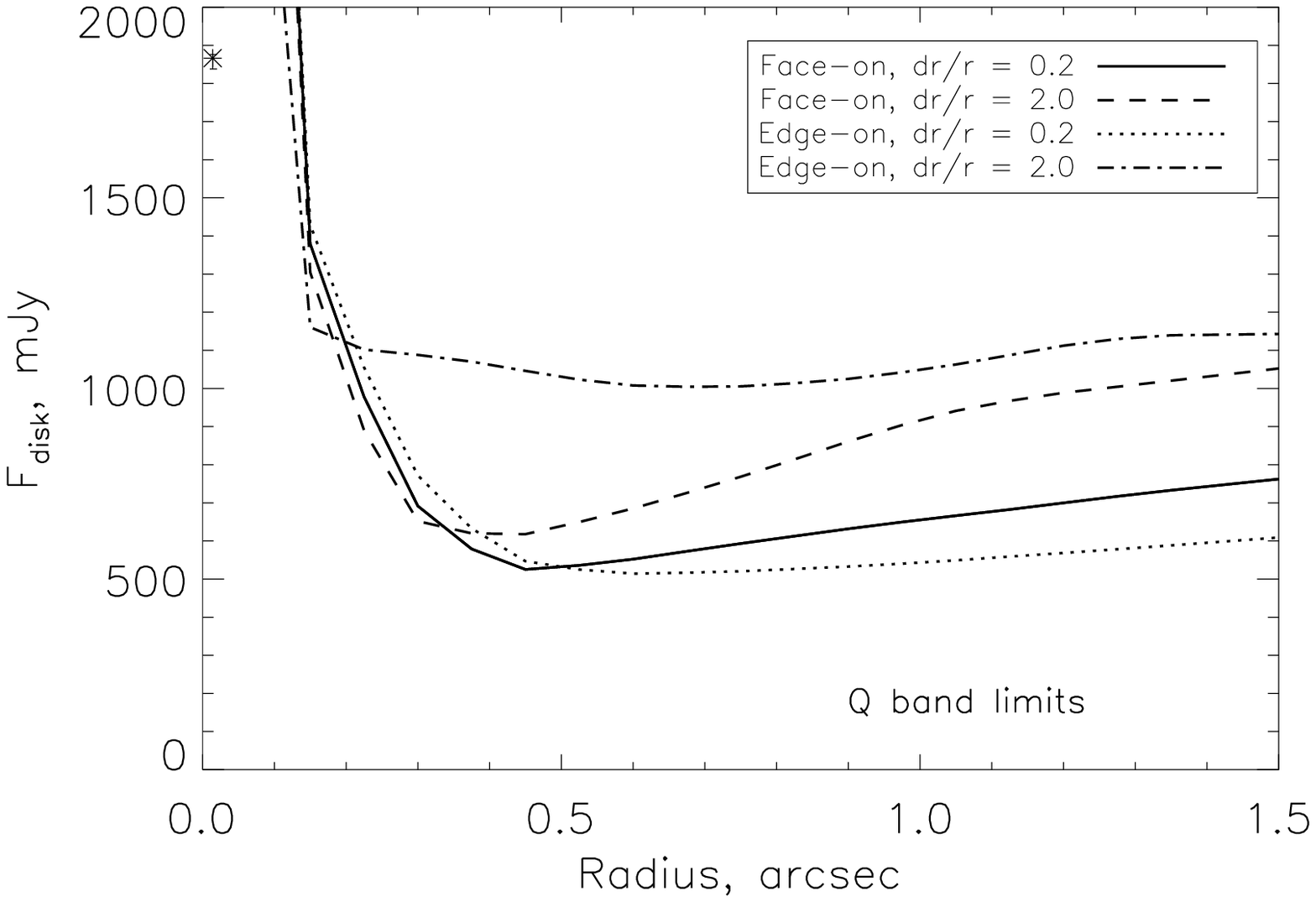}
\caption{\label{fig:ext} The limits on the disc location around
  HD113766 placed by non-detection of extension in our images.
  Different source geometries are indicated by different line styles
  as given in the legend.  
  Intermediate inclinations would have limits
  between the face-on and edge-on discs.  The regions above the lines
  represent discs 
  that would have been detected at the 3$\sigma$ level or higher.
  Thus the regions below the lines represent the possible disc
  parameter space, given our non-detection of extended emission. The
  asterisks mark the mid-point of the predicted disc location
  according to the fit by \citet{lisse08}. At these levels of emission
  (1599 mJy at N and 1867 mJy at Q) we would have expected to detect
  any disc larger than 0\farcs13.}
\end{figure}

\begin{figure*}
\begin{minipage}{1cm}
\hspace{1cm} 
\end{minipage}
\begin{minipage}{4.0cm}
\center{Standard}
\end{minipage}
\begin{minipage}{4.0cm}
\center{HD172555}
\end{minipage}
\begin{minipage}{4.0cm}
\center{Residual} 
\end{minipage}\\
\begin{minipage}{1cm}
N band
\end{minipage}
\begin{minipage}{4.0cm}
\includegraphics[width=4.0cm]{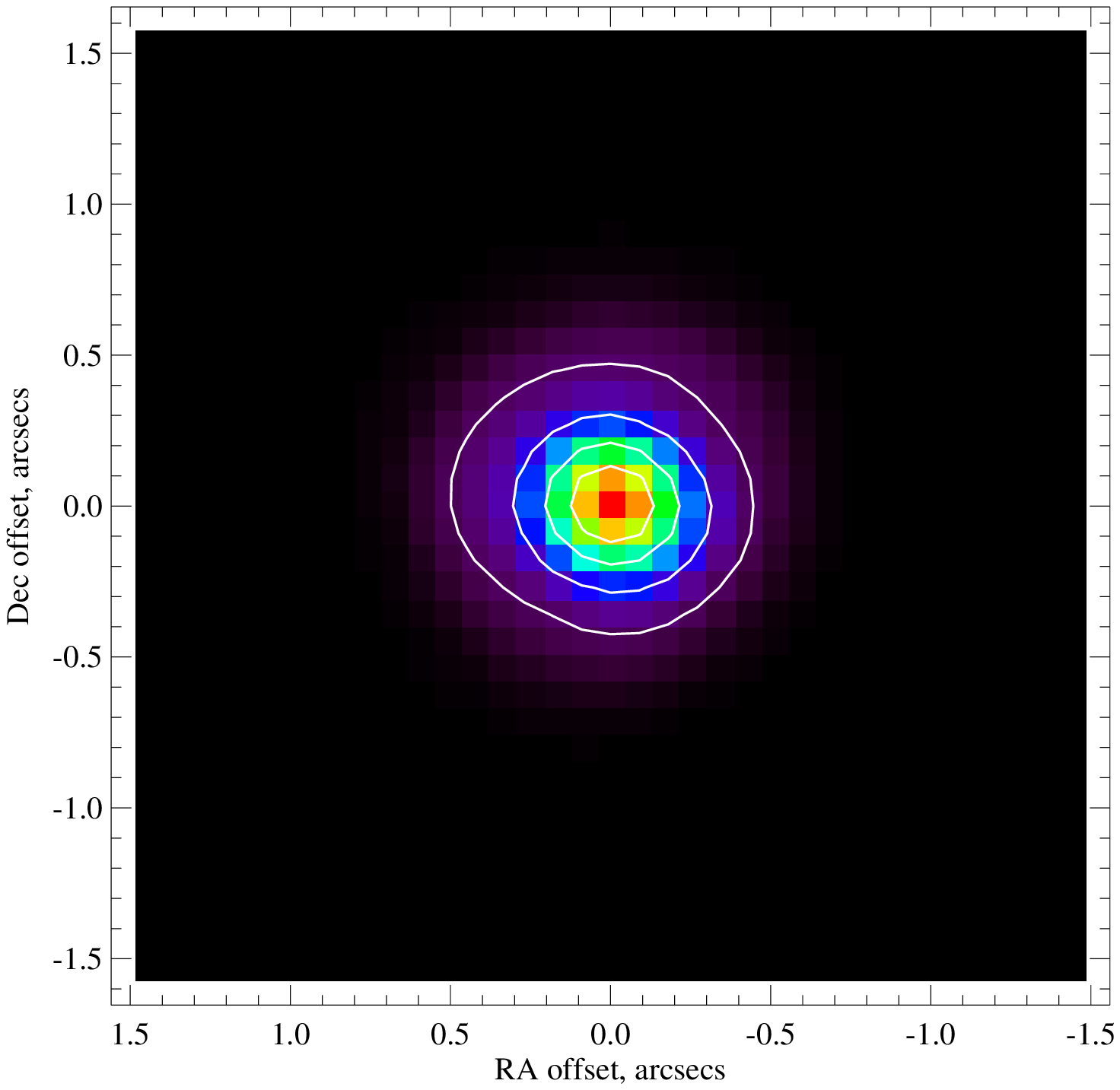}
\end{minipage}
\begin{minipage}{4.0cm}
\includegraphics[width=4.0cm]{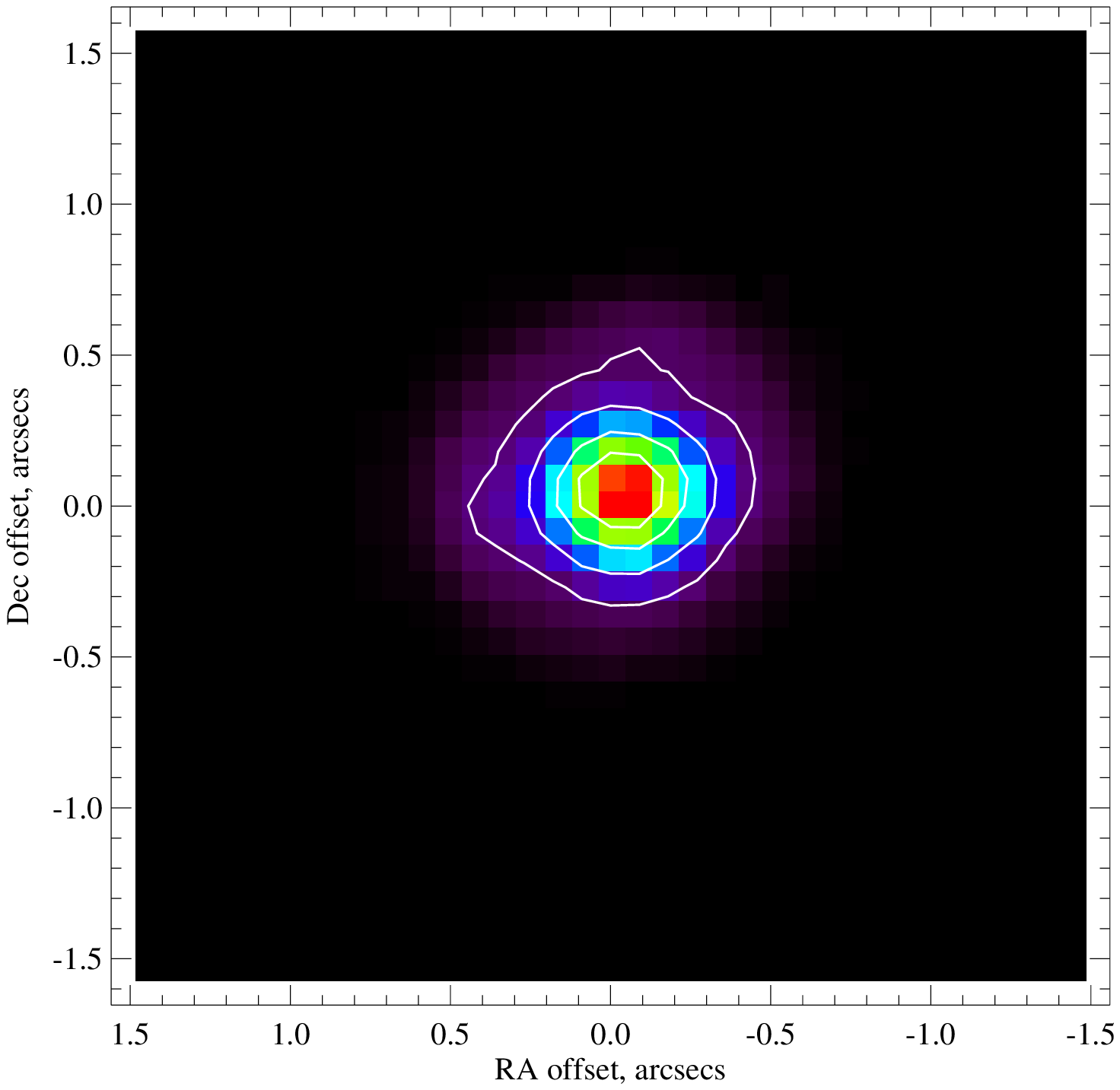}
\end{minipage}
\begin{minipage}{4.0cm}
\includegraphics[width=4.0cm]{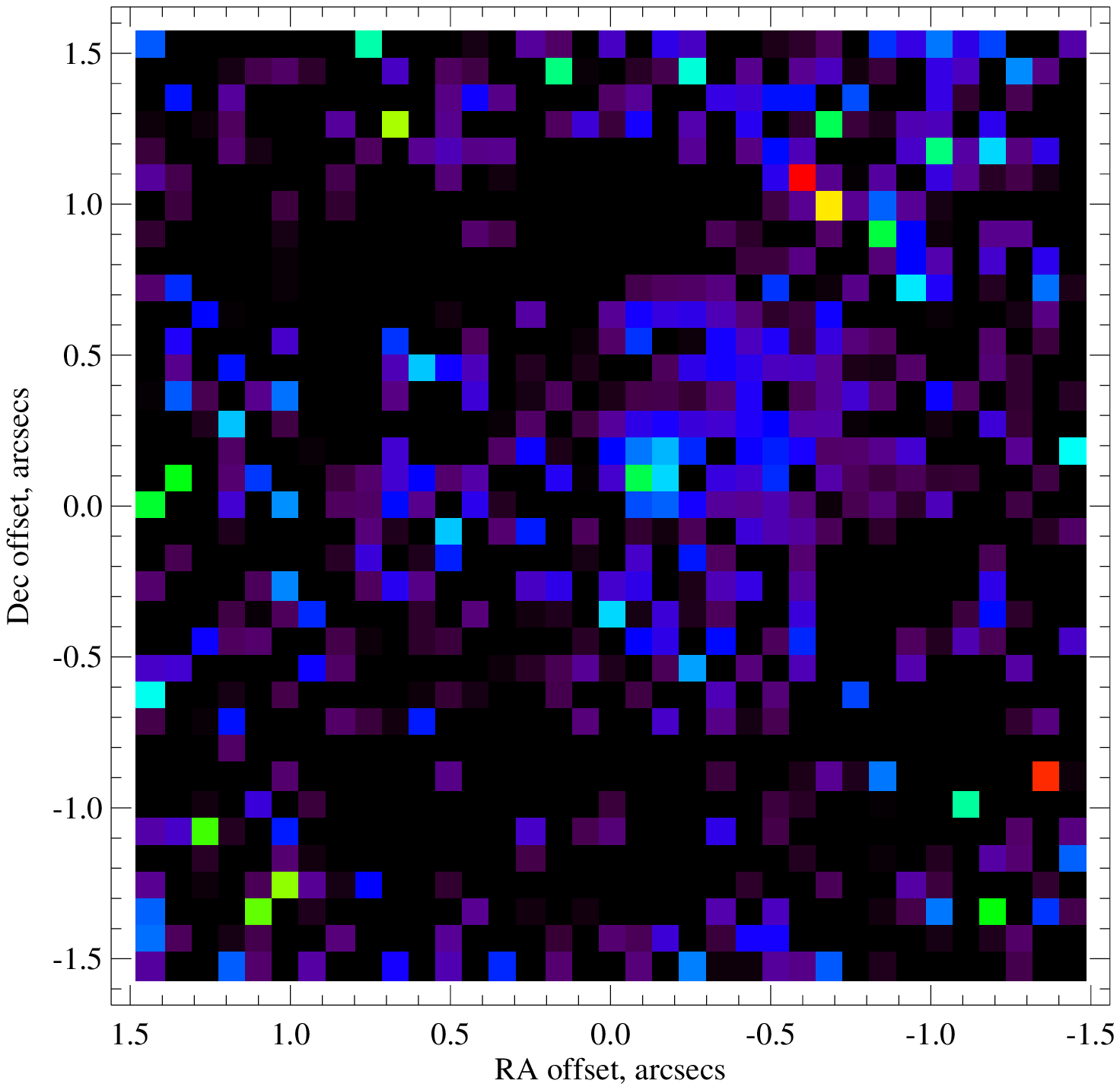}
\end{minipage} \\
\begin{minipage}{1cm}
Q band
\end{minipage}
\begin{minipage}{4.0cm}
\includegraphics[width=4.0cm]{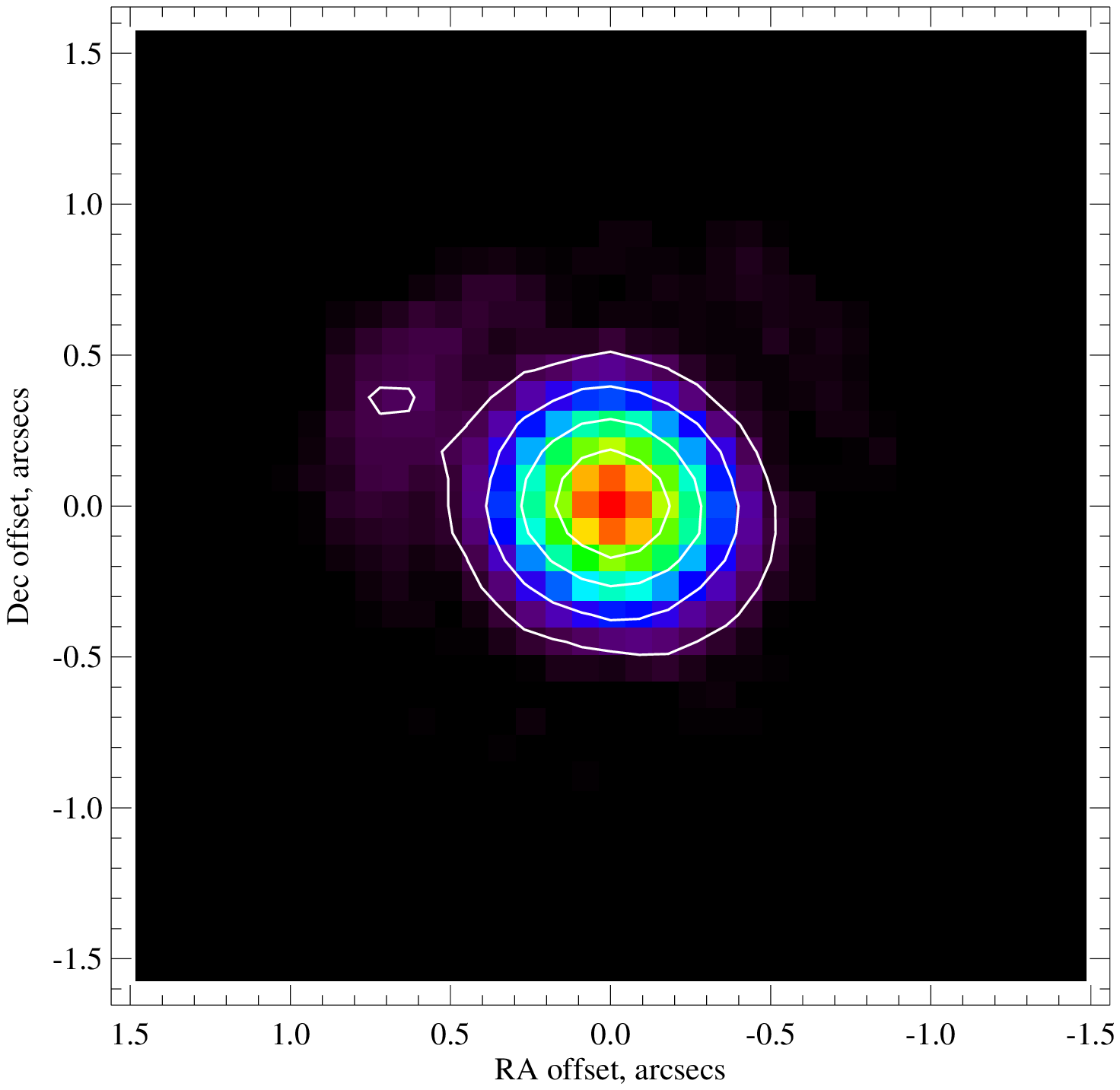}
\end{minipage}
\begin{minipage}{4.0cm}
\includegraphics[width=4.0cm]{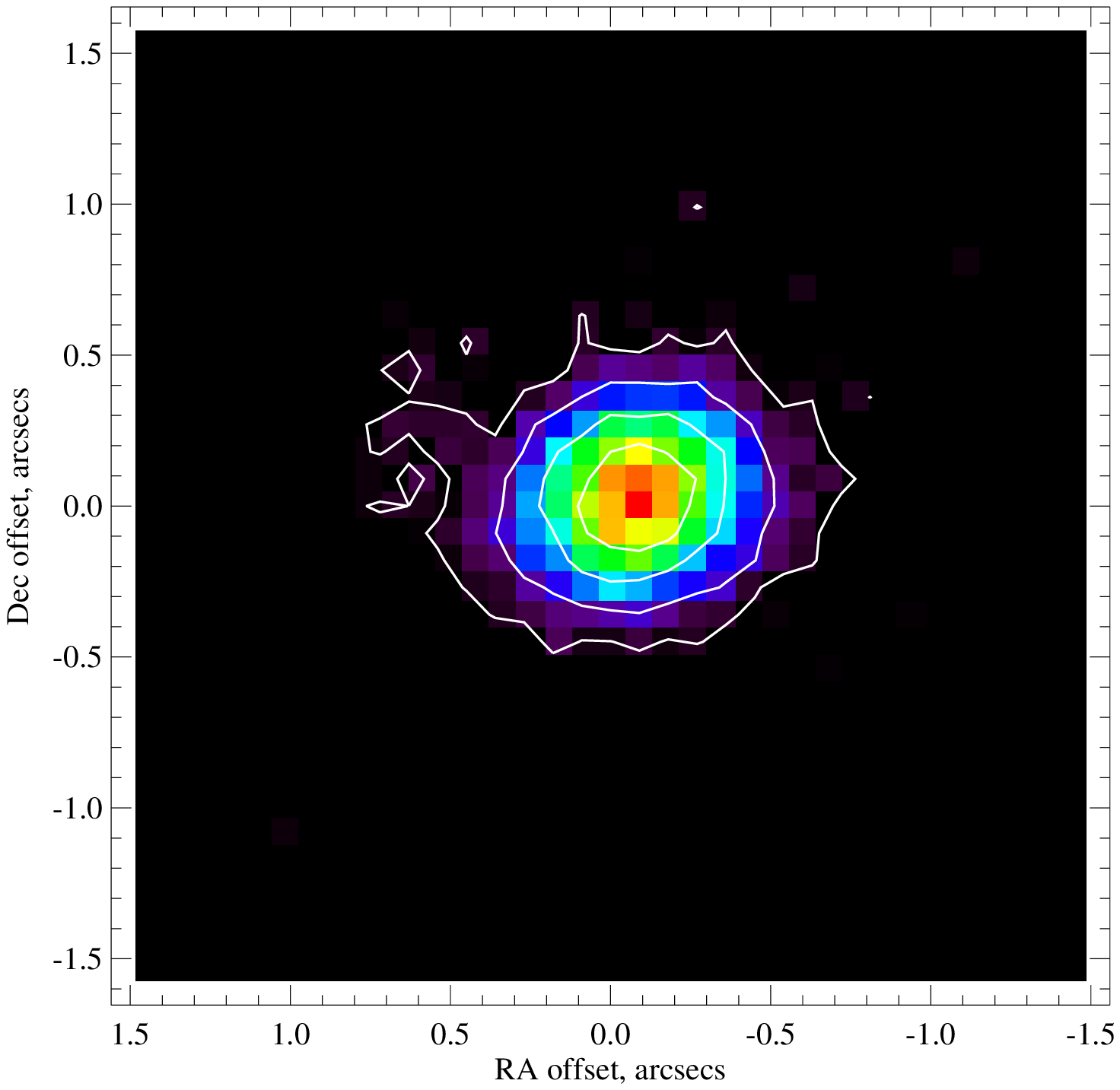}
\end{minipage}
\begin{minipage}{4.0cm}
\includegraphics[width=4.0cm]{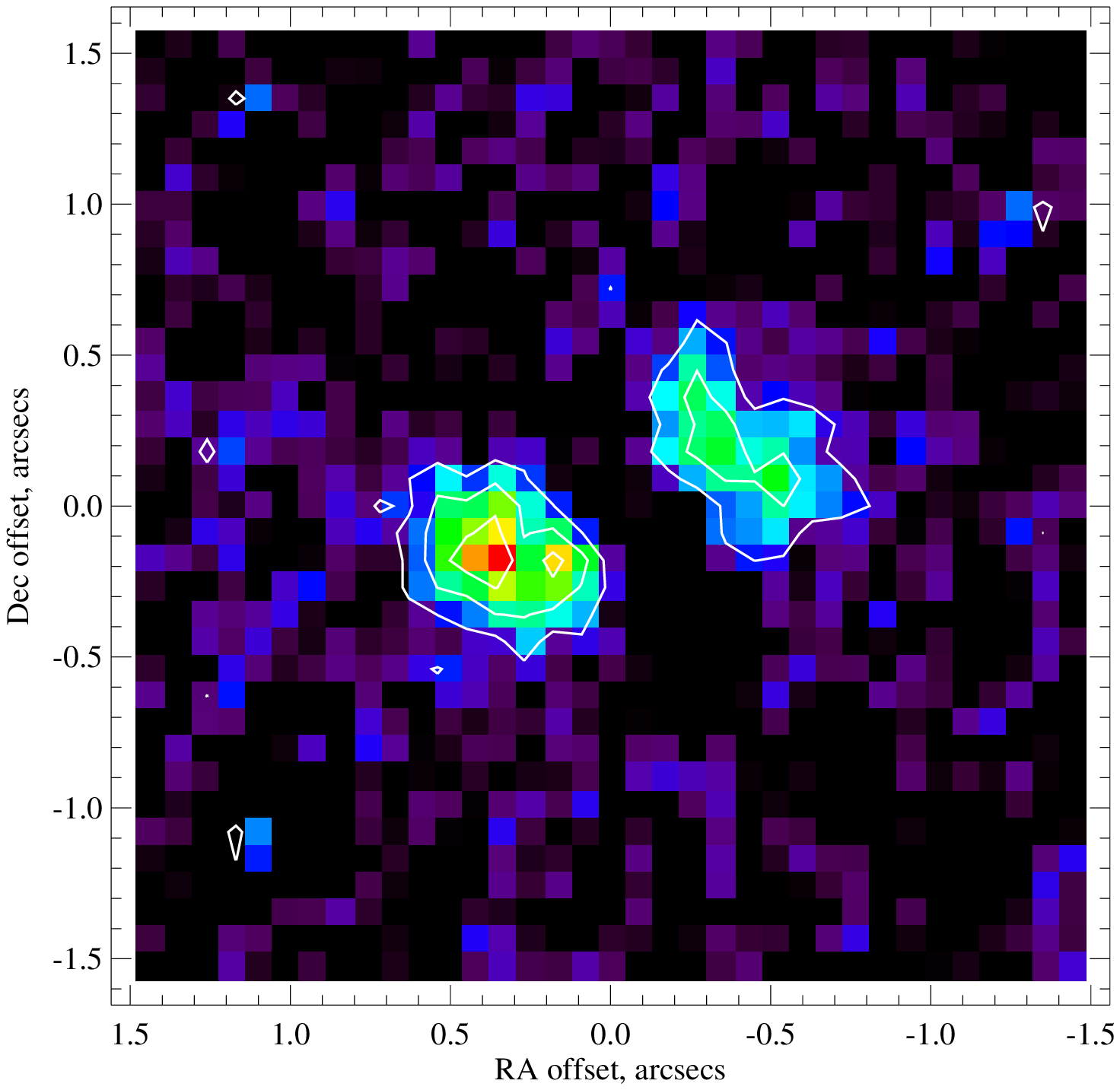}
\end{minipage}\\
\caption{\label{fig:172555trecs} The final co-added images for HD171759
  (the standard star) and HD172555 observed with TReCS.  The contours
  are at 10, 25, 50 and 75\% of the peak and show that there is no
  obvious evidence for extended emission beyond the size of the PSF (colour
  scale is linear with brightness).  The
  small area at $\sim$10\% of the peak offset from the source in the Q
  band standard star image is the peak of the Airy ring, seen also in
  the Q band image of HD172555. In the
  residual image, created by subtracting the final standard star image
  from the HD172555 image after scaling to the peak, there is no
  evidence for significant residual emission which might be a resolved
  disc in the N band, however we do see lobed emission in the Q band
  which could be evidence of an edge-on disc.  In these images black
  pixels are $<$1$\sigma$ per pixel. } 
\end{figure*}

In addition to examination of the images, surface brightness profiles
and FWHM measurements, we use the technique presented in
\citet{smithhot} to assess the limits of our detection capability for
extended emission.  For each photometric band we created a series of
disc models comprising rings of radius $r$ with ``widths'' $dr$ (so
that the inner radius of the disc would be $r - dr/2$, outer radius
$r+dr/2$) and different inclinations to the line of sight.
The rings were assumed to have constant surface brightness.  These
model images were added to point sources, 
representing the stars, and these sums then convolved with a PSF
model.  The final model images of the discs plus
stars were then subjected to the testing detailed in \citet{smithhot}.
In brief, this testing consisted of subtracting the point-like
emission (by scaling the PSF model to the peak of the image in the
same way as the residual images above were created), multiplying the
residual image by a mask, and testing this final image for emission
above the noise level of the image.  The masks blank all pixels apart
from a region of optimal size and shape to detect the disc emission
based on the geometry of the disc (optimal regions were determined by
extensive modelling of optimising disc detectability depending on
source geometry presented in \citealt{smithmidi}).  This
allowed us to determine what disc parameters (and levels of disc flux,
which were also varied) would have led to a detection of extended
emission.  \footnote{The science images
were also tested in the same way, subtracting both HD113766A and
HD113766B by subtracting scaled standard star images.  The emission
in the masked regions (using a full range of masks for a range of disc
geometries) was always found to be consistent with the noise levels in
the images. }
The resulting 3$\sigma$ extension limits on the location of
the excess emission are shown in Figure \ref{fig:ext}.  For a given
radius the limiting flux required for a 3$\sigma$ detection of
extended emission is shown for a disc centered on the star with median
radius at that radius, with a
full radial extent dependent on the geometry of the disc. The limits
suggest that for narrow ring models in the N band, with a flux equal
to the disc flux inferred from the IRS spectrum (1599mJy), any face-on or
edge-on discs greater than approximately 0\farcs13 (17AU) in radius
would have been detected in our data.  Similarly in the Q band, for
discs with 1867mJy of flux (determined by subtracting the predicted
photospheric emission from the Spitzer IRS spectrum of HD113766) any
extended disc larger than $\sim$0\farcs135 (18AU) in radius would
have been visible at the 3$\sigma$ level.  The photospheric
contribution was calculated from a scaled Kurucz model photosphere as
outlined in Table \ref{tab:sources}.  Detailed modelling by
\citet{lisse08} suggests that the warm emission comes from a region
$\sim$1.8AU (13mas)from the primary, and icy grains are situated in a
belt at 4--9AU (31--69mas) from the star.  Additional ice at 30--80AU
(23--61mas) in the \citeauthor{lisse08} model contributes to longer
wavelength excess but has very low emission at the VISIR
wavelengths. The limits from the VISIR imaging presented here are
consistent with the fit of \citet{lisse08}.

The imaging data of HD113766 would have detected extended disc
structure at the levels detected in the IRS spectrum on scales of
$>$17AU.  The lack of evidence for any extended emission in either the
N or Q band images therefore allows us to place an upper limit of 17AU
on the extent of dust emission around the source.  

\begin{table}
\caption{\label{tab:trecsimg} Observations imaging HD 172555 and a
  standard star (HD 171759) with TReCS, first published in
  \citet{moerchenastar}.} 
\begin{tabular}{cccc} \hline Filter & Int. time (s) & Object
  type & Star \\ \hline Si5 & 60 & Cal & HD171759 \\
Qa & 60 & Cal & HD171759 \\ Qa & 640 & Sci & HD172555 \\ Si5 & 640 &
Sci & HD172555 \\ Si5 & 60 & Cal & HD171759 \\ Qa & 60 & Cal &
HD171759 \\ \hline 
\end{tabular}
\end{table}

\subsection{TReCS imaging of HD172555}

\begin{figure*}
\begin{minipage}{6cm}
\includegraphics[width=6cm]{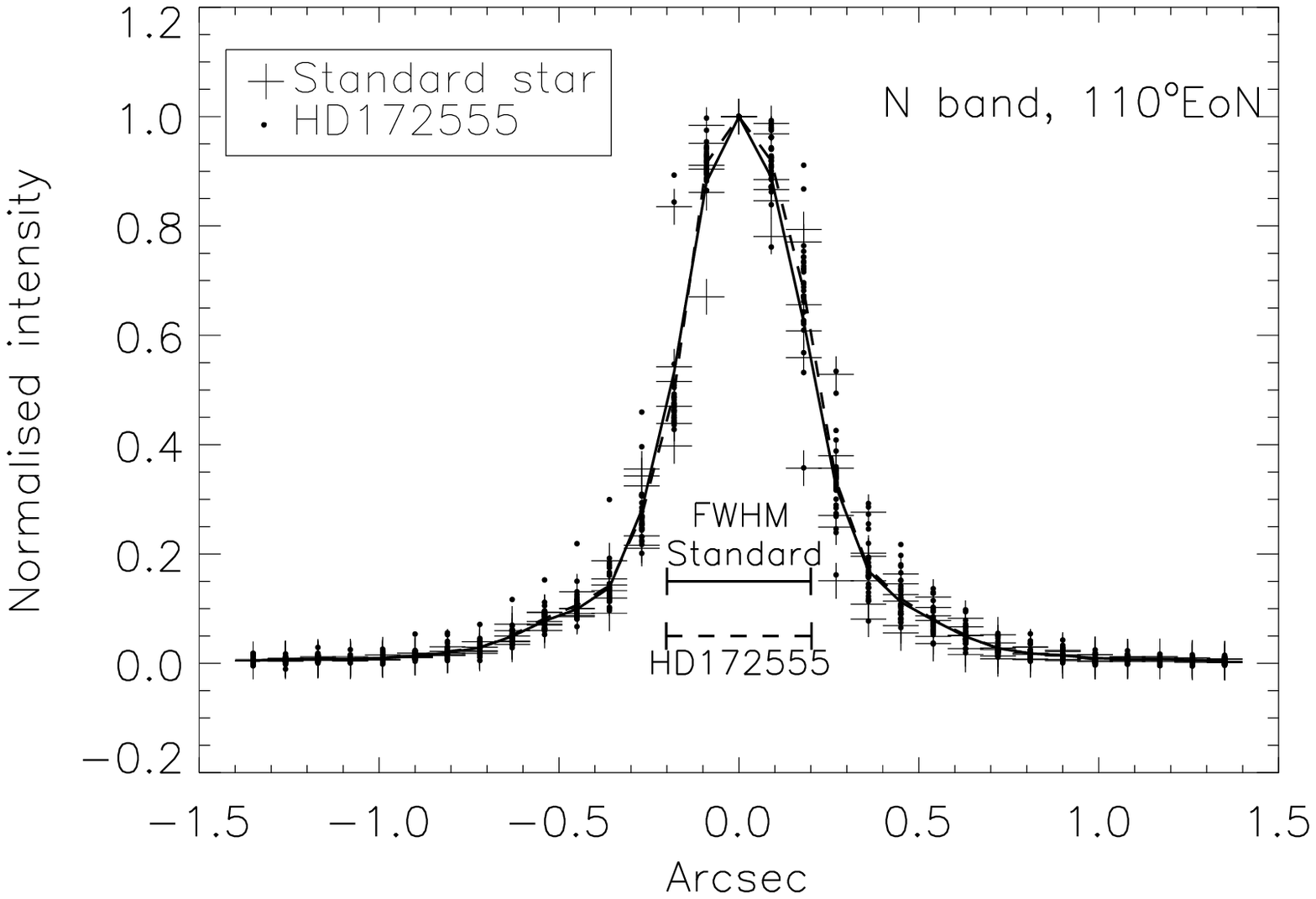} \\
\includegraphics[width=6cm]{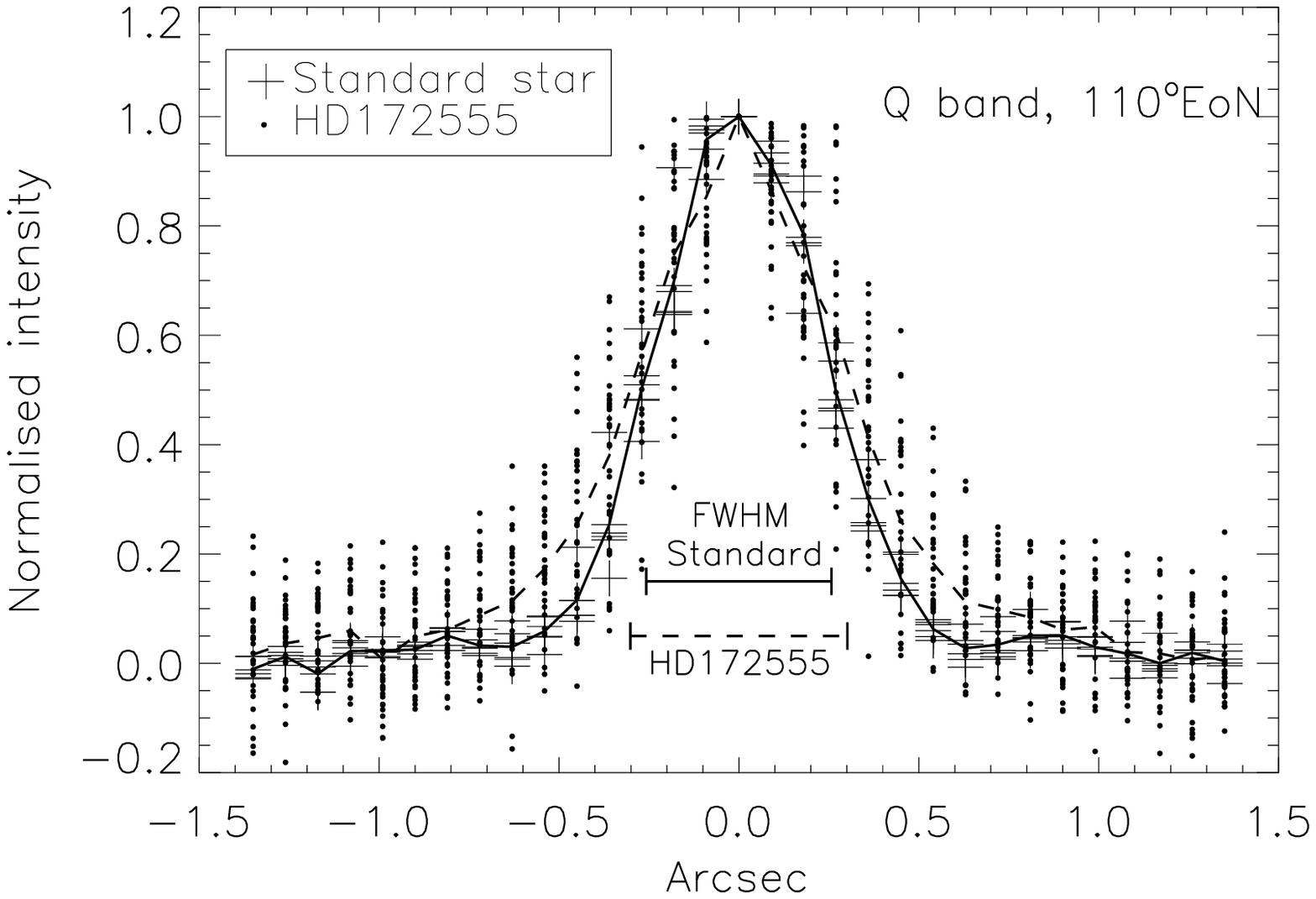} 
\end{minipage}
\begin{minipage}{6cm}
\includegraphics[width=6cm]{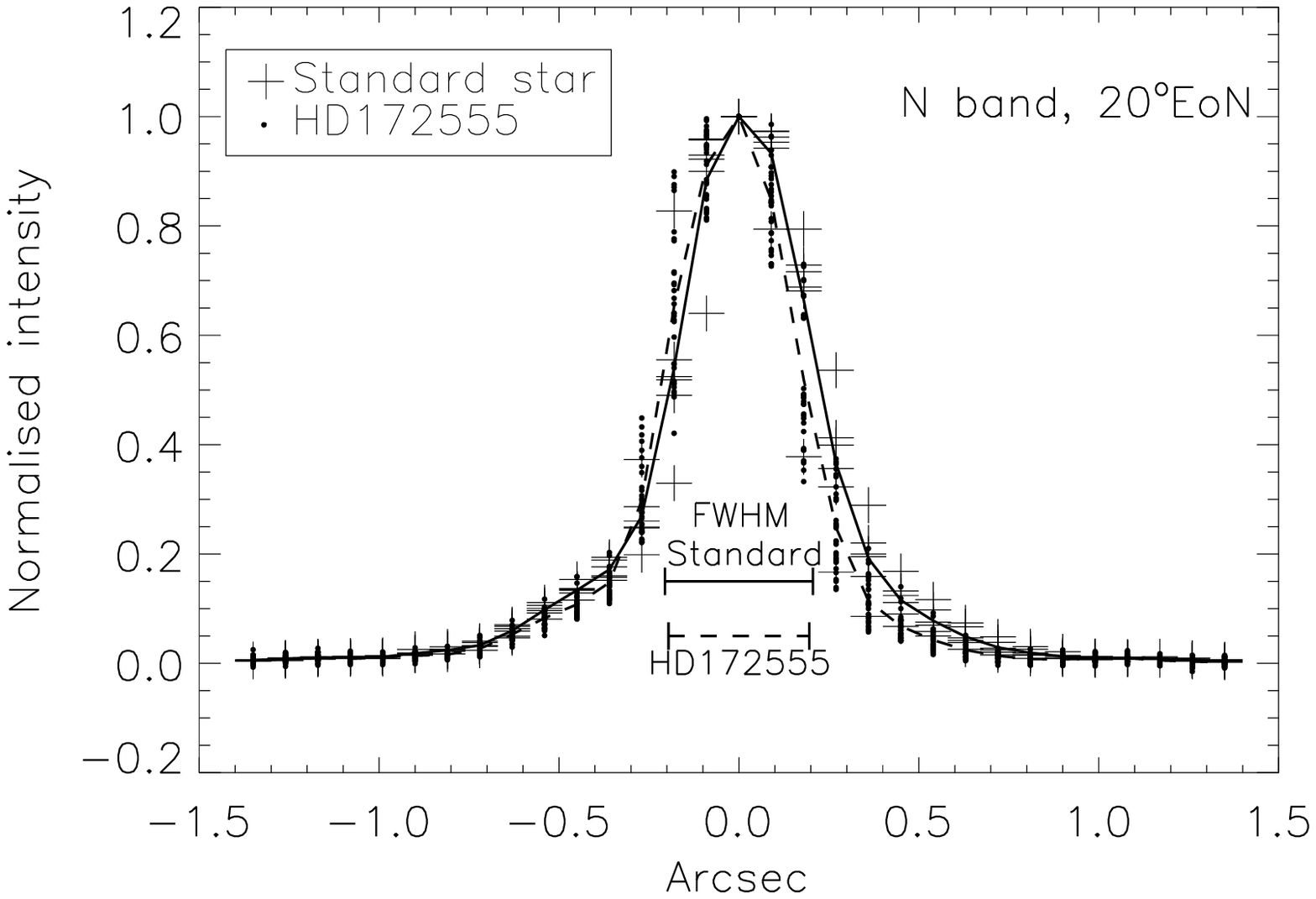} \\
\includegraphics[width=6cm]{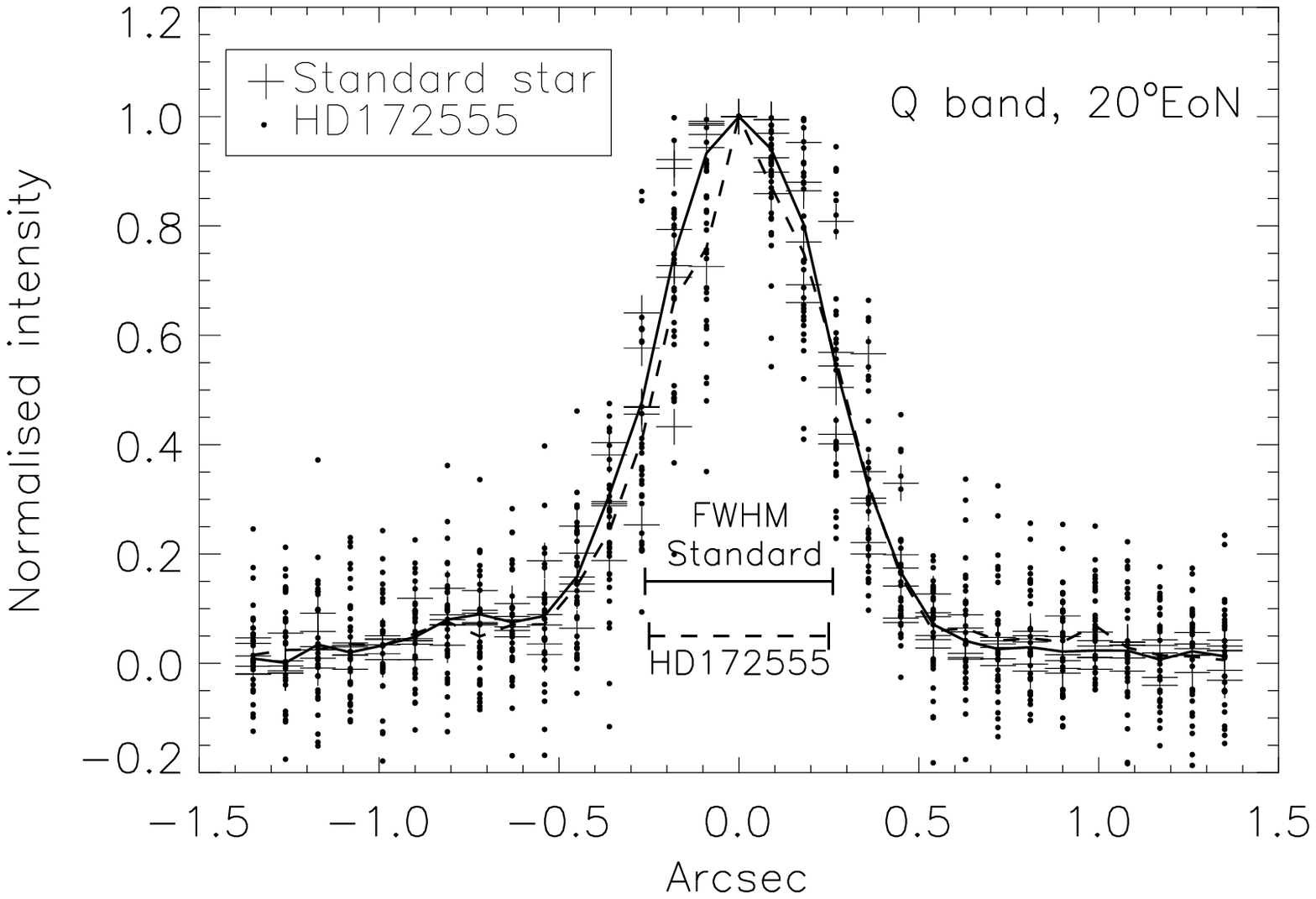} 
\end{minipage}
\begin{minipage}{4cm}
\caption{\label{fig:172555profs}  Profiles of line cuts through
  individual chop-nod integrations in the observations of HD172555 and
standard star HD171759.  The left-hand column shows line cuts at
110$^\circ$EoN, and the right-hand column shows cuts perpendicular (at
20$^\circ$EoN) to this.  There is no difference in the profiles of the
science and standard star target in the N band.  In the Q band the
profiles of the science target show greater variation, but overall the
profile of HD172555 is wider at 110$^\circ$EoN than the standard star
target.  The average profiles shown by solid (standard star) and
dashed lines (HD172555) are the mean taken over all individual
frames.  The FWHM of the average profile (determined through a 1-d
Gaussian fit to the profile) is shown by a bar in the same linestyle.
}
\end{minipage}
\end{figure*}

We do not have VISIR imaging of HD172555, but recently 8m-imaging
data of this target has been presented by \citet{moerchenastar}.  In
680s of on-source integration with TReCS in both the Si-5
($\lambda_c=11.66\mu$m, hereafter N band) and Qa
($\lambda_c=18.3\mu$m, hereafter Q band) filters the authors found no
evidence for extended emission.   However,
  \citet{pantin_172555} recently presented Lucky Imaging of this
  target with VISIR which suggested that HD172555 appears extended in
  the Q band, although not in the N band.  To explore this issue
  further, we have obtained the TReCS raw data presented in
  \citet{moerchenastar} from the Gemini Science Archive.  The data
  were reduced using the same custom procedures used for our VISIR
  imaging of HD113766.  The science observations were calibrated using
observations of the standard star HD171759, taken immediately before
and after the science observations.  A summary of the observations is
given in Table \ref{tab:trecsimg}.  In contrast to the VISIR imaging,
the TReCS observations were performed with a parallel chop-nod
pattern.  As the off-beams are unguided which can effect the shape of
the PSF, we only use the guided images of the targets in our analysis.
Sub-integrations on the targets were co-added using the Gaussian
centering techinque employed for the VISIR observations of HD113766.
Aperture photometry centered on the Gaussian peaks of the final images
was performed using 1\arcsec radius apertures (noise levels were
determined from an annulus with inner radius 2\arcsec and outer radius
4\arcsec). The calibrated flux found in the final images was
1120$\pm$67mJy for HD172555 at N and 1039$\pm$85mJy at Q (errors
include calibration errors of 3\% and 8\% respectively determined from
variation in calibration factors between the two standard star
observations in each filter).  These values are consistent with those
quoted in \citeauthor{moerchenastar} (2010; values were listed as
1155$\pm$116mJy at N and 1094$\pm$164mJy at Q including fiducial 10\%
and 15\% calibration uncertainties).

The final images for the PSF reference (standard star) and
  HD172555 are shown in Figure \ref{fig:172555trecs}.  The final
  images are shown with contours at 10, 25, 50 and 75\% of the peak.
  There is no evidence for extension in the N band image of HD172555.
  In the Q band image we see some evidence of greater ellipticity in
  the image of HD172555 than is seen in the PSF reference image.  This
  is confirmed in the residual image, which is created by subtracting
  from the science image the PSF reference image scaled to
  the peak of the science image.  The residual image is shown in the
  right-hand column of Figure \ref{fig:172555trecs}. We see two clear
  lobes of extended emission in the Q band residual image aligned
  along a position angle of 110$^\circ$EoN, but no
  significant emission in the N band residual image.  The total flux
  subtracted from the Q band image of HD 172555 to obtain the residual
image is 934 mJy, much higher than the predicted flux from the stellar
photosphere in this filter of 202mJy.  

\begin{figure*}
\begin{minipage}{6cm}
\includegraphics[width=6cm]{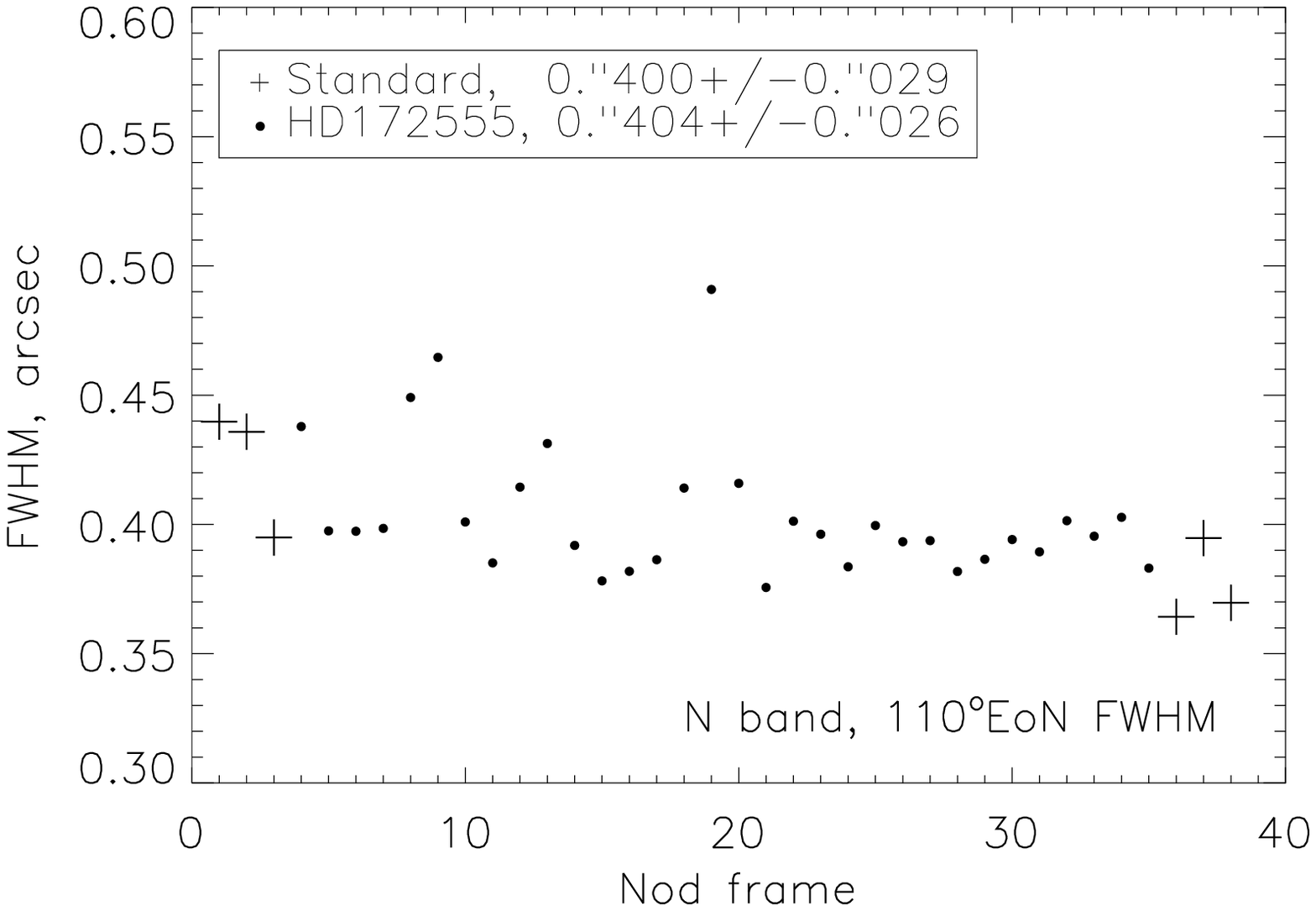} \\
\includegraphics[width=6cm]{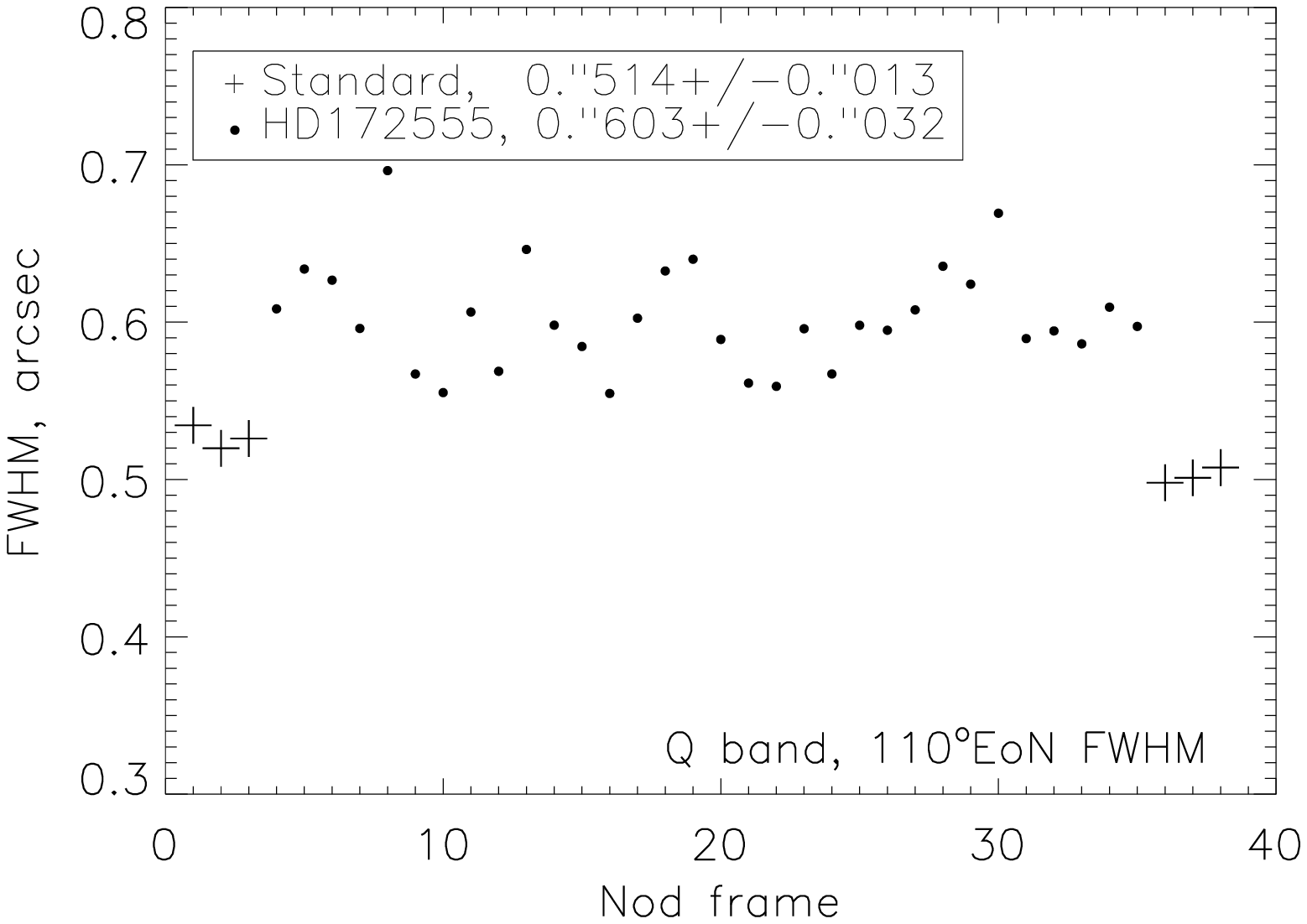} 
\end{minipage}
\begin{minipage}{6cm}
\includegraphics[width=6cm]{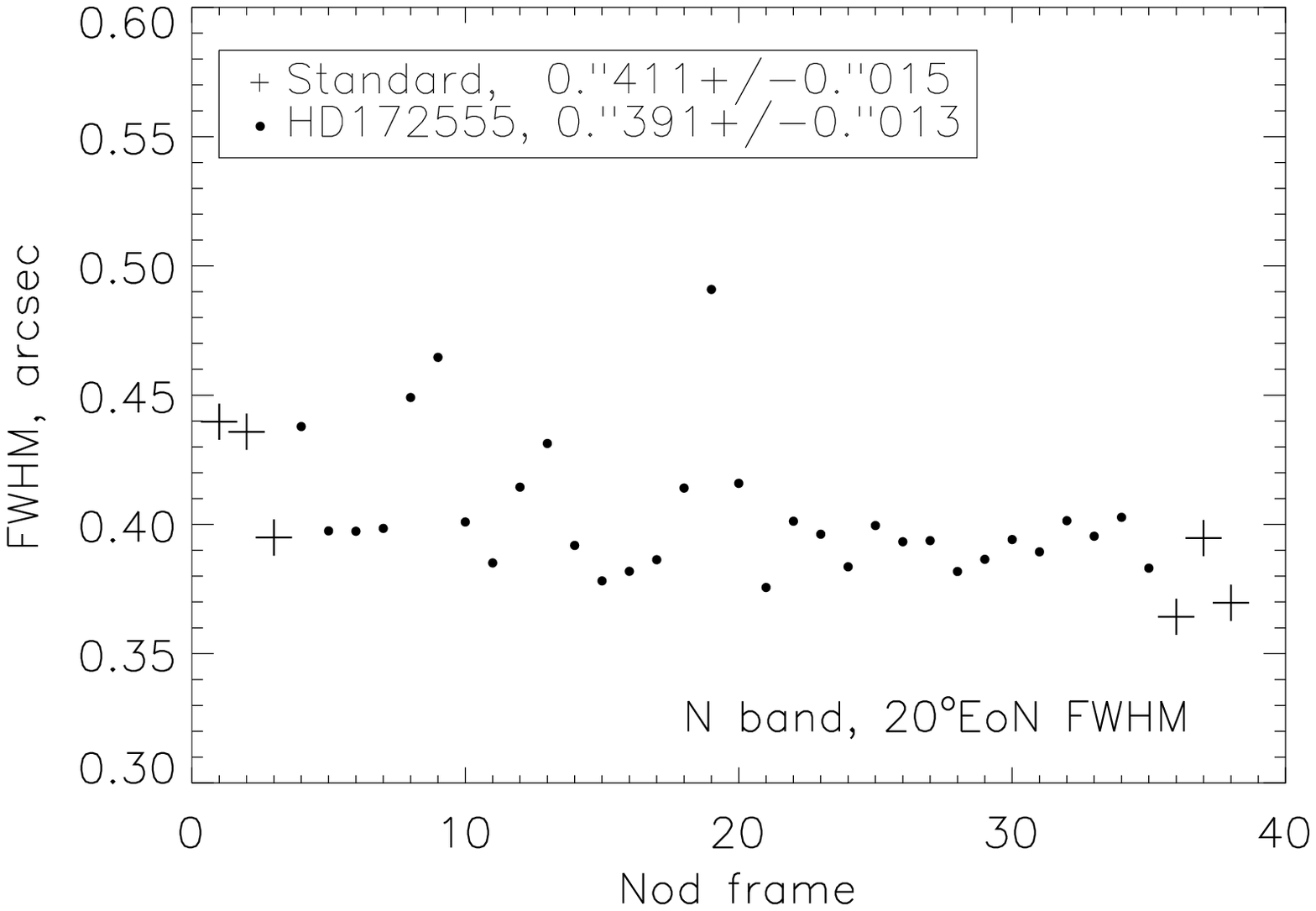} \\
\includegraphics[width=6cm]{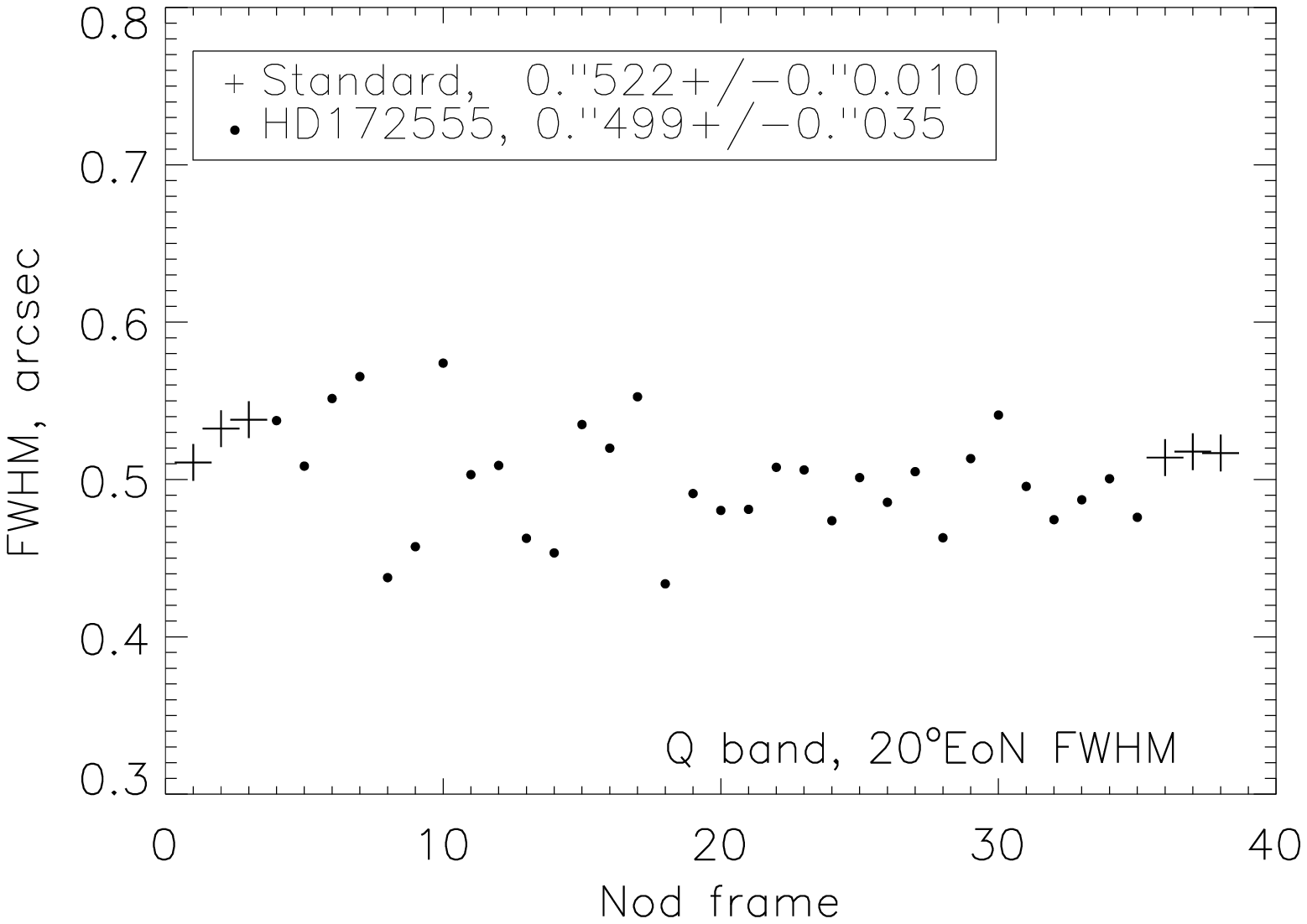} 
\end{minipage}
\begin{minipage}{4cm}
\caption{\label{fig:172555fwhms}  Measurements of the FWHM of the
  image profiles of HD172555 and standard star HD171759 taken from
  line cuts through single chop-nod integrations centered on the
  stars.   The FWHM values were determined from a Gaussian 1-d fit to
  each profile.  Mean and rms errors on the FWHM values are
    listed in the legends. For the Q band profiles at 110$^\circ$EoN
  the science target HD172555 has a larger FWHM in all frames than the
  standard star target.  In the N band and the Q band at 20$^\circ$EoN
  the FWHM of HD172555 are consistent with the FWHMs measured for the
  standard star observations. }
\end{minipage}
\end{figure*}

To test whether the emission we observed is truly significant, or the
result of a varying PSF, we examined the science and standard star
image profiles in sub-integrations (single chop-nod cycles) on the
targets.  We have 6 sub-integrations on the standard star object and
32 on HD172555 in each band.  Taking a strip centered over the peak of
the image of 3 pixels' width
(0\farcs09/pixel) and averaging across this width we create a 1-d
profile of the image for each sub-integration.  These profiles are
taken at 110$^\circ$ EoN (to coincide with the residual emission peaks
at Q) and 20$^\circ$ EoN (perpendicular to the residual peaks).  The
profiles are shown in Figure \ref{fig:172555profs}.  We see that in
the N band there is no evidence that the profiles of HD172555 are more
extended than the profiles of the standard star target.  In the Q band
the image of HD172555 is extended in the 110$^\circ$EoN direction, but
not at 20$^\circ$ EoN.  This suggests that if the extended emission we
are viewing arises from a disc, then we are viewing the disc close to
edge-on, or it is a highly elliptical disc.  The profiles in the Q
band of the science target are quite noisy.  As a final test of the
extension detection, we examine the FWHMs of the sub-integration
profiles.  The FWHMs are determined by fitting a 1-dimensional Gaussian
to each profile.  The results are shown in Figure
\ref{fig:172555fwhms}.  It is clear from this plot that the FWHMs of the
HD172555 images are always larger than the FWHM of the standard star
images at 110$^\circ$EoN in the Q band.  At 20$^\circ$ EoN and at both
angles in the N band, the FWHMs of HD172555 are consistent with those
measured for the standard star target.  

\begin{figure}
\includegraphics[width=7cm]{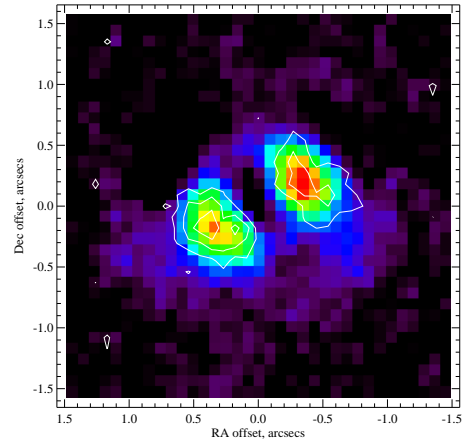}
\caption{\label{fig:172555extmodel} The best fitting disc model for
  the Q band extended emission seen around HD 172555, shown after
  subtraction of the PSF model for direct comparison to the residual
  image in Figure \ref{fig:172555trecs} (colour scale and contour
  levels the same).  The contours overplotted are those from the HD
  172555 residual image, and show that the residual peaks of the model
  are of a similar size and shape.  }
\end{figure}

The disc models used in the extension limits testing for HD
  113766 were then used to determine the approximate size of the
  extended emission.  The models (of varying size, thickness,
  inclination and rotated to different position angles, see previous
  section 2.1 for a description) were added to point sources
  representing the star and any unresolved excess and convolved with
  the PSF (the standard star image).  These model images were compared
  to the image of HD172555 by subtracting the model image from the
  science image (scaling to the peak) and comparing the residuals with
  the noise on the image.  Using a $\chi^2$ calculation to determine
  the best fitting model we find that a disc of radius 0\farcs27 (7.9AU
  at 29.2pc), width $dr = 1.2r$, inclined at 75$^\circ$ to the
  line of sight lying at a position angle of 120$^\circ$ EoN provides
  the best fit to the data.  In this best-fitting model 65\% of the
  total flux arises from the extended emission, suggesting a
  point-like flux of $\sim$ 363mJy, exceeding the 202mJy predicted to
  arise from the star.  The best fitting model image is shown, after
  subtracting the scaled PSF image, in Figure
  \ref{fig:172555extmodel} which can be directly compared to the
  residual image in Figure \ref{fig:172555trecs}.  The contours plotted on
  this figure are from the HD172555 residual image, and show that the
  model does indeed reproduce the main features of the extended
  emission.  To determine limits on the radius of the disc
    model, we sum the $\chi^2$ values over all values of the other
    parameters tested ($dr$, inclination, position angle and flux in
    the disc).  Using the percentage points of the $\chi^2$
    distribution we find that discs with radii 0\farcs09 $< r
    <$0\farcs31 (2.6--9.1AU) fit the observed emission within
    3$\sigma$.  Using the same technique for the other model
    parameters we obtain the following limits: $0.36r < dr < 2r$;
    inclination $>47^\circ$; $40^\circ$ EoN $<$ position angle $<
    130^\circ$ EoN; 14\% $<$ percentage of total flux in disc $<$
    74\%.  


\begin{figure}
\includegraphics[width=7cm]{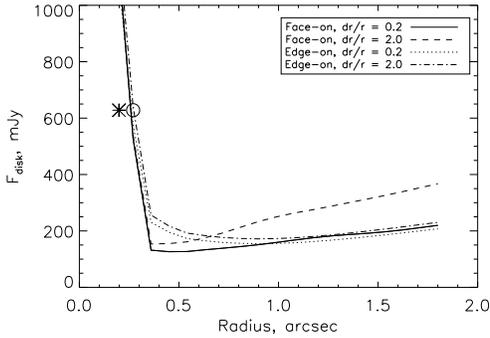}
\caption{\label{fig:ext172555} The limits on the disc location around
  HD172555 in the N band based on our analysis of archival TReCS
  data.  
  Different source geometries are indicated by different line styles
  as given in the legend.  The regions above the lines represent discs
  that would have been detected at the 3$\sigma$ level or higher.
  Thus the regions below the lines represent the possible disc
  parameter space, given our non-detection of extended emission. The
  asterisk marks the predicted disc location according to the fit by
  \citet{lisse09}. At this level of emission (635 mJy at N) 
we would have expected to detect any disc larger than
0\farcs27.  The radial offset of the best fitting model to the
extended emission observed in the Q band is shown by a circle.  If the
material dominating the N band emission was coincident with the
material which dominates the emission at Q we would have expected to
detect extension in the N band image. }
\end{figure}

The focus of this paper is the size and geometry of the dust
emitting at 10$\mu$m.  As we see no extension in the N-band image, we
use the same procedures adopted in the VISIR imaging of HD113766 to
place a limit on the size and geometry of the HD172555 disc in the N
band.  The results are shown in Figure \ref{fig:ext172555}. These
limits suggest that the emission in the N band lies at $<$0\farcs27
(7.9AU) and thus if the material dominating the N band
  emission was coincident with that dominating the Q band extended
  emission we would have expected to detect extension, making this
  result a significant non-detection.
The size limit matches the size of the best fitting model to the
extended emission detected in the Q band image.  As this model showed
some evidence for residual unextended emission, this may suggest that
there are two populations of dust in this system.  In this case the
excess around HD172555 is a possible multiple component disc like that
seen around other main sequence stars (e.g. $\eta$ Tel
\citealt{smitheta}).  However, it is also possible that we are
observing a more extended disc, and that the N and Q band data are
probing the inner and outer parts of the distribution due to their
greater sensitivity to hotter and cooler dust in the disc
respectively.  This would also account for the apparent unresolved
component in the Q band image, as there would still be some emission
arising from inner part of the disc observable in the Q band.

\subsection{VISIR spectroscopy} 

Detailed spectra of both our science targets have been obtained with IRS
on the Spitzer Space Telescope \citep{chen06}.  As discussed in
\citet{smithmidi} the measurements of photometry with MIDI suffer from
poor background subtraction.  It is helpful therefore to use
the IRS spectra as reference total photometry for comparison to the
correlated fluxes measured with MIDI.   However, these spectra
were obtained in a much broader slit than is used in MIDI observations
(3\farcs7 or 4\farcs7 used in IRS; MIDI slit is 0\farcs52 wide).  It
is thus possible that extended emission caught in the IRS slit would
not be observed in the MIDI slit.  The binary companion to HD113766
must also be ruled out as a source of excess emission. 

\begin{figure*}
\begin{minipage}{7cm}
\includegraphics[width=7cm]{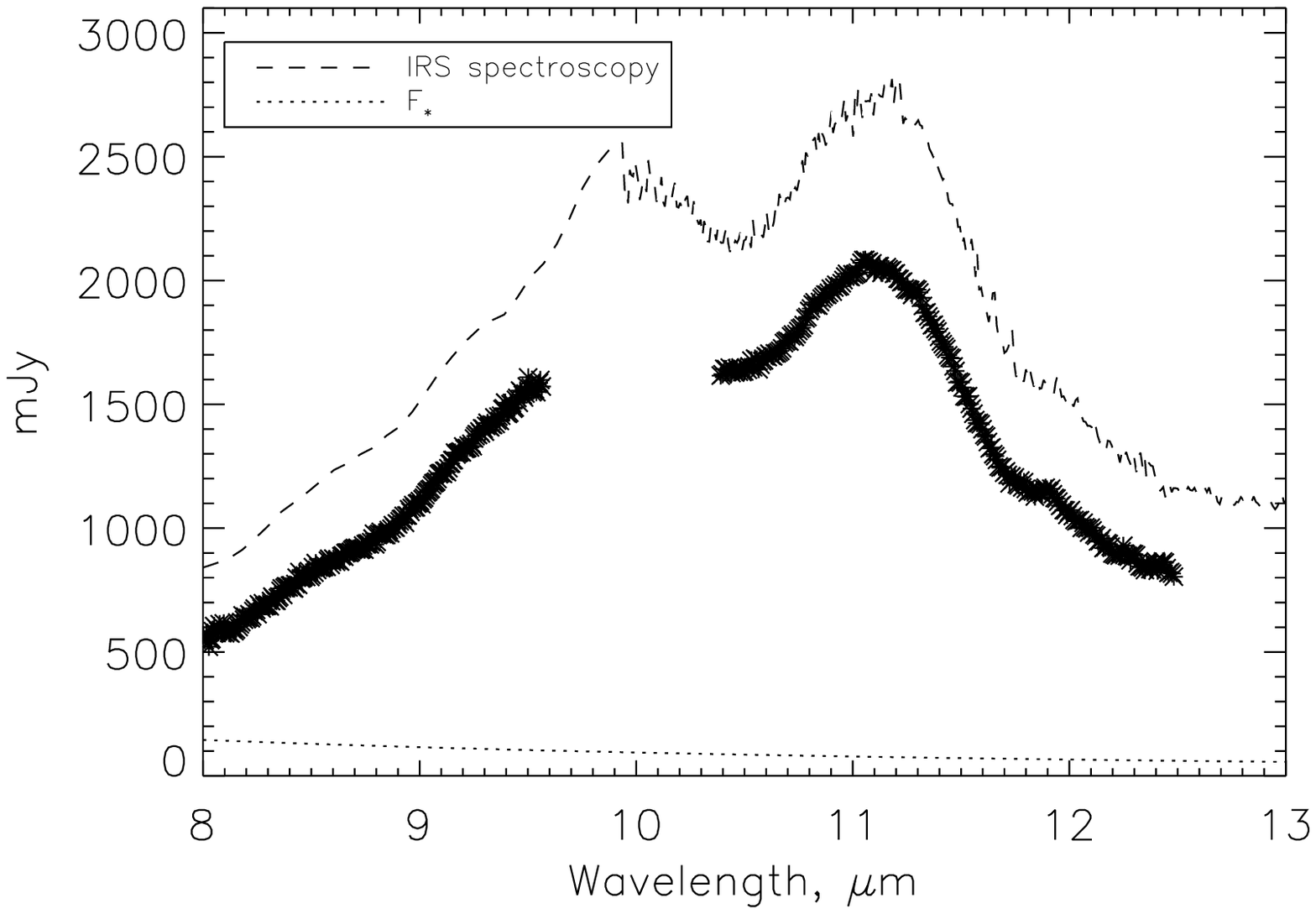}
\end{minipage}
\begin{minipage}{7cm}
\includegraphics[width=7cm]{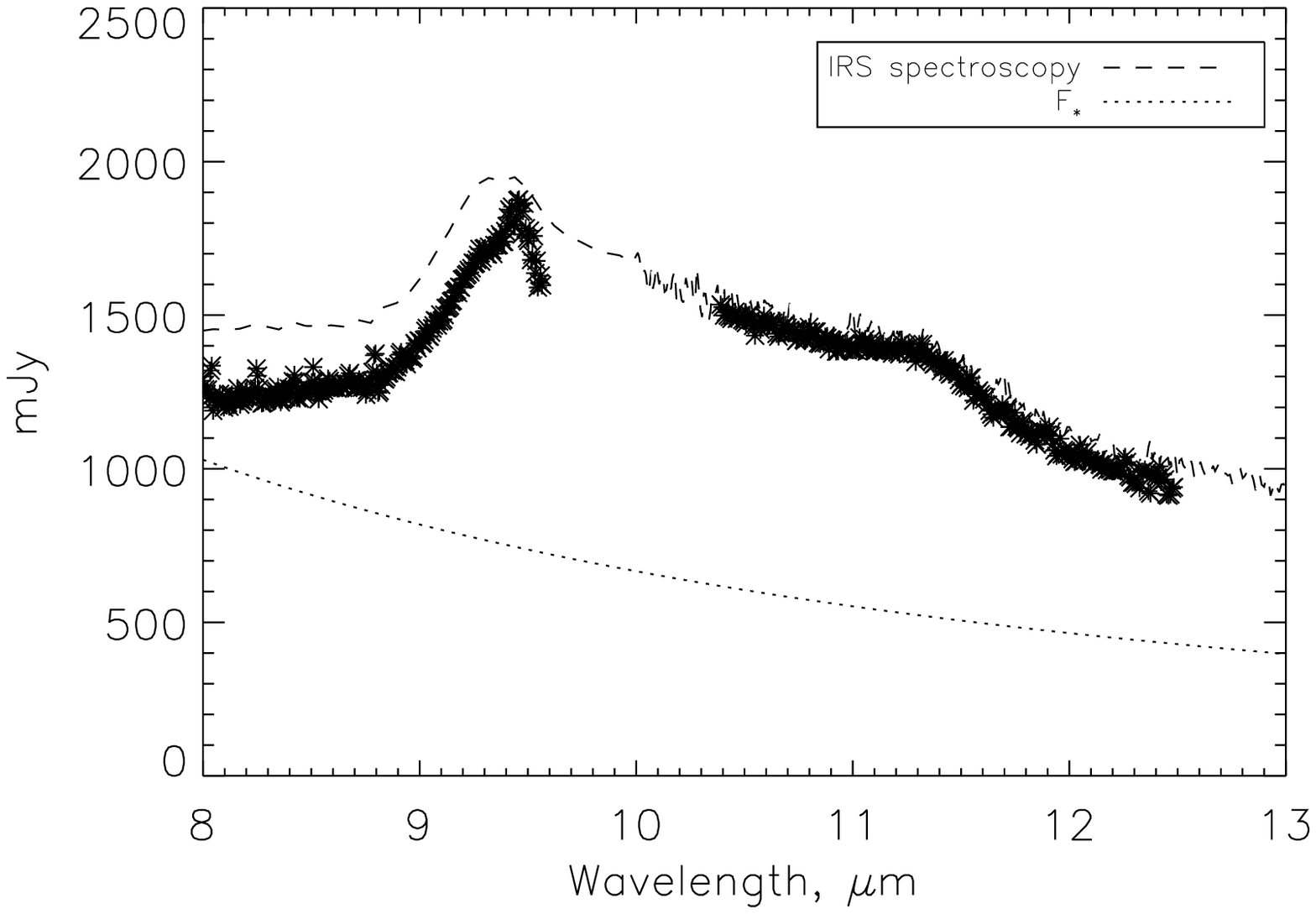}
\end{minipage} \\
\caption{\label{fig:visirspec} The VISIR spectroscopy of HD113766
  (left) and HD172555 (right). The IRS spectra of these targets are
  shown for comparison. The stellar emission as calculated from a
Kurucz model photosphere of appropriate spectral type scaled to the
2MASS K band emission of the star is shown as $F_\star$.}
\end{figure*}

To avoid any biasing of the interferometric visibilities which could
arise from assuming 
a higher total flux than falls within the MIDI slit, we obtained VISIR
spectroscopy of both science targets in low resolution mode
(R$\sim$350 at 10$\mu$m) with a 0\farcs75 slit (program ID
083.C-0775(E)).  Two filters were used to examine the short and longer
wavelength ranges covered by MIDI ($\lambda_c$ = 8.8$\mu$m range
8--9.6$\mu$m, hereafter filter 8.8; $\lambda_c$ = 11.4$\mu$m range
10.43--12.46$\mu$m, hereafter filter 11.4).  The
observations were performed in chop-nod mode with standard star
observations taken immediately before and after each science
observation (Table \ref{tab:visir} summarises the observations).  

\begin{table}
\caption{\label{tab:visir} VISIR low-resolution spectroscopy} 
\begin{tabular}{cccc}\hline Date & Target & Filter & Int. time (s)
  \\ \hline 11th May 2009 & HD111915 & 8.8 & 240 \\ 
11th May 2009 & HD113766 & 8.8 & 900 \\ 
11th May 2009 & HD111915 & 8.8 & 240 \\ 
11th May 2009 & HD111915 & 11.4 & 240 \\ 
11th May 2009 & HD113766 & 11.4 & 600 \\ 
11th May 2009 & HD111915 & 11.4 & 240 \\ \hline
25th May 2009 & HD156277 & 8.8 & 240 \\ 
25th May 2009 & HD172555 & 8.8 & 900 \\ 
25th May 2009 & HD156277 & 8.8 & 240 \\ 
7th July 2009 & HD156277 & 11.4 & 240 \\
7th July 2009 & HD172555 & 11.4 & 600 \\
7th July 2009 & HD156277 & 11.4 & 240 \\ \hline
\end{tabular}
\end{table}

The observations were reduced with the VISIR pipeline procedures
available at
http://www.eso.org/sci/data-processing/software/pipelines/.
Calibration was performed using an average of the two standard star
observations obtained either side of the science observations.  The
standard stars fluxes were taken from \citet{cohen} models for the
mid-infrared standards used.  We plot the observed spectra in Figure 
\ref{fig:visirspec}.  

The VISIR spectroscopy of HD113766 was performed on one night, with
the observation block (standard star HD111915, science target HD113766
and the standard star) in filter 11.4 taken immediately after the
observation block in filter 8.8.  The final observed spectrum looks
very similar in shape to the IRS spectrum of HD113766, although it is lower
everywhere by a factor of 1.3.  This difference is not due to the binary
component which fell within the slit of the IRS spectroscopy but
outside the slit in the VISIR observations, because this source contributes
only an average of 2\% to the total flux across the 8--13$\mu$m range
(from scaled Kurucz model photosphere as described in Table
\ref{tab:sources} and the previous subsection).  \citet{geers} found
that when using Spitzer IRAC photometry to calibrate VISIR N band
spectra errors in absolute calibration could be as large as 30\%,
consistent with the difference between the VISIR and IRS spectra of
HD113766 seen here.  We also obtained archival IRS spectra of
  the standard star HD 111915.  As the spectral ranges of VISIR and
  IRS are different, this allowed a comparison of the VISIR and IRS
  spectra of the standard star in the longer wavelength observation
  with VISIR only.  A comparison of the IRS spectra reduced using
  pipeline routines shows that the calibration factor was varying over
the course of the longer wavelength observation (we cannot test the
stability of the calibration for the shorter wavelength observations
as there is no overlap with the IRS spectra).
The difference between the IRS and VISIR spectra was a factor of 1.07 and
1.15 for the VISIR calibration observations taken before and after the
science observation, although again the shape of the spectra was
consistent between the two.  Although these differences are not as large as
the factor of 1.3 observed for the science target, the difference
between the two indicates that the absolute calibration was quite
unstable during these observations. 
Within the errors of absolute calibration for
VISIR, we see no evidence for extended emission detected in the IRS
spectroscopy that would fall outside the MIDI slit.  It is also worth
noting that we do not see evidence for temporal evolution in the
emission (within the calibration errors).  The prospect of temporal
evolution of the excess emission around HD 69830 (another star with
bright levels of excess in the terrestrial planet region) has recently
been ruled out \citep{beichman69830}, but this remains a possibility
for a star undergoing planet forming collisions.  Photometry from the IRAS
database (obtained in 1983; \citealt{beichman88}) gives a flux of
1590$\pm$80mJy at 12$\mu$m. The Spitzer Space telescope IRS flux
averaged over the finite bandwidth of the VISIR N band filter is 
1803$\pm$92 mJy (data obtained in 2004; 
\citealt{chen06}).  The VISIR photometry taken just 2 years prior to
the VISIR spectroscopy presented in 
section 2.1 (1673$\pm$42mJy for HD113766A at N) is also consistent
with a constant level of excess emission.  

For HD172555 the difference between between the IRS spectrum of the
target and that measured with VISIR is an average of 3\% in filter
11.4, consistent with the variations observed between different
standard star observations (measured to be 5\%).  The observations
in filter 8.8 were taken on a different night,  and the difference
between the IRS spectrum and VISIR spectrum is found to be closer to
13\%.  Taken separately, the two sections of the VISIR spectrum again
appear to be simply scaled versions of the IRS spectrum.  The apparent
sharp slightly offset peak in the VISIR spectrum at $\sim$9.5$\mu$m is
consistent with the flatter peak between 9.2--9.5$\mu$m seen in the
IRS spectrum within the uncertainty on the spectrum.  The
difference in calibration on different nights is the likely cause of
the difference in scale factors for the observations in filters 8.8
and 11.4.  In both cases the differences are within the 30\% absolute
calibration uncertainty found by \citet{geers}.  There are no
  IRS observations of the standard star HD 156277 in the Spitzer
  archive. 

Although the errors in absolute calibration for the VISIR spectroscopy
are large, the shape of the spectra for HD113766 and HD172555 agree
with the shape of the IRS spectra for both targets.  We would expect
cooler emission to be further from the star and thus more likely to be
excluded from the VISIR spectroscopy.  As we do not see a preferential
loss of flux in the VISIR spectra as compared to the IRS spectra at
long wavelengths, there is no evidence on the basis of the spectroscopy
for extended emission that would fall outside the MIDI slit being
detected in the IRS spectra.  The absolute offsets between the VISIR
and IRS spectra are consistent with an expected uncertainty of 30\%,
and there is no evidence for temporal evolution in the flux levels
from near contemporaneous measurements.  
The IRS spectra of both targets shall therefore be used as a measure
of the expected total photometry for comparison to the correlated
fluxes measured with MIDI, however the implications if the VISIR
spectrum had measured the true flux of the HD 113766 system are
discussed briefly in sections 3.2 and 4.3.1.

\section{The MIDI observations}

Observations were taken over several semesters through a combination of 
service and visitor mode observations.  Table \ref{tab:midiobs} lists the 
observing run IDs and dates of all observations, together with the 
baseline configurations used.  All observations used MIDI on the UTs in 
the HIGH-SENS mode where the interferometric fringe exposures are followed 
by separate photometric exposures from each telescope in turn to measure 
the target spectrum through each beamline (for a summary see 
\citealt{smithmidi}; further details can be found in the MIDI instruction 
manual or in \citealt{tristam}).

\begin{table*}
\caption{\label{tab:midiobs} Observations of science and calibrator
  targets with MIDI}
\begin{tabular}{llcccllccc} \hline Date & Observing & Baseline &
  Baseline & Baseline & Target & Target & Seeing & $\tau_0$ & Flux
  \\ & ID & configurations & length (m) & Pos. Angle ($^\circ$) & name
  & type & \arcsec & ms & RMS \\ \hline 08/03/2007 & 078.D-0808(D) &
  UT1-UT3 & 102.2 & 36.52 & HD116870 & Cal & 1.76 & 2.0 & 0.0085 \\ 
08/03/2007 & 078.D-0808(D) & UT1-UT3 & 92.8 & 0.97 & HD172555 & Sci &
0.70 & 6.2 & 0.0034 \\ \hline
09/04/2007 & 079.C-0259(G) & UT1-UT3 & 101.8 & 2.8 & HD113766 & Sci &
0.72 & 3.1 & 0.0024 \\
09/04/2007 & 079.C-0259(G) & UT1-UT3 & 102.2 & 9.0 & HD112213 & Cal &
0.83 & 2.6 & 0.0053 \\
10/04/2007 & 079.C-0259(G) & UT1-UT3 & 101.8 & -2.7 & HD113766
& Sci & 0.59 & 3.0 & 0.0362 \\
10/04/2007 & 079.C-0259(G) & UT1-UT3 & 102.3 & 3.0 & HD112213 & Cal &
0.74 & 2.8 & 0.0028 \\ \hline
30/05/2007 & 079.C-0259(F) & UT1-UT2 & 48.7 & 41.7 & HD112213 & Cal &
1.52 & 1.0 & 0.0129 \\ 
30/05/2007 & 079.C-0259(F) & UT1-UT2 & 42.6 & 49.0 & HD113766 & Sci &
1.73 & 0.7 & 0.0297 \\ \hline
18/03/2008 & 080.C-0737(C) & UT3-UT4 & 57.9 & 85.8 & HD156277 & Cal &
0.81 & 5.7 & 0.0034 \\ 
18/03/2008 & 080.C-0737(C) & UT3-UT4 & 56.1 & 78.8 & HD172555 & Sci &
0.59 & 7.6 & 0.0020 \\ \hline
20/03/2008 & 080.C-0373(D) & UT1-UT3 & 88.9 & 20.4 & HD156277 & Cal &
0.88 & 6.4 & 0.0029 \\
20/03/2008 & 080.C-0373(D) & UT1-UT3 & 92.6 & 7.8 & HD172555 &
Sci & 0.61 & 9.7 & 0.0021 \\
20/03/2008 & 080.C-0373(D) & UT1-UT3 & 100.7 & 12.8 & HD169767 & Cal &
0.51 & 9.0 & 0.0026 \\
20/03/2008 & 080.C-0373(D) & UT1-UT3 & 92.0 & 14.5 & HD172555 &
Sci & 0.49 & 10.3 & 0.0025 \\ \hline
21/03/2008 & 080.C-0737(E) & UT2-UT4 & 89.3 & 60.8 & HD156277 & Cal &
0.82 & 4.6 & 0.0029 \\
21/03/2008 & 080.C-0737(E) & UT2-UT4 & 88.1 & 45.6 & HD172555 &
Sci & 0.71 & 4.8 & 0.0034 \\
21/03/2008 & 080.C-0737(E) & UT2-UT4 & 89.4 & 70.7 & HD156277 & Cal &
0.79 & 5.8 & 0.0023 \\ \hline
07/05/2009 & 083.C-0775(C) & UT3-UT4 & 58.7 & 90.0 & HD171759 & Cal &
1.29 & 1.8 & 0.0037 \\
07/05/2009 & 083.C-0775(C) & UT3-UT4 & 59.3 & 95.5 & HD172555 &
Sci & 1.10 & 1.8 & 0.0044 \\
07/05/2009 & 083.C-0775(C) & UT3-UT4 & 60.8 & 103.9 & HD171212 & Cal &
1.10 & 1.7 & 0.0040 \\
07/05/2009 & 083.C-0775(C) & UT3-UT4 & 60.2 & 102.7 & HD171759 & Cal &
1.14 & 2.0 & 0.0036 \\
07/05/2009 & 083.C-0775(C) & UT3-UT4 & 60.9 & 108.0 & HD172555 &
Sci & 0.95 & 2.0 & 0.0044 \\
07/05/2009 & 083.C-0775(C) & UT3-UT4 & 62.4 & 135.4 & HD152186 & Cal &
0.82 & 2.0 & 0.0035 \\
07/05/2009 & 083.C-0775(C) & UT3-UT4 & 61.2 & 114.7 & HD171759 & Cal &
0.98 & 2.3 & 0.0027 \\
07/05/2009 & 083.C-0775(C) & UT3-UT4 & 62.1 & 121.4 & HD172555 &
Sci & 1.49 & 1.5 & 0.0039 \\
07/05/2009 & 083.C-0775(C) & UT3-UT4 & 62.4 & 128.2 & HD171212 & Cal &
1.14 & 1.9 & 0.0024 \\
07/05/2009 & 083.C-0775(C) & UT3-UT4 & 62.0 & 130.2 & HD171759 & Cal &
0.76 & 2.9 & 0.0026 \\ \hline
08/05/2009 & 083.C-0775(D) & UT1-UT3 & 86.2 & 49.6 & HD112213 & Cal &
0.47 & 6.2 & 0.0017 \\
08/05/2009 & 083.C-0775(D) & UT1-UT3 & 82.2 & 52.6 & HD113766 &
Sci & 0.46 & 6.7 & 0.0026 \\
08/05/2009 & 083.C-0775(D) & UT1-UT3 & 75.8 & 55.7 & HD110253 & Cal &
0.52 & 6.8 & 0.0015 \\
08/05/2009 & 083.C-0775(D) & UT1-UT3 & 75.7 & 55.2 & HD112213 & Cal &
0.51 & 5.8 & 0.0021 \\
08/05/2009 & 083.C-0775(D) & UT1-UT3 & 72.9 & 57.9 & HD113766 &
Sci & 0.49 & 6.2 & 0.0020 \\ \hline
09/05/2009 & 083.C-0775(B) & UT2-UT4 & 74.6 & 113.0 & HD112213 & Cal &
0.97 & 3.0 & 0.0027 \\
09/05/2009 & 083.C-0775(B) & UT2-UT4 & 74.9 & 116.2 & HD113766
& Sci & 1.16 & 2.6 & 0.0020 \\
09/05/2009 & 083.C-0775(B) & UT2-UT4 & 67.6 & 126.0 & HD110253 & Cal &
0.71 & 4.9 & 0.0022 \\
09/05/2009 & 083.C-0775(B) & UT2-UT4 & 69.9 & 125.2 & HD113766
& Sci & 0.76 & 4.9 & 0.0022 \\
09/05/2009 & 083.C-0775(B) & UT2-UT4 & 62.6 & 132.9 & HD112213 & Cal &
0.71 & 4.8 & 0.0020 \\
09/05/2009 & 083.C-0775(B) & UT2-UT4 & 89.4 & 65.9 & HD171759 & Cal &
0.56 & 5.9 & 0.0020 \\
09/05/2009 & 083.C-0775(B) & UT2-UT4 & 89.4 & 68.8 & HD172555
& Sci & 0.62 & 5.6 & 0.0025 \\
09/05/2009 & 083.C-0775(B) & UT2-UT4 & 89.4 & 74.6 & HD171212 & Cal &
0.80 & 4.5 & 0.0030 \\
09/05/2009 & 083.C-0775(B) & UT2-UT4 & 89.2 & 78.0 & HD171759 & Cal &
0.56 & 5.6 & 0.0034 \\ \hline
\end{tabular}
\begin{flushleft}
\small{Seeing, $\tau_0$ (coherence time) and flux RMS are taken from
  the ESO ambient conditions database
  (http://archive.eso.org/cms/eso-data/ambient-conditions) for the
  time at which the interferometric stage of the observations was
  taken.  Seeing is defined as the FWHM of a stellar image observed
  with an infinitely large telescope at 500nm wavelength and at the zenith.
  Flux RMS gives a measure of the background, with levels $>$0.05
  indicating cloud cover and levels $>$0.02 indicating possible cloud
  cover.  Coherence times ($\tau_0$) of less than 3ms are considered very
  fast, and indicate the presence of rapid atmospheric fluctuations that
   are likely to have degraded the interferometric signal-to-noise.
  \newline
  Science observations were calibrated using the bright
  standard star observations obtained before and after the science
  observation where possible (i.e. when the science observation occured
  between two standard star observations close both in time and
  on-sky position).  Otherwise only the observation of a bright
  standard star (standard taken from the CalVin tool, see Table
  \ref{tab:sources}) closest in time to the science observation was used
  for calibration purposes. }
\end{flushleft}
\end{table*}

\begin{figure*}
\begin{minipage}{7cm}
\includegraphics[width=7cm]{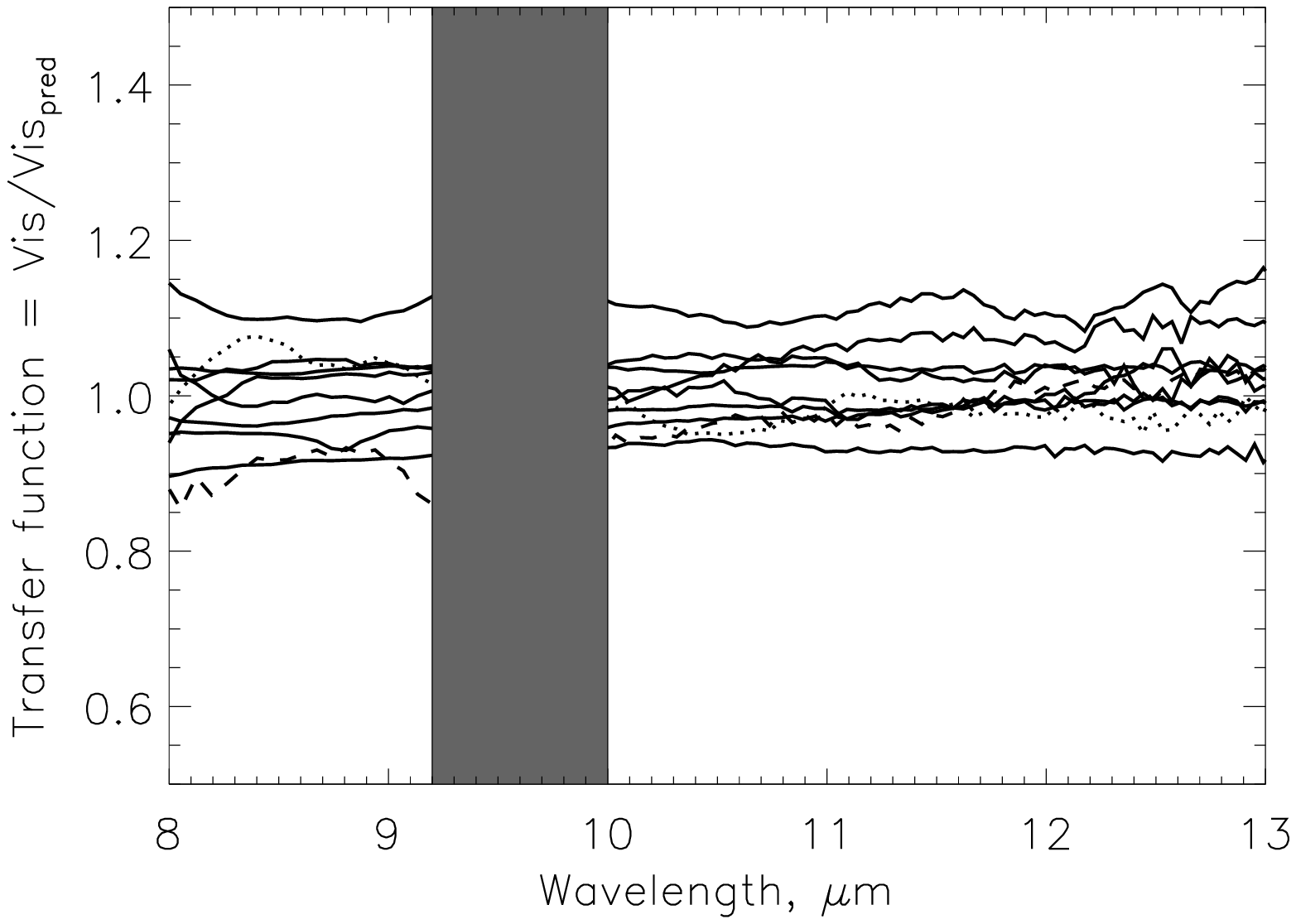}
\end{minipage}
\begin{minipage}{7cm}
\includegraphics[width=7cm]{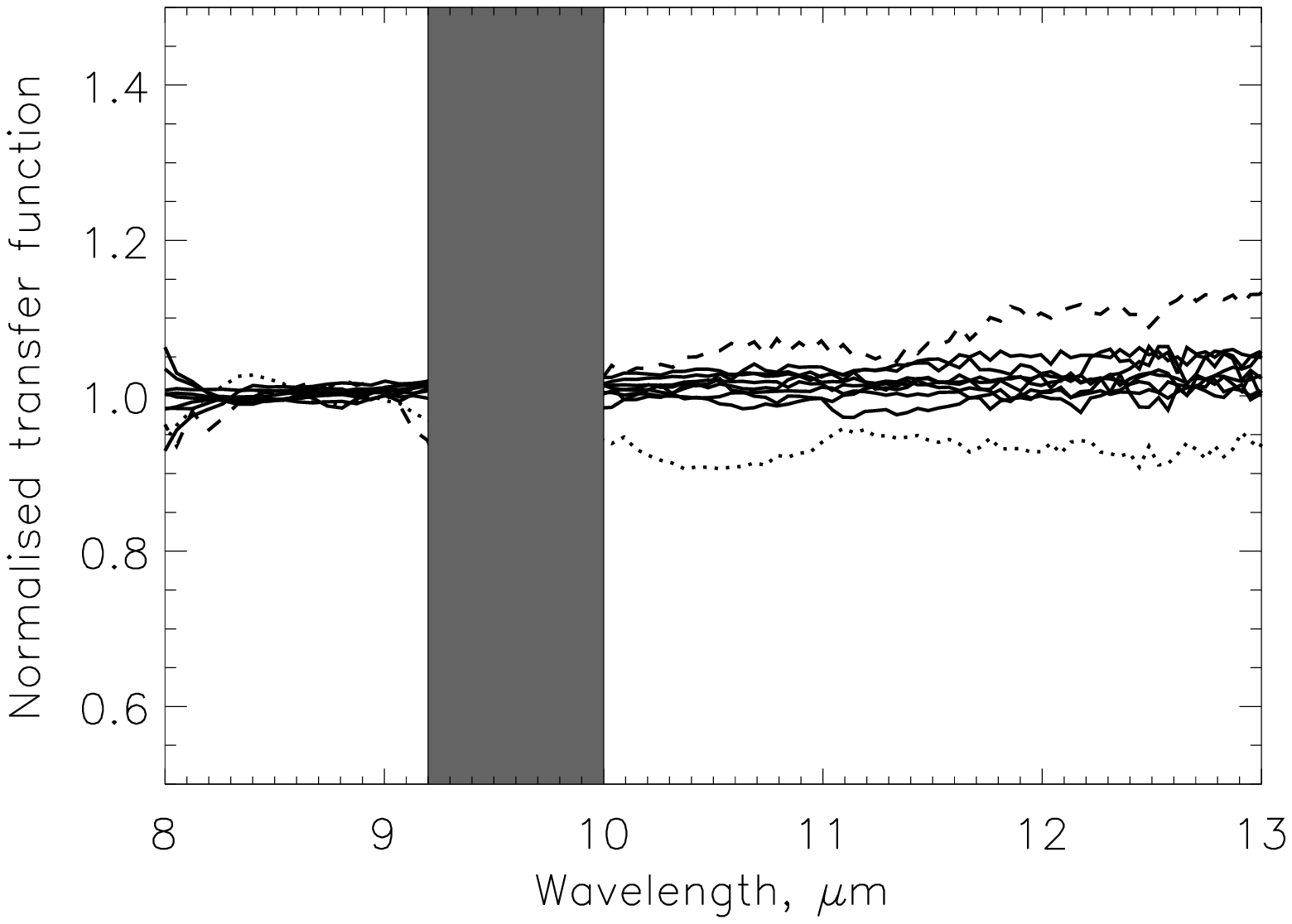}
\end{minipage}
\caption{\label{fig:transf} The visibility transfer function as
  measured on bright standard star targets calibrated by other bright
  standard star observations.  The region between 9.2--10$\mu$m is
  subject to high ozone absorption and is therefore excluded from the
  analysis.  The dashed lines represents an observation of HD112213
   taken under proposal 079.C-0259(G) calibrated by an observation of
   the same target taken on the same baseline configuration but on the
   following night. The  dotted line shows an observation of HD112213
   observed on 083.C-0775(B) calibrated by an observation of HD171759,
   which is $>90^{\circ}$ away, but
   taken immediately following on the same baseline configuration.  In
   the right hand panel the transfer functions are scaled to 1 in the
   wavelength range 8--9.2$\mu$m to show how the transfer function
   changes with wavelength.  These observations are discussed in the text.  }
\end{figure*}

\begin{figure*}
\begin{minipage}{7cm}
\includegraphics[width=7cm]{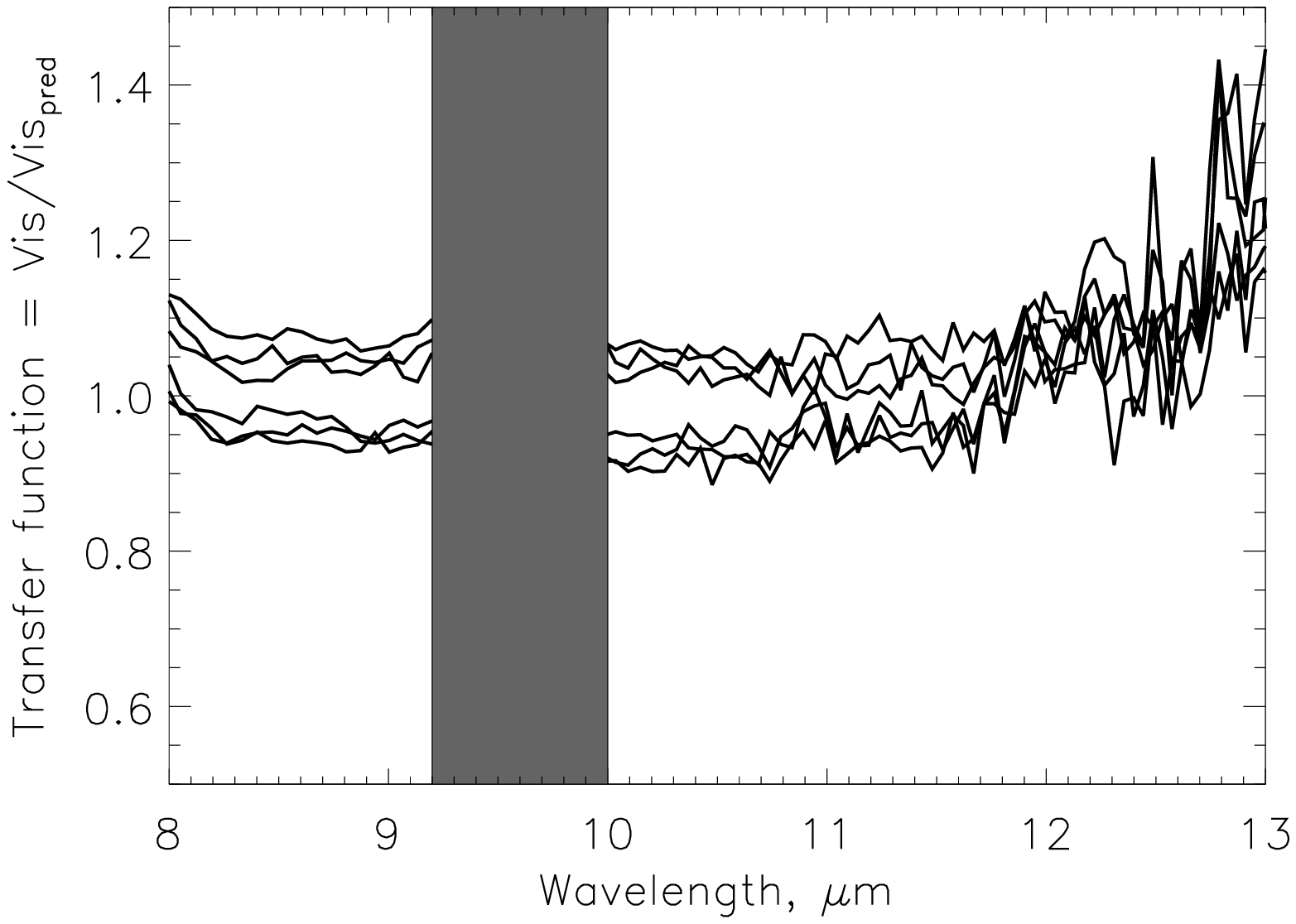}
\end{minipage}
\begin{minipage}{7cm}
\includegraphics[width=7cm]{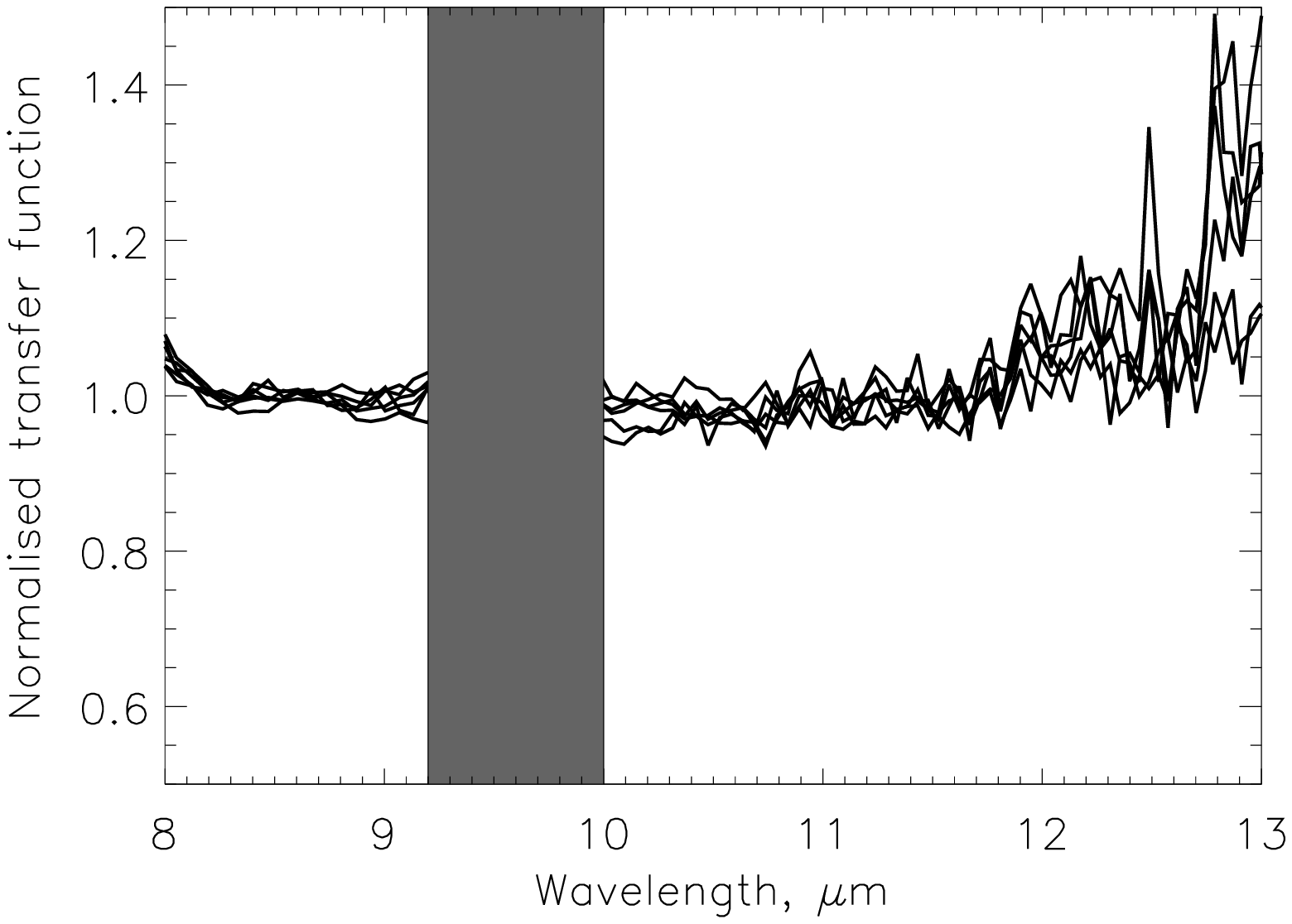}
\end{minipage}
\caption{\label{fig:transf_fnt} The visibility transfer function as
  measured on faint standard star targets calibrated by bright
  standard star observations.  The presentation of the data is the
  same as the previous figure. For these fainter targets, the transfer
  function is flat between 8.5 and 12 microns, but then shows a
  systematic increase at longer wavelengths, where in addition the
  signal-to-noise is starting to degrade.}
\end{figure*}

Reduction of the MIDI data was performed using the EWS software,
available as part of the MIA+EWS package (see
http://www.strw.leiden.nl/$\sim$nevec/MIDI/index.html).  Reduction
followed the standard EWS routines.  A summary of these steps is given
below.  Further details can be found in the manual available at the
above link. 
\begin{itemize}
\item Frames were multiplied by a mask and compressed in the direction
   perpendicular to the spectral dispersion to obtain a one-dimensional
   fringe intensity spectrum.  Following \citet{smithmidi} we
   used masks determined from a fit to the total intensity (photometry)
   frames, as provided by the MIA reduction package, to best
   exclude source-free background pixels and thereby reduce noise
   levels. 
\item The interferometric fringe data were aligned in time using an analysis
   of the measured \emph{group delay} to remove components of the delay
   arising from instrumental and atmospheric effects.
   For further details of this procedure the reader should consult the
   EWS manual or \citet{tristam}.
   Fringes were then averaged in time to produce a \emph{correlated flux}
   (or more correctly correlated intensity $I_{\rm{corr}}$ as no
   flux calibration had been determined at this point).  The correlated flux
   $F_{\rm{corr}}$ was then compared to the total source flux
   $F_{\rm{tot}}$ to give the source visibility $V =
   F_{\rm{corr}}/F_{\rm{tot}}$. 
\end{itemize}
In principle, the total source flux ($F_{\rm{tot}}$) could have been
determined from the photometry frames observed following the fringe
exposures with MIDI.  The MIDI photometric data are consistent
  with the IRS spectra, however the variation between individual
  observations of the science target photometry with MIDI was as high
  as 30--40\% across the full wavelength range.  We have found that
photometric measurements of faint targets with MIDI can often be
rather noisy (see \citealt{smithmidi}) and so instead we used the IRS
spectra --- which are not significantly different  from the VISIR
measurements (see Section 2) --- to provide these data. 

\subsection{Visibility calibration}

Two types of standard stars were used to calibrate the observations of
the science targets.  Bright standards were selected with the ESO
CalVin tool (see http://www.eso.org/instruments/midi/tools). Much
fainter standards were identified by searching for sources in the IRAS
catalogue, within 25$^\circ$ of the science targets, that had similar
12$\mu$m fluxes and that showed no evidence of binarity or
variability. Our decision to utilise both faint and bright calibrators
was motivated by a finding from our earlier studies of HD69830 and
$\eta$ Corvi \citep{smithmidi} which showed tentative evidence of a
loss of correlated flux at shorter wavelengths for faint targets.
Our goal was to observe faint standards {\em as though they were faint
science targets} to check for any bias in their measured visibilities.
For these tests, we ensured that the 12$\mu$m fluxes of the faint
standards were consistent with the emission predicted from scaled
\citet{cohen} photospheres, and were confident that they showed no
evidence of any excess emission (see next paragraph).

For those targets in the \citet{cohen} catalogue of mid-infrared
standard stars (HDs 111915, 116870, 156277, 112213) we used the Cohen
spectrum of the target as the target flux.  For the remaining targets
we used Cohen templates for stars of the same spectral type scaled to
the 10$\mu$m flux listed in the CalVin tool (for HDs 169767 and
171759) or for the faint standard stars (HDs 171212, 152186 and
110253) scaled to the source's listed 2MASS K band flux.  The
diameters for the bright standard stars used for the MIDI observations
were taken directly from the CalVin tool.  For the faint standards
source diameters were determined by assuming that the stars had a
diameter typical for their spectral type (taken from \citealt{allen})
and using the Hipparcos listed parallax to determine their distances.
Our inferred source diameters for the standards are listed in Table
\ref{tab:sources}.

The calibration of the target correlated flux was determined using the 
following equation: 
\begin{equation}\label{eq:fcorr} F_{\rm{corr,tar}} = (I_{\rm{corr,tar}}
       / I_{\rm{corr,cal}}) \times F_{\rm{tot,cal}} \times
       V_{\rm{cal}}, \end{equation} 
where $I_{\rm{corr,tar}}$ and
$I_{\rm{corr,cal}}$ were the correlated intensities measured in the
fringe exposures of the `target' and `calibrator' respectively and 
$V_{\rm{cal}}$ was the visibility of the calibrator, assumed to be
that of a uniform disc with the diameter given in Table
\ref{tab:sources}. For $F_{\rm{tot,cal}}$ we used the total flux of
the calibrator taken from the Cohen spectrum of the standard (or the
scaled Cohen spectrum, see paragraph above). The visibility of the
target was then evaluated as 
\begin{equation}\label{eq:vtar} 
V_{\rm{tar}} =  F_{\rm{corr,tar}}/F_{\rm{tot,tar}},
\end{equation} 
where $F_{\rm{tot,tar}}$ was the total flux of the target, which in the
case of the science targets was taken from the Spitzer IRS spectrum of
the source.

To determine the accuracy of our derived visibilities, we first
examined the visibilities of the bright standard star targets when
calibrated by other bright standard stars.  By using pairs of standard
star observations taken as close as possible in time, we were able to
generate 10 independent visibility functions.  These pairs of standard
star observations were in general taken either side of science
observations, and so were separated in time by roughly twice the time
between a science and standard star observation.  When corrected for the
sizes of the standard stars used, these produced the visibility
transfer functions shown in Figure \ref{fig:transf}.  Note that the
bandpass between 9.2--10$\mu$m is subject to high levels of
uncertainty due to ozone absorption, and so this region has been
ignored in our analysis. The variation in transfer function level
in the left hand panel of Figure \ref{fig:transf} is indicative of the
range of seeing mismatch between the observations of source and
calibrator, with the values nearest to unity being associated with the
use of calibrator stars closest in time and space.  The weighted mean
of the transfer functions across the whole spectral range gave a mean
value of observed visibility/predicted visibility of 1.002$\pm$0.057
(or $\sim$ 6\%).  To calculate the
weighted mean we used weights derived from the errors on the
correlated flux.  These errors came from the variance found in the EWS
reduction by splitting the fringe observations into 5
sub-integrations.  We then calculated the weighted mean over the MIDI
spectral range (excluding the 9.2--10$\mu$m region which suffers from
ozone absorption), and took a mean over all 10 pairs of standard stars
to get the figure above.

In order to remove this source of variation, we scaled each transfer
function so that its weighted mean in the range 8--9$\mu$m was unity.
These normalised transfer functions are presented in the right-hand
panel of 
Figure \ref{fig:transf}. Most of the normalised transfer functions are
very similar, but two of them deserve mention. The first, identified
by a dashed line, is from an observation of HD112213 taken on
09/04/2007 (under observing ID 079.C-0259(G)) calibrated by an
observation of the same target taken under the same observing
configuration \emph{the next night}. The second was derived from an
observation of HD112213 taken on 09/05/2009 (under proposal
083.C-0775(B)) but has been calibrated using a ``standard'' \emph{located
over 90$^\circ$ away on the sky} (HD171759). These aberrant transfer
functions highlight the need for very careful calibration
strategies. Overall, we found the ratio of the transfer function in
the range 10.5--11.5$\mu$m (12--13$\mu$m) to that at 8--9$\mu$m for
the remaining bright-bright pairings to be 1.010$\pm$0.013
(1.016$\pm$0.018).  These data confirm that for our data the
visibility functions for bright targets are likely to be calibrated to
within $\sim$6\%, and also that the differential visibility (the
visibility with reference to that at a fixed wavelength) is a factor
of 2--3 times more accurate than the absolute value of the visibility.

A similar analysis for our faint unresolved targets calibrated by
bright standard stars is shown in Figure \ref{fig:transf_fnt}. The
behaviour of these transfer functions is broadly similar, but there is
definite evidence of curvature of the functions at the extremes of the
MIDI bandpass. This is limited to the very first few spectral bins at
the short wavelength end, but is more noticeable beyond a wavelength
of 12$\mu$m, where the data are increasingly noisy. We found that
the weighted mean absolute values of the transfer functions across the
MIDI wavelength range (excluding the 9.2--10$\mu$m region) was
1.035$\pm$0.072, and that the mean normalised transfer function
between 10.5--11.5$\mu$m relative to that at 8--9$\mu$m was
0.984$\pm$0.030. The equivalent value for the 12--13$\mu$m bandpass
was 1.064$\pm$0.047. These data suggest that the visibilities of our
faint targets can be measured to better than 10\% (i.e. at a level
consistent with what has been presented before, see
e.g. \citealt{chesneau}), and that the differential visibility across
the whole wavelength range is stable to better than 5\%. Somewhat
better accuracy can be expected if data at wavelengths $>$12$\mu$m is
excluded, typically by a factor of two.

\subsection{MIDI observations of HD113766}

The observations of HD113766 were calibrated using two bright standard
star observations where possible (see Table \ref{tab:midiobs}).  In
these instances the average of the calibrated correlated flux using
two standard star observations 
was used.  The standard deviation between the two calibrations was
added in quadrature to the error on the correlated flux as estimated
from the time variation in the correlated intensity.  This was derived
by splitting the fringe integration into 5 sub-integrations and
determining the standard deviation between the different
sub-integration datasets.

\begin{figure*}
\begin{minipage}{7cm}
\includegraphics[width=7cm]{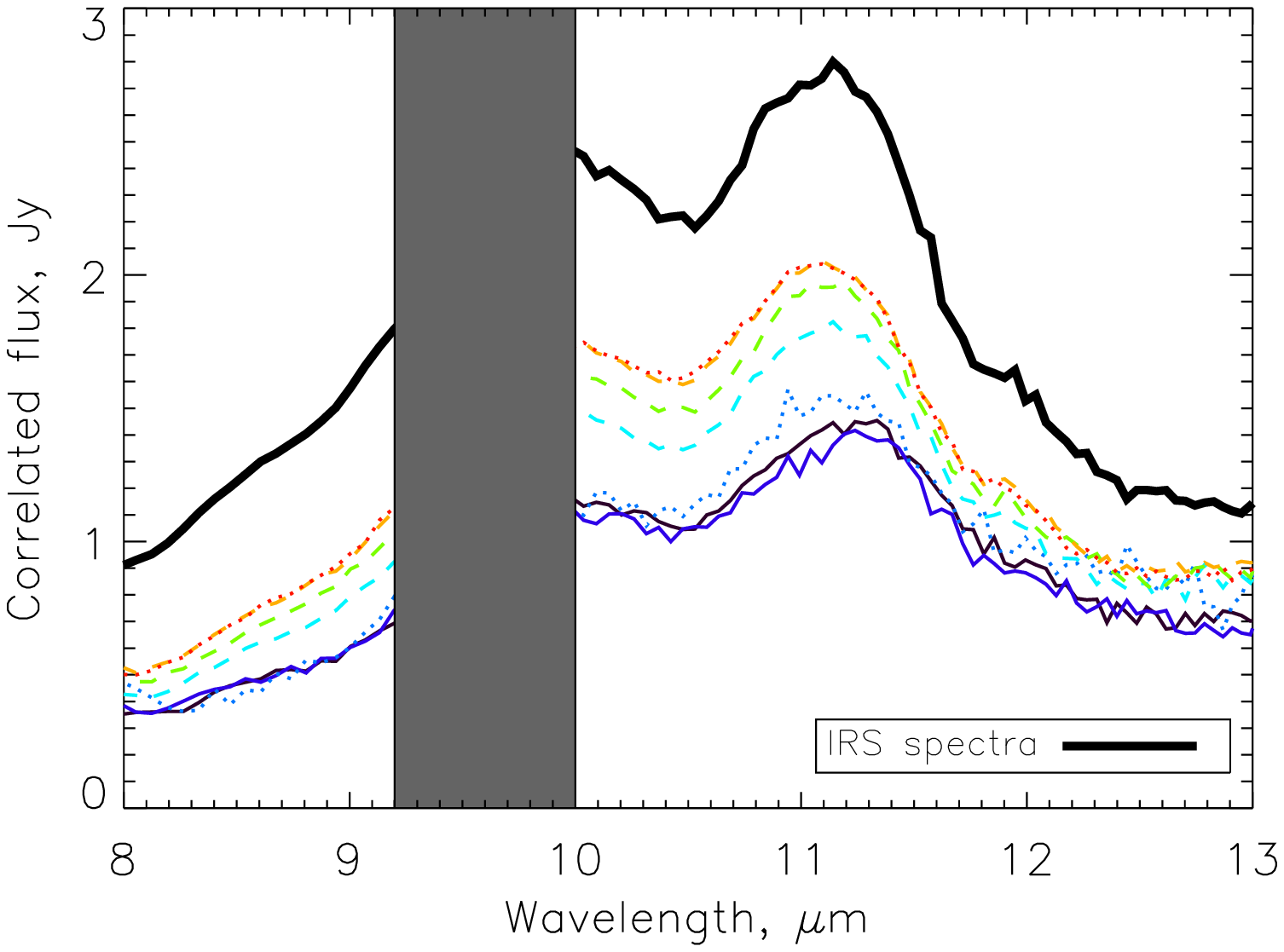}
\end{minipage}
\begin{minipage}{7cm}
\includegraphics[width=7cm]{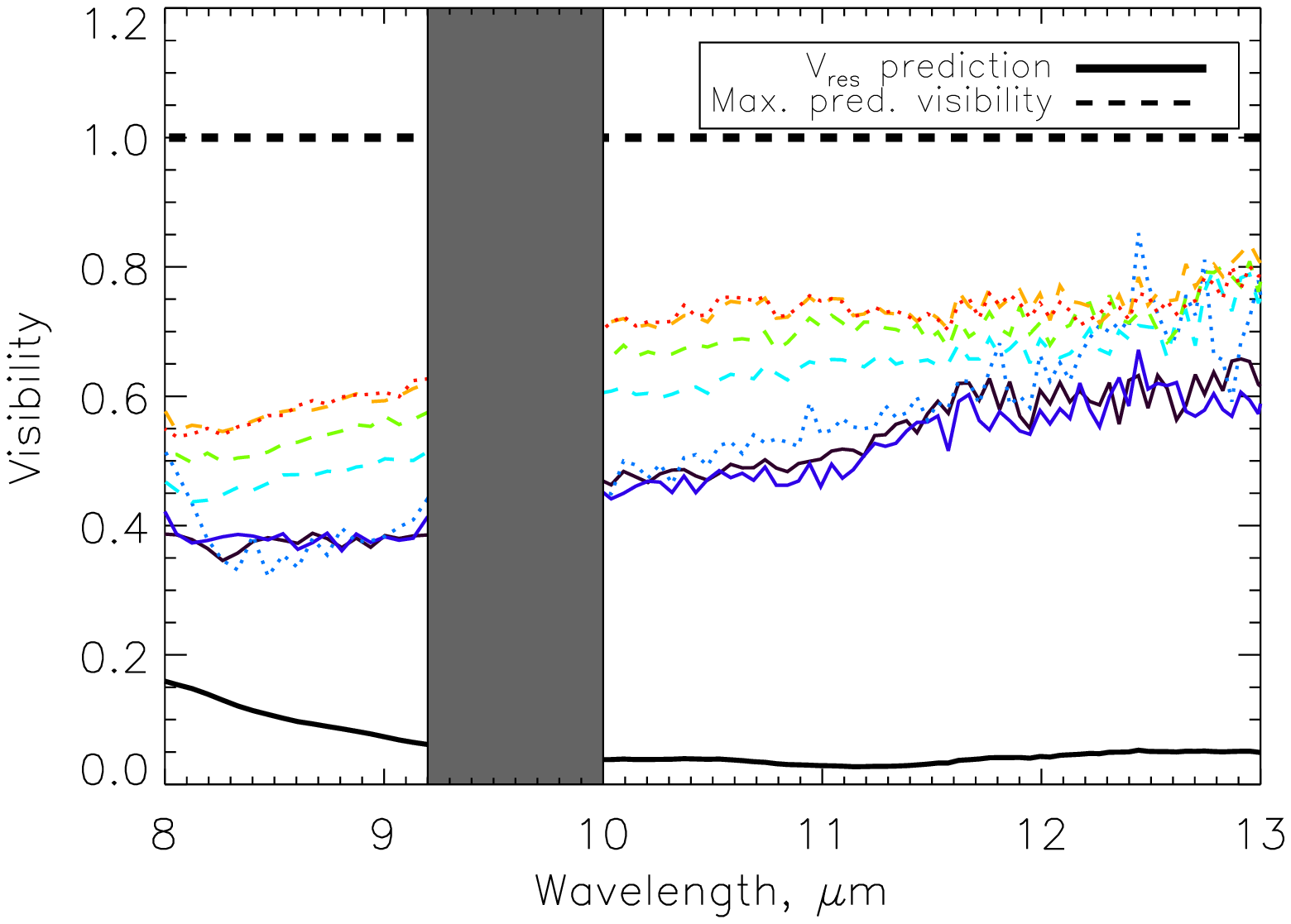}
\end{minipage} \\
\includegraphics[width=12cm]{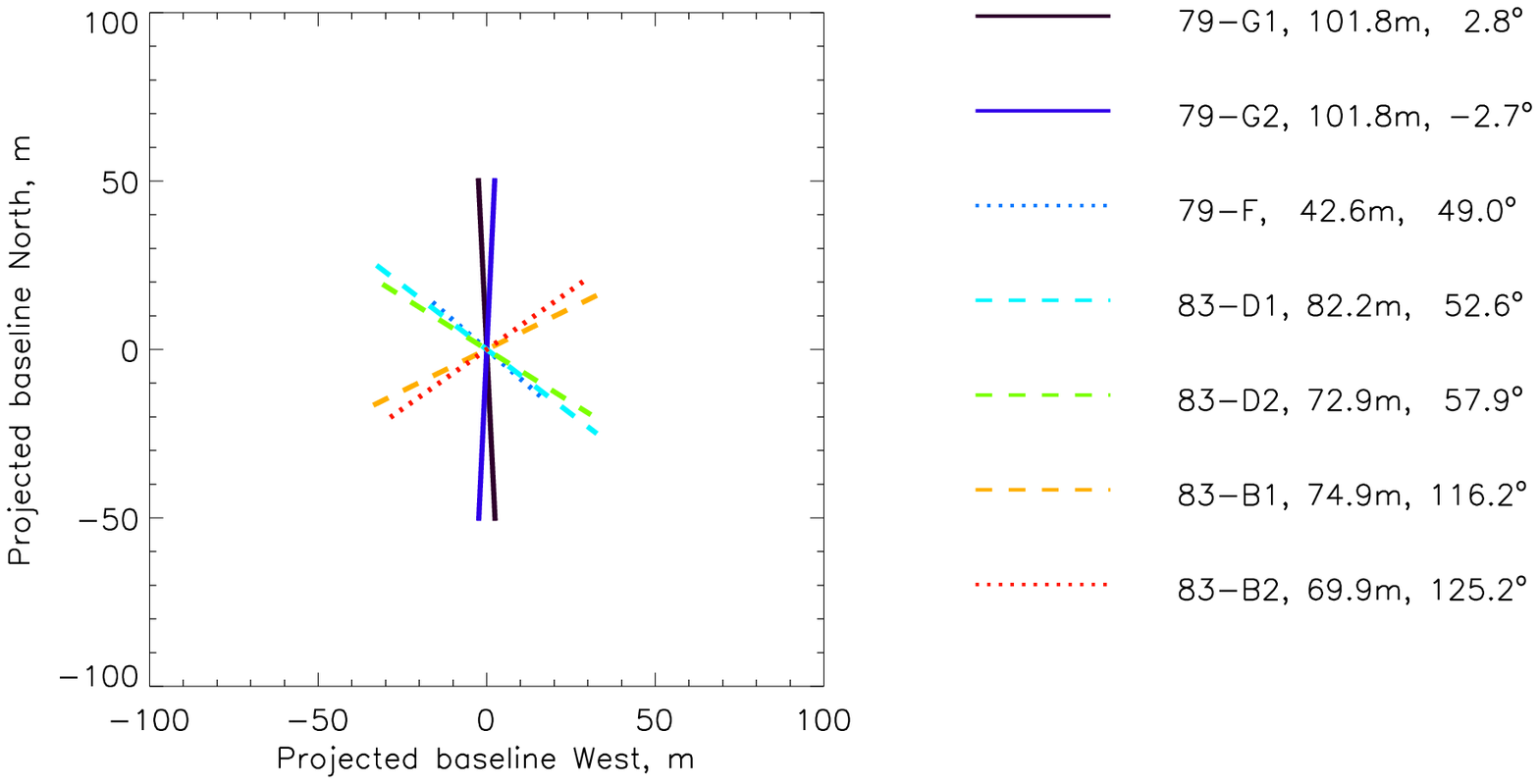}
\caption{\label{fig:113766midi} The correlated flux measurements (top
  left) and subsequent visibility curves (top right) for the MIDI
  observations of HD113766.  In the left hand top panel the IRS
  spectrum (thick black line) is compared to the correlated flux
  measurements for reference.  For the 
  visibility curves the $V_{\rm{res}}$ prediction assumes that all the
  excess emission detected has been resolved, and that the only
  contribution to the correlated flux would be emission from the star
  itself.  This is not a flat line as the relative contribution of
  star and excess emission to the total flux from the system changes
  through the MIDI wavelength range (see Figure \ref{fig:visirspec}). 
  The maximum predicted visibility (a flat line at 1) assumes
  none of the emission has been resolved.  The labels for the
  different observations are coded according to the semester during
  which the observations were taken, the run letter and a number
  indicating which observation on a particular run it is, so for
  example the first observation of HD113766 taken under proposal
  079.C-0259(G) is given the label 79-G1.  The baselines of all
  observations of HD113766 are shown in the bottom panel.  For the
  observational results the same colours and line-styles are used for
  all three plots.  }
\end{figure*}

The correlated flux measurements for our 7 observations of HD113766
taken on various baselines (see Table \ref{tab:midiobs}) are shown in
Figure \ref{fig:113766midi}, where the solid line is the IRS spectrum
of the target. It is immediately clear that there is a significant
difference between the correlated flux and the IRS spectrum. There
also appears to be some difference between the observations taken
under proposal 079.C-0259 and proposal 083.C-0775.  This is further
explored through comparison of their visibility functions.

The visibilities calculated from the correlated flux measurement
are shown in Figure \ref{fig:113766midi}.  Interestingly, the
visibility on baseline 79-F is somewhat lower than that measured on
the {\em longer} but approximately parallel baseline 83-D (see caption
to Figure \ref{fig:113766midi} for description of labels and
Table \ref{tab:midiobs}).  This might indicate that rather than having
a distribution like a Gaussian, the source emission is more complex.
A ring for example would have an oscillating visibility function which
would be higher on some longer baselines (see, e.g., Figure 5 
of \citealt{dullemondaraa} for an example of this).  However, the
observations on 79-F were taken under the poorest observing conditions
in the study --- the ``flux RMS'' and coherence time diagnostic
metrics were particularly high --- and under these conditions it is
likely that the measured visibility was biased to a lower value. As a
result, we did not use this visibility measurement as a
constraint in the modelling of the HD113766 data presented in Section
4. 

\begin{figure*}
\begin{minipage}{7cm}
\includegraphics[width=7cm]{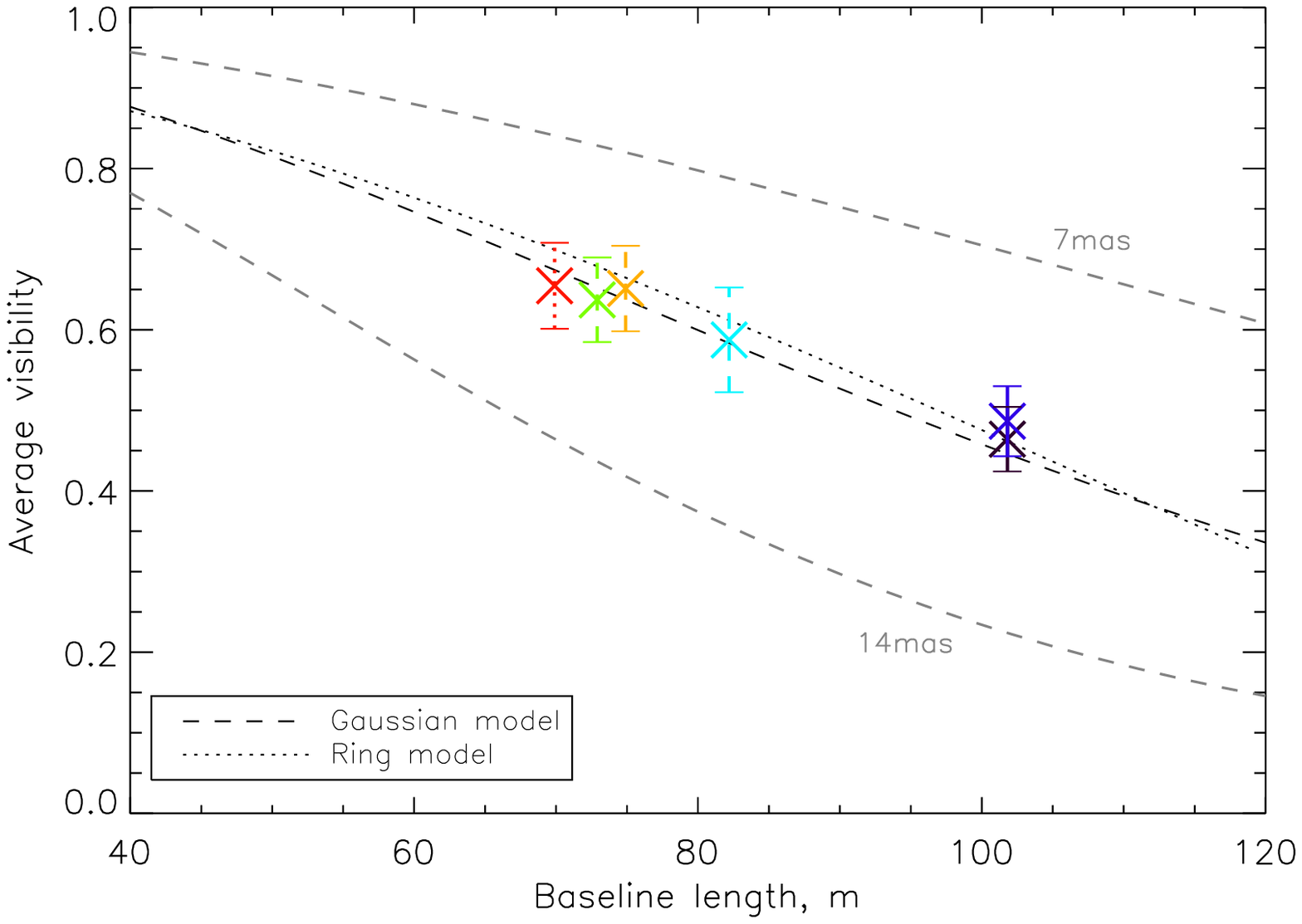}
\end{minipage}
\begin{minipage}{7cm}
\includegraphics[width=7cm]{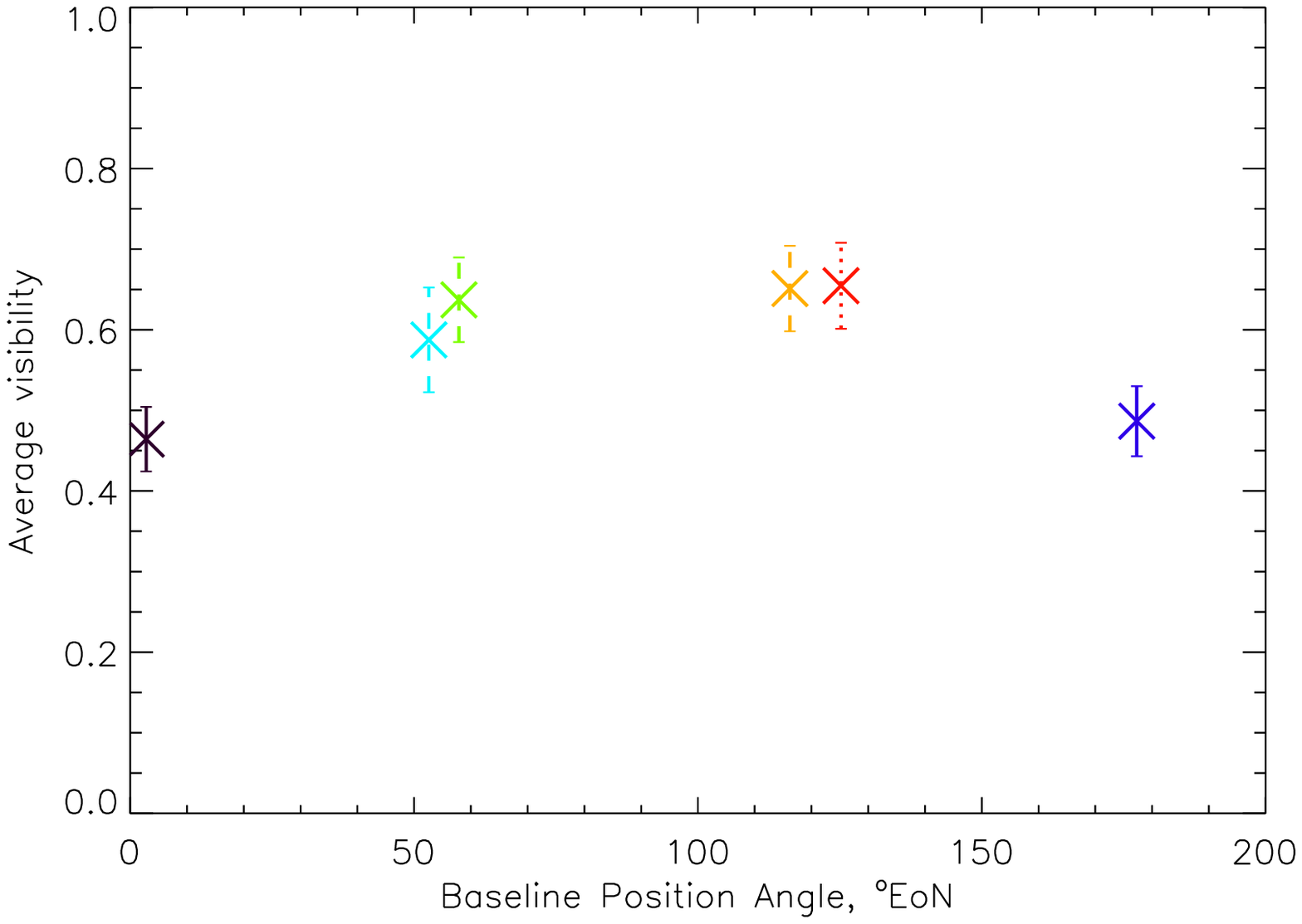}
\end{minipage}
\caption{\label{fig:113766bybase} The weighted mean of the visibility
  functions (over the MIDI wavelength range) measured for HD113766
  shown against baseline length (left) and position angle (right).
  Means are weighted by errors added in quadrature from
  the statistical error on the correlated flux, error on the
  calibrator correlated flux, error on the calibrator total flux and
  error on the IRS spectrum used to derive the visibility from the
  correlated flux (see Section 4.2 for more details).  Symbol colours
  are given in the key of Figure \ref{fig:113766midi}.  The
  observation on baseline 79-F is excluded from this figure due to
  the high uncertainty associated with this data point (see text for
  details).  The dashed line marks the visibility expected for a model
in which the excess emission is distributed as a circularly symmetric
Gaussian with FWHM 10mas.  This gives a first order approximation for
the size of the emitting region. Also plotted in grey dashed lines are
5$\sigma$ limits to the Gaussian model (see section 4.3).  The best
fitting ring-like model for the distribution is also shown by a dotted
line.  See section 4.3 for a detailed discussion of these fits to the
observed visibilities.  }
\end{figure*}

In the plot of the visibility function (Figure \ref{fig:113766midi}
top right) we have shown the behaviour 
for two extremal classes of targets, i.e. those that are completely
unresolved and then those whose dust emission is fully resolved.  For
the first of these cases, we expect that the correlated flux is always
equal to the total flux, and hence that the visibility is equal to
unity at all wavelengths. In the second case, the situation is
slightly more complicated, since both the potentially unresolved
stellar contribution and the resolved dust emission need to be considered. 
For such targets, the visibility of the source will be given by
\begin{equation}\label{eq:vtot} 
V =  \frac{F_\star}{F_{\rm{tot}}}V_\star +
       \frac{F_{\rm{disc}}}{F_{\rm{tot}}}V_{\rm{disc}} , 
\end{equation} 
where $V_{\rm{disc}}$ and $V_\star$ are the visibility of the disc and
the star respectively, $F_{\rm{disc}}$ and $F_\star$ are the fluxes of
the disc and star respectively, and $F_{\rm{tot}}$ is the total flux
(i.e. $F_{\rm{tot}} = F_{\rm{disc}} + F_\star$).  The stellar emission
component in HD113766 is likely to be completely unresolved. It is a
F3V-type star at a Hipparcos distance of 131pc, and so with an
expected radius of 1.38$R_\odot$ would subtend an angle of roughly
$0.048$mas, well beyond the resolving power of VLTI/MIDI.
The minimum expected visibility, given that the star
is expected to be completely unresolved ($V_\star = 1$), is therefore
the visibility in the event that the disc flux is completely resolved,
$V_{\rm{disc}} = 0$, $\Rightarrow V = V_{\rm{res}} =
F_\star/F_{\rm{tot}}$.  We have labelled this visibility
$V_{\rm{res}}$ in Figure \ref{fig:113766midi}.  As it is
clear that for all baselines observed the visibility function lies
between this and the maximum visibility of 1, the disc appears
partially resolved on all baselines.

The visibility functions measured on similar baselines are consistent
within the error levels expected for the visibility of a faint target
($\sim$10\%, see Section 3.1), The visibility functions measured on
79-G1 and 79-G2 are very similar to one another, as are the visibility
functions measured on 83-B1 and 83-B2.  It is also clear that the
visibility functions measured on longer baselines are lower than those
measured on shorter baselines (with the exception of 79-F as discussed
above, and which we treat as unreliable).  Solid lines are used in
Figure \ref{fig:113766midi} to denote the longest baselines, dotted
lines the shortest and dashed lines intermediate baseline lengths.

To display more clearly how the visibilities are varying with baseline
length and position angle we show the average (weighted mean over the
MIDI spectral range excluding the ozone dominated 9.2--10$\mu$m
region) visibility of HD113766 plotted against baseline length and
position angle in Figure \ref{fig:113766bybase}.  Although the
visibility is seen to change with wavelength (Figure
\ref{fig:113766midi} top right), to first order this effect can be
attributed to the decreasing resolution of MIDI with increasing
wavelength, and so we are unlikely to be biasing the data through this
averaging.  Excluding baseline 79-F, there is a drop in 
visibility with increasing baseline length, consistent with a simple
source geometry such as a Gaussian. \footnote{\label{foot:vis}If we
  had instead used 
  the VISIR spectrum as a model for the total flux of the HD 113766
  system, then the calculated visibilities would be higher, but still
  exhibit the same behavior, i.e. decreasing with baseline length.  In this
case the points would lie close to the 7mas model shown in Figure
\ref{fig:113766bybase}.  } Also shown in this figure are the
visibilities plotted as a function of baseline position angle.
Although the visibilities closest to 0 or 180$^\circ$ appear lower,
these are the longest observed baselines.  If we consider only
baselines of similar lengths (83-D1, 83-D2, 83-B1 and 83-B2) there is
no evidence for a change in visibility with baseline position angle
that would indicate a non-circularly symmetric source distribution.
More details on the limits we can place on the source geometry with
these observations are discussed in Section 4.

\subsection{MIDI observations of HD172555}

Our observations of HD172555 have been taken over several semesters.
Where possible, two bright standard star observations were used
to calibrate the correlated flux measurements of this target (see Table
\ref{tab:midiobs}). Errors were calculated in the same way as
described for HD113766 in the previous subsection.

\begin{figure*}
\begin{minipage}{7cm}
\includegraphics[width=7cm]{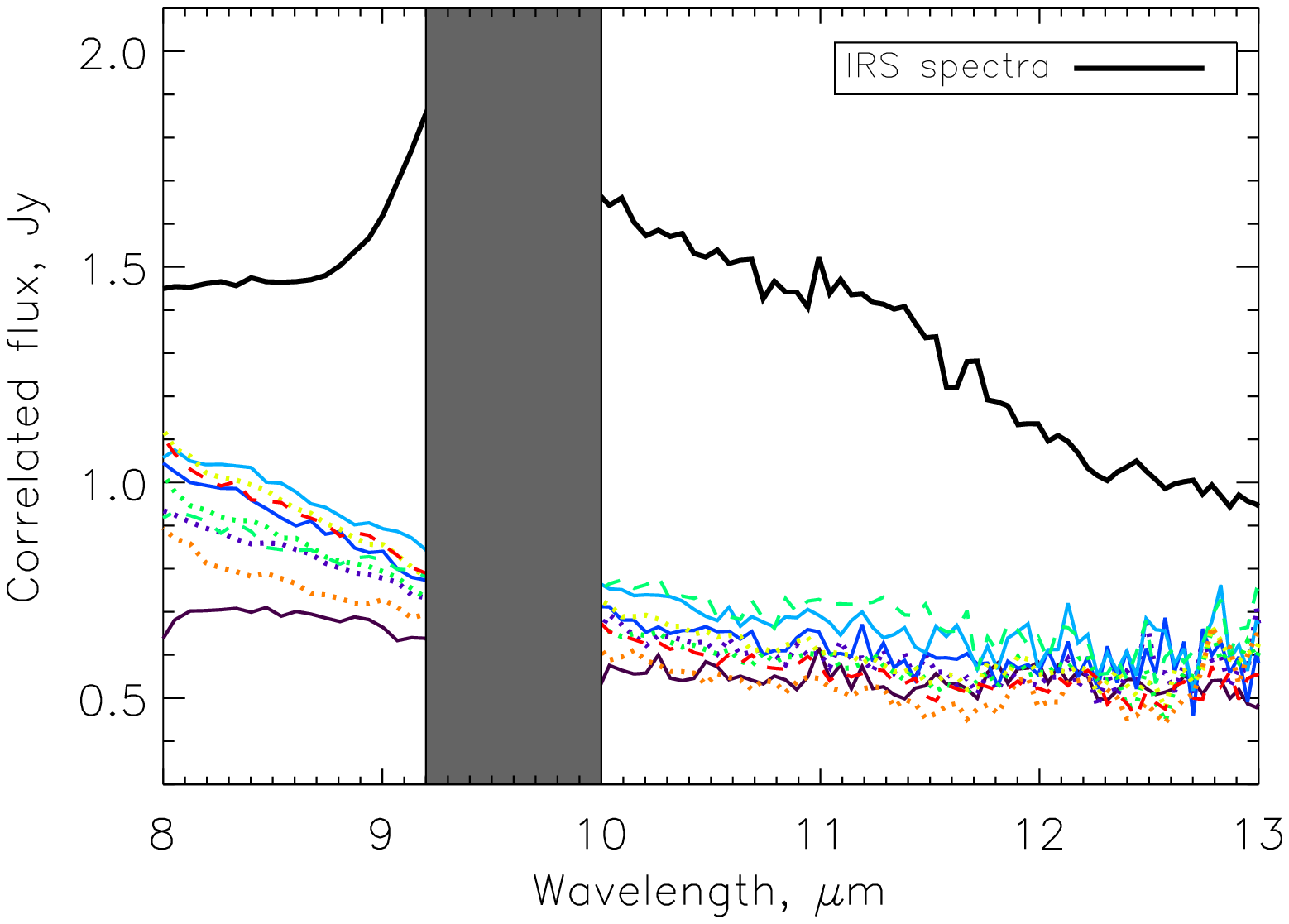}
\end{minipage}
\begin{minipage}{7cm}
\includegraphics[width=7cm]{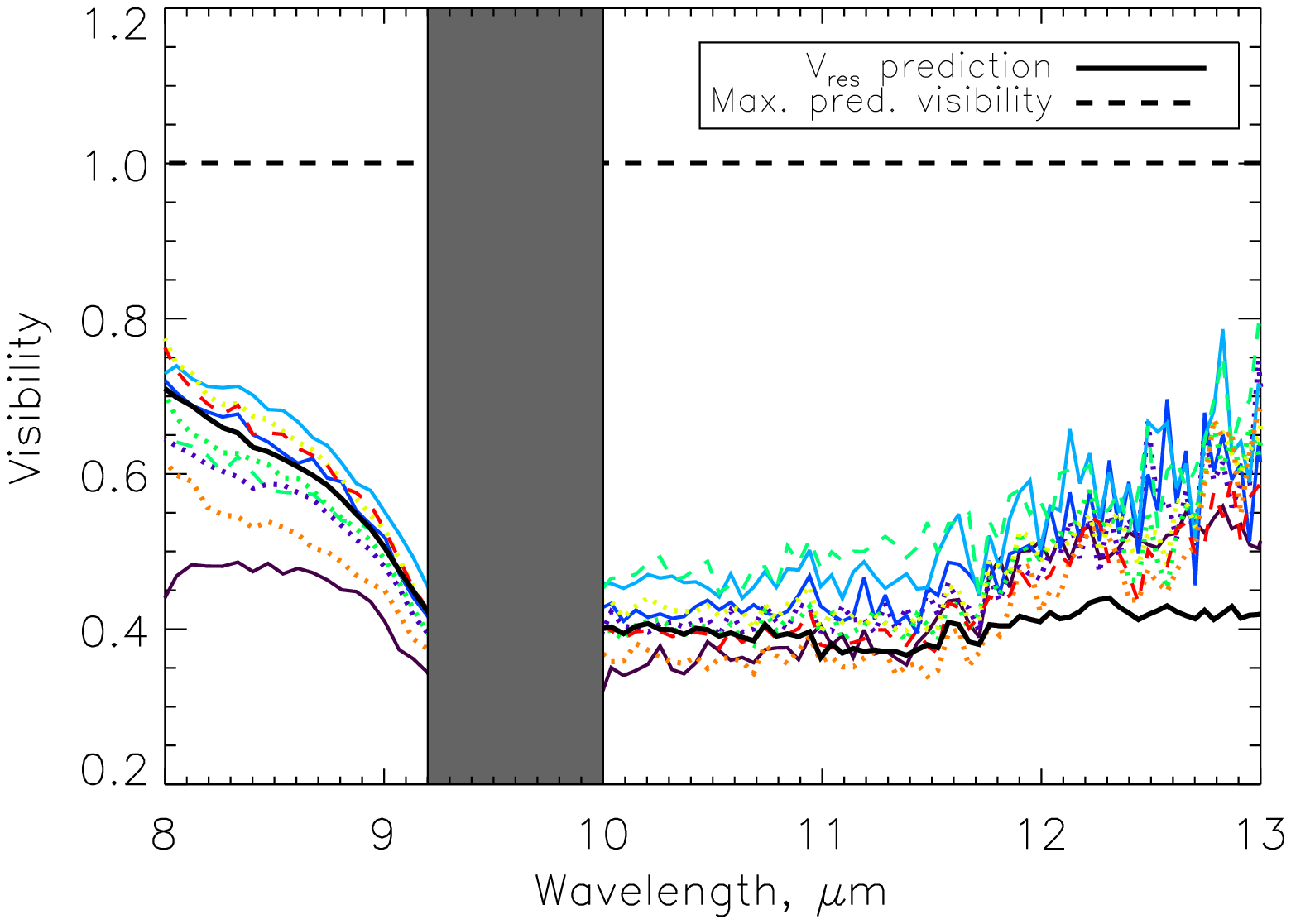}
\end{minipage} \\
\includegraphics[width=12cm]{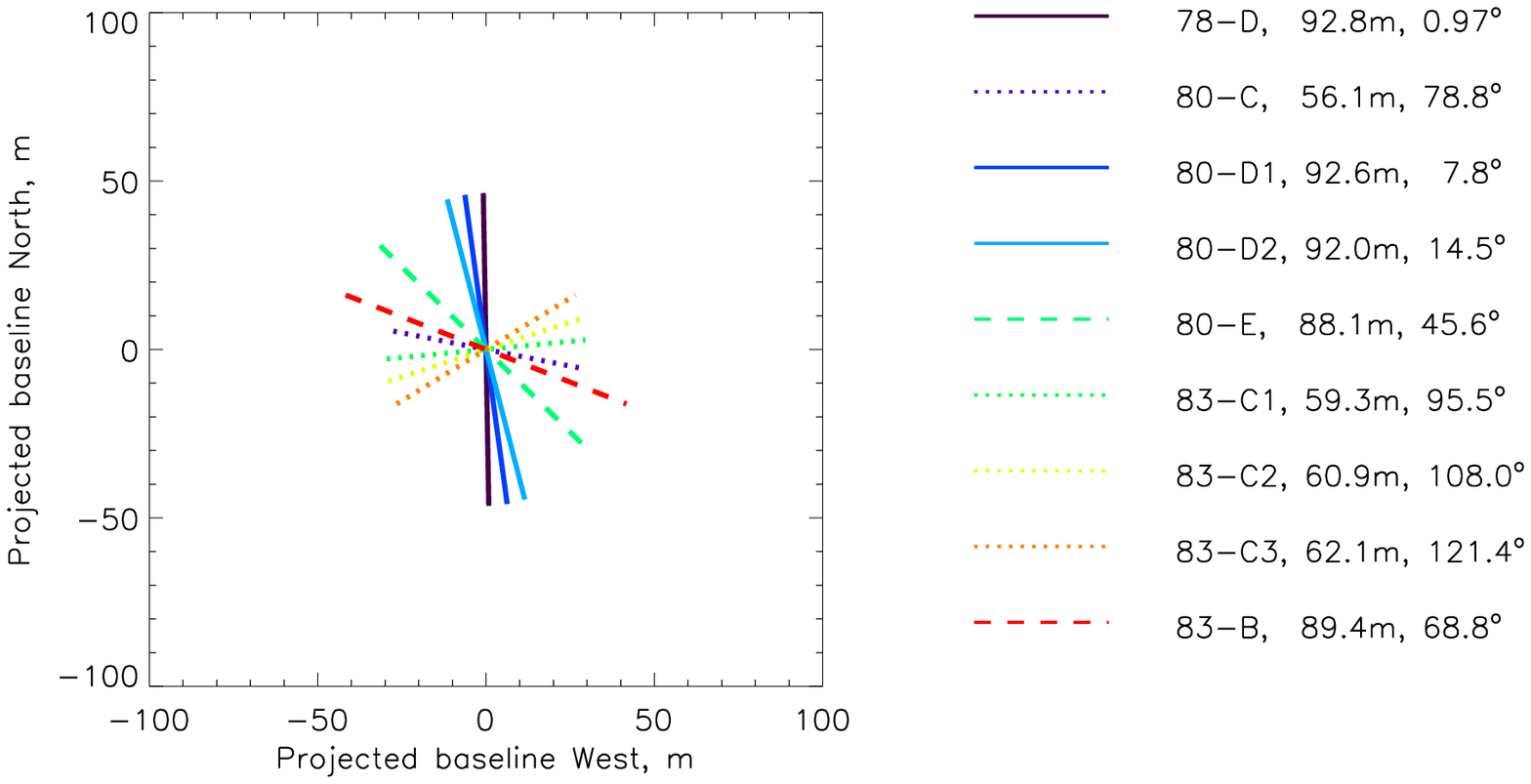}
\caption{\label{fig:172555midi} The correlated flux measurements (top
  left) and subsequent visibility curves (top right) for the MIDI
  observations of HD172555.  In the top left panel the IRS spectrum
  (thick black line) is 
  compared to the correlated flux measurements for reference.  For the
  visibility curves the $V_{\rm{res}}$ prediction assumes that all the
  excess emission detected has been resolved, and that the only
  contribution to the correlated flux would be emission from the star
  itself.  This is not a flat line as the relative contributions of
  the star and excess emission to the total emission vary with
  wavelength (see Figure \ref{fig:visirspec}).  The maximum predicted
  visibility (a flat line at 1) assumes that none of the emission has
  been resolved.  Different runs are labelled following the convention
  described in Figure \ref{fig:113766midi}.  The baselines of all
  observations of HD172555 are shown in the bottom panel.  For the
  observational results the same colours and line-styles are used for
  all three plots.  }
\end{figure*}

The calibrated correlated flux measurements from our 9 observations of
HD172555 with MIDI (see Table \ref{tab:midiobs}) are shown in Figure
\ref{fig:172555midi}. As for HD113766, the IRS spectrum has been
over-plotted for comparison.  These plots show rather similar
correlated flux levels for all baselines. The degree of resolution of
the target is better seen in the visibility functions of the target,
also shown in Figure \ref{fig:172555midi} in the right hand panel.  As
for HD113766, we have over-plotted the predicted visibilities for
targets with fully unresolved ($V=1$ across all wavelengths) and fully
resolved ($V_{\rm{res}}$) disks for reference.

\begin{figure*}
\begin{minipage}{7cm}
\includegraphics[width=7cm]{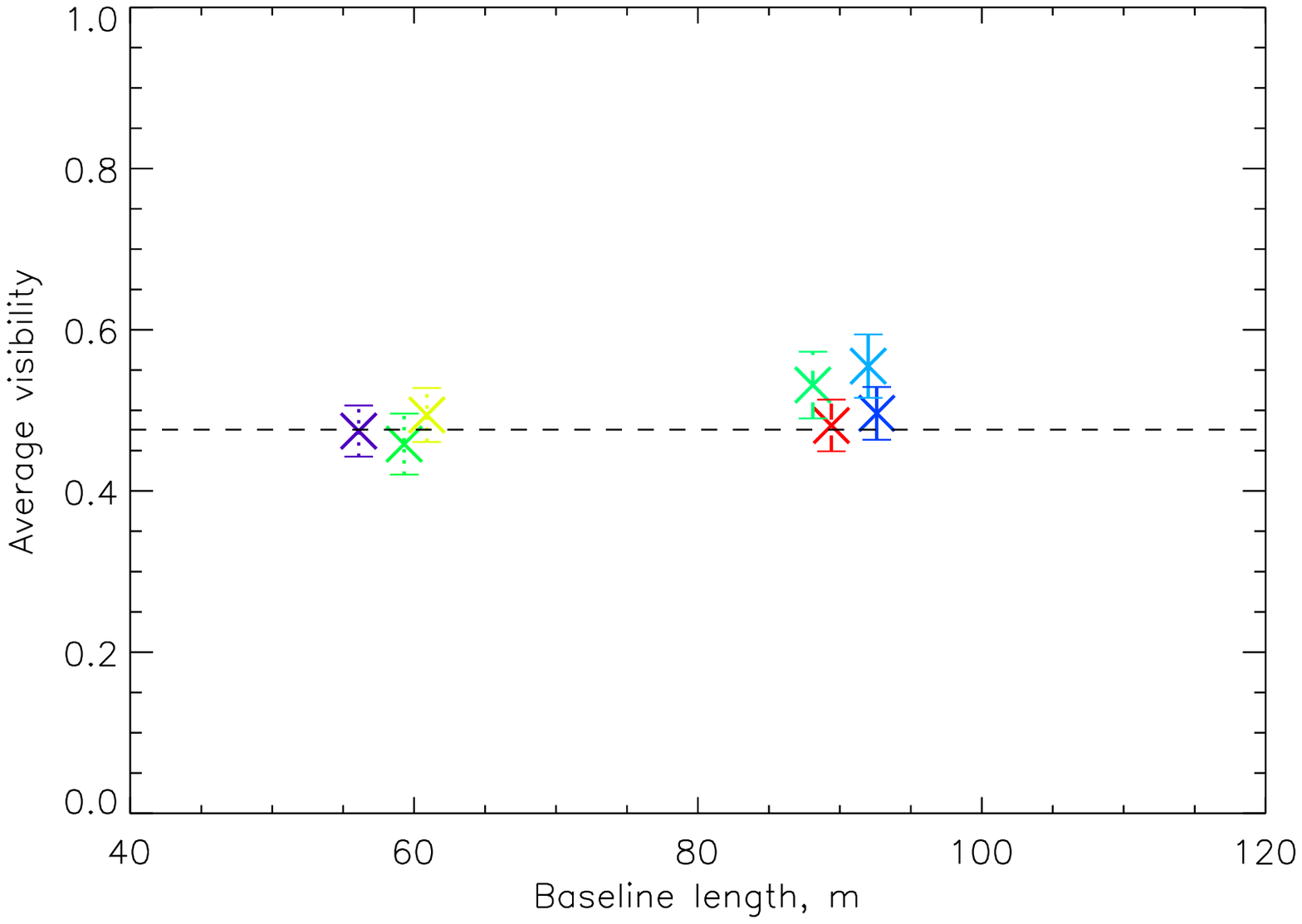}
\end{minipage}
\begin{minipage}{7cm}
\includegraphics[width=7cm]{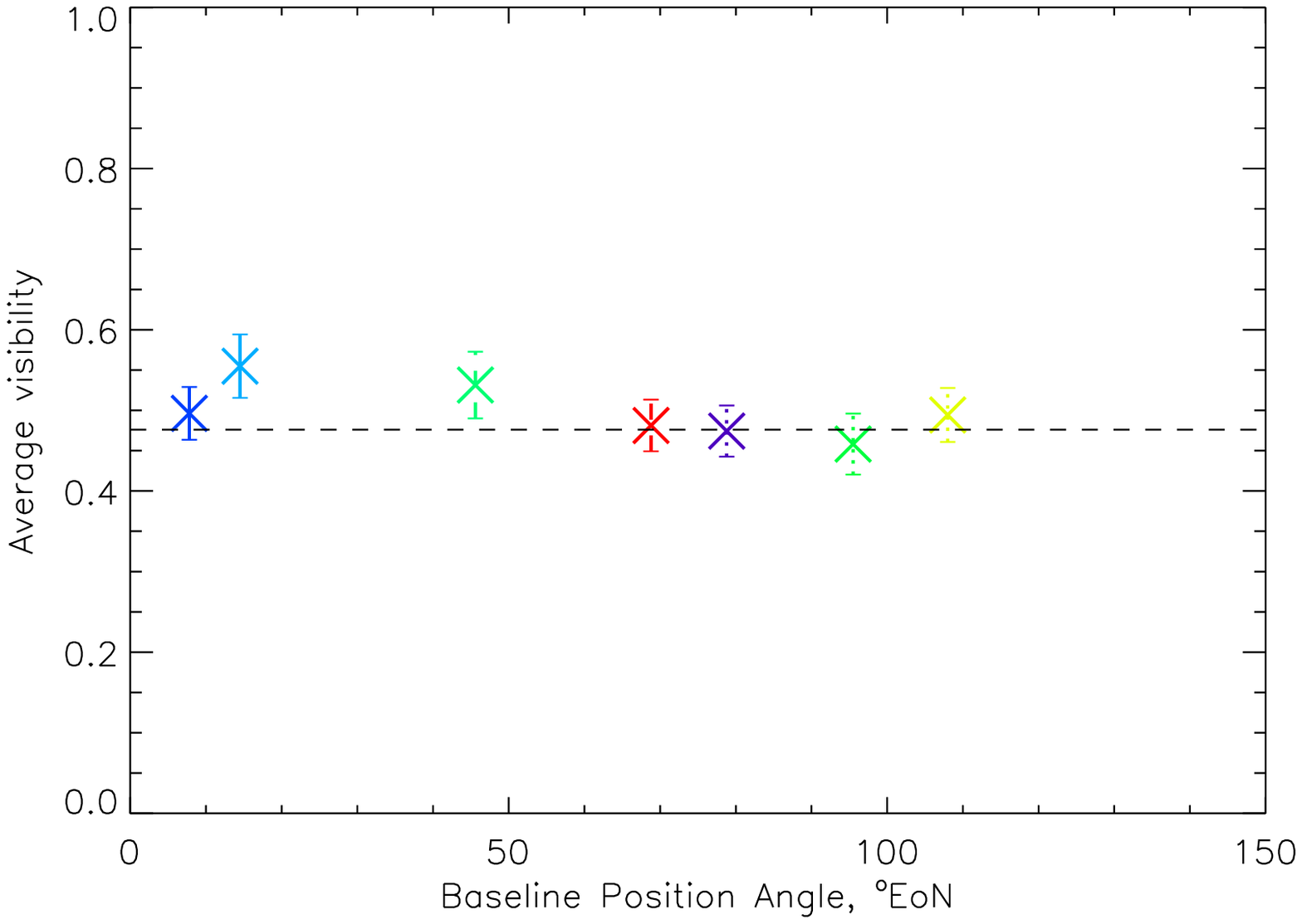}
\end{minipage}
\caption{\label{fig:172555bybase} The weighted mean of the visibility
  functions (over the MIDI wavelength range) measured for HD172555
  shown against baseline length (left) and position angle (right).
  Means are weighted by errors added in quadrature from
  the statistical error on the correlated flux, error on the
  calibrator correlated flux, error on the calibrator total flux and
  error on the IRS spectrum used to derive the visibility from the
  correlated flux (see Section 4.2 for more details).  Symbol colours
  are given in the key of Figure \ref{fig:172555midi}.  The data from
  baselines 78-D and 83-C3 are excluded from this plot due to high
  uncertainty associated with these observations (see text for
  details).  The dashed line indicates the mean visibility (across the
  MIDI wavelength range) in the case that the excess emission is
  completely resolved ($V_{\rm{res}}$), indicating that a model in
  which the disc has a large spatial extent is likely to provide a
  good fit to the data.  There is tentative evidence (at the
  1--2$\sigma$ level) that the disc
  emission is not fully resolved on baselines 80-D2 and 80-E,
  suggesting the emission may not be circularly symmetric.  }
\end{figure*}

These plots show that for observations 78-D and 83-C3 the observed
visibilities are more than 10\% lower than the predicted 
visibility in the case that the disc is fully resolved
($V_{\rm{res}}$).   Although values lower than $V_{\rm{res}}$ can be
observed on intermediate baselines (when $V_{\rm{disc}}<0$ due to the
phase jumps in its pattern, see Figure 5 of \citealt{dullemondaraa}
for an example of this behavior), the visibilities observed on
similar baselines (e.g. 81-D1 and 83-C2 respectively) do not fall
significantly below $V_{\rm{res}}$.
We believe that the results on baselines 78-D and 83-C3
can be explained by a combination of calibration errors.  The
coherence times for the observations on baseline 83-C were very short,
particularly for the third observation of HD172555 which has a
significantly shorter coherence time than the observations of standard
stars used to calibrate the observation (see Table \ref{tab:midiobs}).
This could be the cause of the low visibility as we believe was the
case for our observation on baseline 79-F for HD113766.  Secondly, the
observation on baseline 78-D was taken at the end of a night towards
twilight and had to be calibrated by observations of a standard star
offset from HD172555 by 95$^\circ$.  As demonstrated in Section 3.1
large offsets between a science target and standard star are an obvious 
source of miscalibration, which we believe has occurred here.

We can see that for all our baselines the observed visibility function
for HD172555 is consistent with a completely resolved disc.  High
visibilities at $>$11.5$\mu$m are likely due to the bias seen in
observations of faint targets and can be compared with those seen for
the faint calibrators (Figure \ref{fig:transf_fnt} right).  There is
no evidence of a systematic decrease in visibility with increasing
baseline length, suggesting that the excess emission is already
completely resolved on the shortest baselines.  There is tentative
evidence that the mean observed visibility changes slightly with
baseline position angle, which can be seen more clearly in Figure
\ref{fig:172555bybase}.  Such a change with baseline
position angle could reveal evidence of a clumpy structure, as may be
expected as the result of a recent massive collision in the disc, the
model favoured by \citet{lisse09}.  However, as the visibility seems
to be smoothly changing with position angle (increasing with
increasing position angle from 0--14$^\circ$ on baselines of similar
length 80-D1 and 80-D2, and decreasing with increasing position
angle from 46$^\circ$ on baselines of similar length 80-E and 83-B)
this suggests a smoother, perhaps, elliptical structure for the
emission. An ellipse with major axis oriented at 120$^\circ$ would
have its lowest visibility on baselines at 120$^\circ$ EoN and its
highest visibility on baselines at 30$^\circ$ EoN. Such an emission
morphology could arise from an inclined circular disc or a truly
elliptical disc.  As we have complete resolution of the disc on the
shortest baselines (56m, 80-C) we would expect such a disc to have a
minimum radial size of at least $\sim$40mas (based on the resolving
power of 56m aperture telescope at 10.5$\mu$m). More detailed
modelling of the observed visibility functions is presented in the
following section.

\section{Modelling the visibility functions}

\subsection{Visibility calculation}

We consider several source geometries to try to fit the observed
visibility functions around HD113766 and HD172555.  The van-Cittert
Vernicke theorem states that the normalised visibility function of a
source is the normalised Fourier transform of the brightness
distribution of the source.  The simplest source geometry we consider
is a circularly symmetric Gaussian.  The source geometry is given by
\begin{equation} \label{eq:Igauss} I(\alpha,\beta) =
  \frac{1}{\sqrt{\pi/(4 \ln 2)} \Theta} \exp\left({\frac{-4 \ln
      2\rho^2}{\Theta^2}}\right) \end{equation}
where $\alpha$ and $\beta$ are angular on-sky coordinates (in radians),
$\Theta$ is the full-width at half-maximum (FWHM) and $\rho =
\sqrt{\alpha^2 + \beta^2}$.  The visibility function of this source is
then 
\begin{equation} \label{eq:Vgauss} V(u,v) =
  \exp\left(-\frac{(\pi\Theta \sqrt{u^2+v^2})^2}{4 \ln 2}
  \right) \end{equation} where $u$ and $v$ are coordinates describing
the spatial frequency of the brightness distribution such that $u =
B_u/\lambda$, $v = B_v/\lambda$ where $B_u$ and $B_v$ are the
projections of the baseline vector on the two axes and $\lambda$ is
the wavelength of the observation \citep{berger}.  For an elliptical
Gaussian we parameterise the ellipticity by $I$ where $v_I = v
\cos{I}$ (and $u_I = u$).  

The Gaussian model provides a good model to an envelope, and offers a
simple approximation to the overall source size.  However, debris disc
emission is expected to be ring-like in distribution.  For a thin ring
the source geometry is given by
\begin{equation} \label{eq:Iring} I(\rho) = \frac{1}{2\pi\rho_0}
  \delta(\rho-\rho_0) \end{equation}
(where $\delta$ is the Dirac delta function), then the visibility of
such an object is given by
\begin{equation} \label{eq:Vring} V(u,v) = J_0(2\pi\rho_0
  r) \end{equation} where $J_0$ is the 0th-order Bessel function and
$r=\sqrt{u^2+v^2}$. For our debris disc models we follow the example
of \citet{malbet} and use a sum of thin rings to model the
distribution and visibility function of a ring of finite thickness.
The emission is distributed as the integration of all ring
contributions given by equation \ref{eq:Iring} from $\rho_0 =
\rho_{\rm{min}}$ to $\rho_0 = \rho_{\rm{max}}$, and the visibility
function of this finite thickness ring is similarly the integral of
the corresponding Fourier transforms (visibility functions).  For a
disc inclined to the line of sight at an angle $I$ and at a position
angle $\theta$ we simply consider the case where $r_{\theta,I} =
\sqrt{u_\theta^2 + v_\theta^2\cos^2(I)}$ which represents the projected
baseline in a new $(u_\theta,v_\theta)$ reference frame corresponding
to a rotation of the array frame by the position angle $\theta$ with a
compression factor of $\cos(I)$.   To include the temperature
distribution of the disc, we make the simplifying assumption that the
disc is optically thin and the 
dust behaves like a blackbody, and so the temperature of each ring is
dependent only on the distance from the star (as the stellar
luminosity is fixed).  Then 
\begin{equation} \label{eq:Tring} T(r) = 278.3
  \sqrt{\frac{\sqrt{L_\star/L_\odot}}{r}} \end{equation} and the flux
we can expect from our disc model is the integration of all the rings
modified by the distance of the source from the observer,
$d$, \begin{equation} \label{eq:Fringtot} F_\lambda = \frac{2\pi}{d^2}
  \int_{r_{\rm{min}}}^{r_{\rm{max}}} rB_\lambda(T(r))\Sigma_0
  \rm{d}r. \end{equation} We assume the disc surface density is flat
and therefore $\Sigma_0$ is a constant.  The visibility function
corresponding to this model is therefore simply the normalised
integration of the visibility function for the thin rings weighted by
their flux.    

These models are simple descriptions for the visibility function of
the excess emission, $V_{\rm{disc}}$.  In our comparison to the data
the final visibility model includes the contribution from the star
which is completely unresolved by the interferometer ($V_\star = 1$,
see Section 3.2).  The final value of $V_{\rm{mod}}$ is calculated
according to equation \ref{eq:vtot}.

\citet{lisse09} suggested that a possible origin for the emission
around HD172555 is a recent massive collision between two large
planetesimals/proto-planets.  In this scenario we might expect the
emission to arise from a clump offset from the central star.  As the
clump is offset from the central star, we cannot simply add the
visibility functions scaled by their relative flux levels.  For a
multi-component function components at positions $\alpha_i$, $\beta_i$
in the plane of the sky with visibilities $V_i$, the normalised
visibility of the full function
is \begin{equation} \label{eq:multi_vis} V(u,v) = \frac{\sum^{n}_{i=1}
    F_iV_i(u,v)\exp(2\pi i(u\alpha_i + v\beta_i))}{\sum^{n}_{i=1}F_i},
\end{equation} where $F_i$ is the flux of the component $i$ and $n$ is
the total number of components.  For a two-component model, the
normalised squared visibility reduces
to \begin{equation}\label{eq:two_vis} \frac{F_1^2V_1^2\!+\!F_2^2V_2^2\! +\!
    2F_1F_2V_1V_2\cos(2\pi(u(\alpha_1\!-\!\alpha_2)\!+\!
    v(\beta_1\!-\!\beta_2)))}{(F_1+F_2)^2}, \end{equation} which reduces
to the familiar $(F_1V_1+F_2V_2)^2/(F_1+F_2)^2$ in the case that
$\alpha_1 = \alpha_2$ and $\beta_1=\beta_2$.

\subsection{Error calculation} 

To determine the goodness of fit of the models, and thereby calculate
the best fitting model parameters, the calculation of the error terms
on the visibility must be carefully considered.  From equation
\ref{eq:fcorr}, the error on the correlated flux arises from the terms
$I_{\rm{corr,tar}}$, $I_{\rm{corr,cal}}$ and $F_{\rm{tot,cal}}$.  Here
we have assumed that the error on the visibility of the calibrator is
negligible.  Errors on the calibrator diameters can be up to $\pm$5\%
\citep{verhoelst}, and so at its highest the error on $V_{\rm{cal}}$
(from observation of HD112213 on baseline 79-G) is 0.5\%, much lower
than the other sources of uncertainty.  The uncertainty on the calibrator
flux is assumed to be 2\% ($\delta F_{\rm{tot,cal}}/F_{\rm{tot,cal}} =
0.02$, see \citealt{cohen}).  

The error on $I_{\rm{corr,tar}}$ is determined by splitting the fringe
observation of the target into 5 sub-integrations and determining the
standard deviation from the mean value at each wavelength.  The error
on $I_{\rm{corr,cal}}$ is the error from calibration.  Here we use the
visibility transfer functions (Section 3.1) to determine the error.
At each wavelength value sampled in the MIDI range, we determine the
mean and standard deviation of the visibility transfer functions for
the faint standard stars.  We check that the mean value is compatible
with a value of 1 within the error given by the standard deviation,
and adopt this standard deviation as our error on
$I_{\rm{corr,cal}}$.  The average value of this error across the MIDI
range (excluding the ozone dominated 9.2-10$\mu$m region) is 0.057.
The error on the correlated flux is then given by 
\begin{eqnarray} \label{errCorr} \left(\frac{
\delta F_{\rm{corr,tar}}}{F_{\rm{corr,tar}}}
\right)^2  & = & \left(\frac{\delta
      I_{\rm{corr,tar}}}{I_{\rm{corr,tar}}}\right)^2 +
    \left(\frac{\delta I_{\rm{corr,cal}}}{I_{\rm{corr,cal}}}\right)^2
    +  \nonumber \\ & & 
\; \; \; \; \; \; \left(\frac{\delta
      F_{\rm{tot,cal}}}{F_{\rm{tot,cal}}}\right)^2. 
\end{eqnarray}  

To calculate the error on the visibility we can see from equation
\ref{eq:vtar} that we need to add the error on the total flux,
$F_{\rm{tot,tar}}$.  As we have adopted the IRS spectra for our
science targets, this is the error on the IRS spectra which itself is
composed of the statistical uncertainty and calibration uncertainty on
these observations.  As the spectral sampling is different between IRS
and MIDI, we use linear interpolation to determine the errors at the
central wavelengths of the spectral bins of the MIDI data.  Then the
final error on the visibility is given
by \begin{equation} \label{eq:errV} \left(\frac{\delta
    V_{\rm{tar}}}{V_{\rm{tar}}} \right)^2 = \left( \frac{\delta
    F_{\rm{corr,tar}}}{F_{\rm{corr,tar}}}\right)^2 + \left(
  \frac{\delta F_{\rm{tot,tar}}}{F_{\rm{tot,tar}}} \right)^2
  , \end{equation} with typical values of 6\% (7\%) for $\delta
    F_{\rm{corr,tar}}/F_{\rm{corr,tar}}$ and 4\% (3\%) for $\delta
    F_{\rm{corr,tar}}/F_{\rm{corr,tar}}$ for HD113766 (HD172555).

For the model visibilities equation \ref{eq:vtot} holds.  We assume
that the star has a fixed visibility of 1 ($V_\star$) and has no
error.  We also assume that $V_{\rm{disc}}$ is perfectly known for any
particular model.  The error on $F_{\rm{tot,tar}}$ is discussed
in the above paragraph.  The errors on $F_\star$ and $F_{\rm{disc}}$
are more complex.  They arise in part from the uncertainty on the
total flux, and also from uncertainty in the relative contributions of
the star and disc to the total flux.  As a change in the value of
$F_{\rm{tot,tar}}$ would result in a systematic change in $F_\star$
and $F_{\rm{disc}}$, similarly a change in $F_\star$ with no
corresponding change in $F_{\rm{tot,tar}}$ would result in a
reciprocal change in $F_{\rm{disc}}$.  We therefore model the error on
the term $F_\star + F_{\rm{disc}} V_{\rm{disc}}$ in a Monte Carlo
manner.  The scaling of the Kurucz model profile used to model the
photosphere is determined by a $\chi^2$ minimisation over a range of
scalings to find a best fit to the B and V band magnitudes of the
stars as listed in the Hipparcos catalogue and the J H and K band
magnitudes from the 2MASS catalogue.  The percentage points of the
$\chi^2$ distribution are used to determine a 1$\sigma$ limit
on this scaling.  The scaling used in the Monte Carlo modelling is
taken from a Normal distribution with mean given by the best fitting
scaling and standard deviation given by the level of the 1$\sigma$
error.  The value of $F_{\rm{tot}}$ is also varied between the
1$\sigma$ error limits on the IRS spectrum (or rather the interpolated
spectrum, see above).  The value of $F_{\rm{disc}}$ is taken to
be $F_{\rm{tot}} - F_\star$.  The error term on $F_\star +
F_{\rm{disc}} V_{\rm{disc}}$ is then taken from the standard
deviation of 1000 random samplings, and is found to be typically at
the 4\% level for HD113766 and 6\% for HD172555 averaged over the whole
wavelength range. The final error on the model visibility is then  
\begin{equation} \label{eq:errVmod} \left(\frac{\delta
    V_{\rm{mod}}}{V_{\rm{mod}}} \right)^2 = \left(\frac{\delta
    (F_\star + F_{\rm{disc}}V_{\rm{disc}})}{F_\star +
    F_{\rm{disc}}V_{\rm{disc}}} \right)^2 + \left(\frac{\delta
    F_{\rm{tot,tar}}}{F_{\rm{tot,tar}}} \right)^2 . \end{equation}

To test how well each model reproduces the observed visibilities we
use the $\chi^2$ goodness-of-fit test.  Typically this takes the form
$\sum_{n=1}^{N} (D-M)^2/\sigma^2$ where $N$ is the number of data
points, $D$ is some observed data with associated error $\sigma$ and
$M$ is a model with no error.  As both our observational data
$V_{\rm{obs}}$ and model $V_{\rm{mod}}$ have associated errors we use
the ratio of the observed to model visibility and compare this to 1
(as if the model is a good fit to the data then $V_{\rm{mod}} \approx
V_{\rm{obs}}$).  The errors for both $V_{\rm{obs}}$ and $V_{\rm{mod}}$
must be included in $\sigma$.  Our $\chi^2$ calculation then
becomes \begin{equation} \label{eq:chisq} \chi^2 =
  \sum_{n=1}^{N}\sum_{w=1}^{W}
  \left(\frac{V_{\rm{obs}}(w,n)/V_{\rm{mod}}(w,n) - 1}{\sigma(w,n)}
  \right)^2. \end{equation} Here $N$ is the number of observations we
have for a source and $W$ is the number of spectral channels.  To
calculate a reduced $\chi^2$ we divide the above by the number of data
points (observations $\times$ spectral channels) minus the number of
free parameters in the model (1, the FWHM, for a circularly symmetric
Gaussian; 4, the radius, thickness, inclination and position angle,
for a ring).  If we
consider equations \ref{eq:vtar} and \ref{eq:vtot} we can see that in 
dividing $V_{\rm{obs}}$ by $V_{\rm{mod}}$ we cancel the term in
$F_{\rm{tot,tar}}$.  The errors from the IRS spectroscopy do not
therefore appear explicitly in the calculation of $\sigma$.  These
errors do appear implicitly in the Monte Carlo calculation of the
error on $F_\star+F_{\rm{disc}}V_{\rm{disc}}$ through the calibration
and statistical errors.  The final value of $\sigma$ is given by the
addition in quadrature of all the remaining error terms.  As
there is evidence of a bias in the visibility of low flux sources at
wavelengths $>$12$\mu$m (see Figure \ref{fig:transf_fnt} and section
3.1) we exclude the $>$12$\mu$m data from the model fitting.

\subsection{Best fitting models for HD113766}

The observed visibility functions from the MIDI observations of
HD113766 are consistent with a partial resolution of the excess
emission.  As a first attempt to model the distribution of the excess
emission we tested circularly symmetric Gaussian models with a range
of FWHM $\Theta$.  We tested $1 \le \Theta \le 100$mas in 60
logarithmically-spaced steps.  The visibility of the model was
calculated according to the method described in Section 4.1, with the
errors on the comparison of the model to the data and the goodness of
fit of the model calculated as described in Section 4.2.  The best
fitting model is that with $\Theta$=10mas, which taken over the
8--12$\mu$m range and the 6 baselines considered (excluding 79-F, see
Section 3.2) has a reduced $\chi^2$ of 0.32.

Adopting an elliptical Gaussian model gives an additional two
parameters to fit.  As well as the FWHM $\Theta$, we also consider the
ellipticity of the Gaussian $I$, where $v_I = v\cos(I)$, and the
rotation of the model $\theta$ which represents the angle of the major
axis of the elliptical Gaussian in degrees East of North.   The ranges
tested for each of these parameters were $\Theta \in (1,150)$mas in 21
logarithmically-spaced steps, $I \in (0^\circ,90^\circ)$ in steps of
$18^\circ$, and $\theta \in (0^\circ, 180^\circ)$ in steps of
$10^\circ$.  The best fitting model ellipticity is $I=0^\circ$, so a
circular Gaussian model (with FWHM $\Theta = 10$mas) provides the best
fit.  

Finally, we tested ring-like distributions for the excess.  The
parameters for these models were the radius of the center of the ring
$\rho_0$ and the width of the ring $d\rho_0$.  As the Gaussian
modelling showed no evidence for inclined structure, we do not include
inclination and position angle in our parameters for ring
models.  The parameter ranges tested were $\rho_0\in(1,100)$mas and 
$d\rho_0 \in (0.2, 2.0)\rho_0$ (where $d\rho_0 = 2\rho_0$ describes a
ring from 0--2$\rho_0$).  The best fitting model parameters are
$\rho_0 = 6$mas (so ring diameter is 12mas, similar to the best
fitting Gaussian FWHM) and $d\rho_0 = 0.6\rho_0$. The reduced $\chi^2$ for
this model is 0.31.    

\subsubsection{Preferred model for HD113766}

\begin{figure}
\includegraphics[width=7cm]{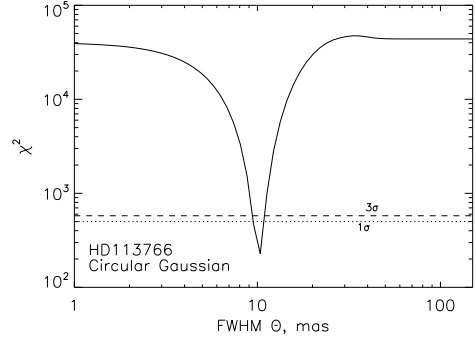}
\caption{\label{fig:circgauss} The goodness-of-fit for circularly
  symmetric Gaussian models for the distribution of excess emission
  around HD113766.  The percentage points of the $\chi^2$ distribution are
  shown and indicate that only a narrow range of FWHM, $9 < \Theta < 12$mas,
  provide a fit to the data within 3$\sigma$.  }
\end{figure}

Despite the increased complexity, the ring models do not offer a
much better fit to the data than the Gaussian models.  A simple
circularly symmetric Gaussian model with a FWHM $\Theta=10$mas offers
a good fit to the data.  This is in keeping with the initial first order
approximation to the size of the emitting region shown in Figure
\ref{fig:113766bybase}.  We see no evidence for a change in
visibility with position angle (once the difference in baseline
length is taken into account), and so a circularly symmetric
distribution is to be expected to offer a good fit to the data.  
Using the percentage points of the $\chi^2$ distribution
the 3$\sigma$ limits suggest a good fit can be achieved with $9 <
\Theta < 12$mas (see Figure \ref{fig:circgauss}).   The visibility of
the best fitting Gaussian model, and the visibilities of models at
the 5$\sigma$ limits ($\Theta = 7$mas and 14mas), are plotted on
Figure \ref{fig:113766bybase} (left).  We also show for comparison the
visibility of the best fitting ring model. \footnote{If we had instead
used the VISIR spectrum as our total flux in the analysis of the MIDI
data, the visibility functions would be higher (see footnote
\ref{foot:vis}.  The resulting best fitting Gaussian model would have
$\Theta = 7$mas. }  

\begin{figure}
\includegraphics[width=7cm]{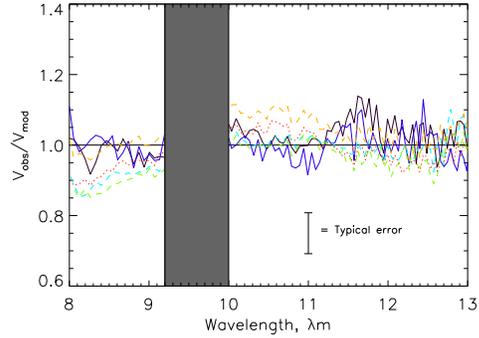}
\caption{\label{fig:hd113766_bestcomp} A comparison of the observed
  and best fitting model visibilities for HD113766.  A value of 1
  (solid black line) would represent a perfect fit of the model to the
  data.  Line colours and
  styles are as described in Figure \ref{fig:113766midi}.  For all
  baselines (excluding 79-F) the model of a circularly symmetric
  Gaussian with FWHM = 10mas provides a good fit to the observed data
  (errors on $V_{\rm{mod}}/V_{\rm{obs}}$ are typically around 10\%
  across all baselines and the full wavelength range and are
  calculated as outlined in Section 4.2).}
\end{figure} 

It is worth noting that for the best fitting models we achieve a
reduced $\chi^2$ of $<1$.  This would normally be termed an over-fit,
and could be taken as evidence that the errors assigned to the data
are too large.  In this case the sources of error have been carefully
considered in turn (see section 4.2).  A more likely possibility is that
the number of degrees of freedom is too high.  We have used the number
of wavelength bins multiplied by the number of observations as the
number of data-points, and subtracted the number of free parameters
for each model considered to determine the degrees of freedom.  This
calculation involves the implicit assumption that the data-points are
independent, which for neighbouring wavelength bins for a particular
observation will not be the case.  This would mean that the values of
reduced $\chi^2$ are underestimated. Calculation of degrees-of-freedom
in such cases is complex (see \citealt{andrae} for a recent discussion).  A
better idea of the true absolute goodness-of-fit might be achieved by
examination of $V_{\rm{obs}}/V_{\rm{mod}}$ for the best fitting
model.  This function is shown in Figure \ref{fig:hd113766_bestcomp}.
For all baselines (excluding 79-F) the circular Gaussian with FWHM
$\Theta$ = 10mas provides a good fit to the observed visibility
function.

\subsection{Best fit models for HD172555}

The observed visibility functions from MIDI observations of HD172555
suggest a completely resolved disc on all baselines.  We first
attempt to model the distribution of the excess emission using
circularly symmetric Gaussian models with a range of FWHM $\Theta$.  We
tested $1\le \Theta \le 200$mas (the predicted radius of the dust is larger
for this target than for HD113766, see Table \ref{tab:sources}) in 60
logarithmically-spaced steps.  The visibility of the model was
calculated using the method described in Section 4.1, with the errors
on the comparison of the model to the data and the goodness of fit of
the model calculated as described in Section 4.2.  The best fitting
model is that with $\Theta = 30$mas, which taken over the 8--12$\mu$m
wavelength range (excluding the ozone absorption region) and 7
baselines (excluding problematic observations 78-D and 83-C3, see
Section 3.3) has a reduced $\chi^2$ of 1.0.  

Testing elliptical Gaussian models suggests that an elliptical 
model provides a better fit.  The best fitting parameters
are found to be $\Theta=48$mas, $I=75^\circ$ and $\theta =
120^\circ$EoN.  Such a configuration gives a slightly higher disc
visibility along baselines at position angles close to 30$^\circ$EoN,
and slightly lower visibility at baselines close to 120$^\circ$EoN.  This
orientation can be compared with Figure \ref{fig:172555bybase}, which
shows that observations at baselines similar to 30$^\circ$ have a
slightly higher visibility.  Confidence limits on these best
  fitting parameters are discussed further in subsection 4.4.1. The
reduced $\chi^2$ for this best fitting model is 0.36. 

We also tested ring-like distributions for the excess.  The
parameters for these models were the radius of the center of the ring
$\rho_0$, the width of the ring $d\rho_0$, the inclination of the ring to the
line of sight $I$ and the position angle of the major axis $\theta$.
The parameter ranges tested were $\rho_0 \in (6, 198)$mas, $d\rho_0 \in
(0.2,2.0)\rho_0$, $I \in (0^\circ, 90^\circ)$ and $\theta \in (0^\circ,
180^\circ)$ EoN.  The best fitting model parameters are $\rho_0 = 80$mas,
$d\rho_0=0.5\rho_0$, $I=75^\circ$ and $\theta=120^\circ$EoN.  The reduced
$\chi^2$ for this model is 0.38.  This visibility model produces higher
$V_{\rm{mod}}$ on baselines close to 30$^\circ$EoN and low
$V_{\rm{mod}}$ ($V_{\rm{mod}} \sim V_{\rm{res}}$) on baselines close
to 120$^\circ$EoN (compare with Figure \ref{fig:172555bybase}).  

The model of \citet{lisse09} suggests that the emission from HD172555
might be the result of a recent massive collision in the system.  If
so, the emitting material might be expected to be concentrated in a
clump rather than spread in a ring-like distribution.  Dust emitting
from a circumplanetary region would also look like a clump orbiting
the star.  A point-like clump at an offset from the point-like star
would have a visibility of 1 on baselines along a position angle
90$^\circ$ from the direction of the offset of the clump from the star
(e.g. if a point-like clump lay directly north of the star, then
observations on baselines lying East-West, or at PAs of 90$^\circ$EoN,
would have visibilities of 1).  In the observed data all visibilities
are $<$1.  Along baselines in the direction of a point-like offset
clump the visibility would be seen to oscillate in a sinusoidal
manner, with maxima of 1, minima of $\sim0.05$ (averaged over the MIDI
wavelength range when $\cos(2\pi(u(\alpha_1-\alpha_2)+
v(\beta_1-\beta_2)))=-1$, see equation \ref{eq:two_vis}), and period
dependent on the distance between the clump and the central star.  As
all visibilities are compatible with a resolved disc at the 2$\sigma$
level (or better), and no visibilities are compatible with an
unresolved source (at the $>11\sigma$ level averaging over the
8--12$\mu$m MIDI wavelength range excluding the 9.2--10$\mu$m ozone
absorption range), we can rule out a point-like offset clump as the
source of the excess emission.    

\begin{figure}
\includegraphics[width=7cm]{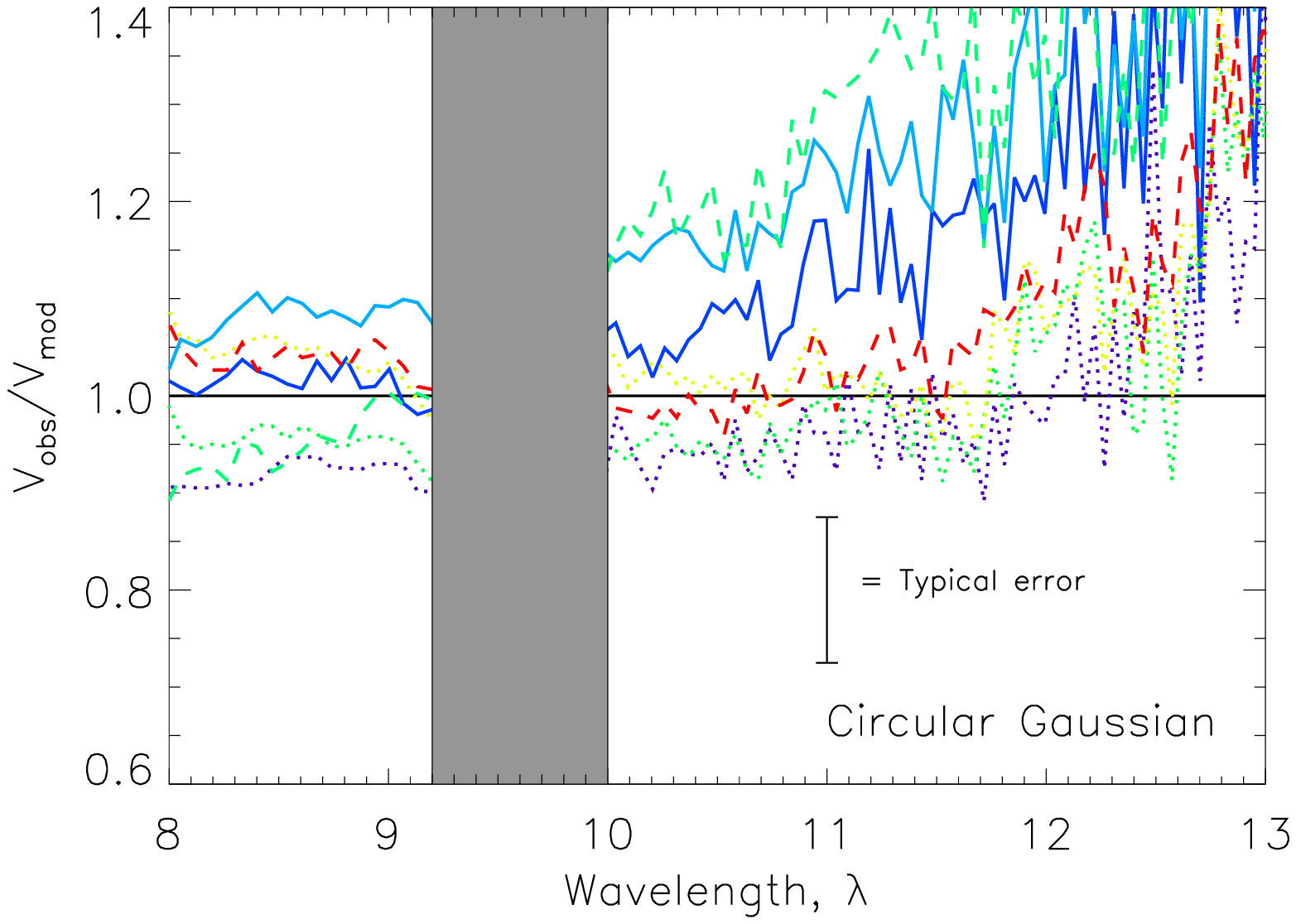}
\includegraphics[width=7cm]{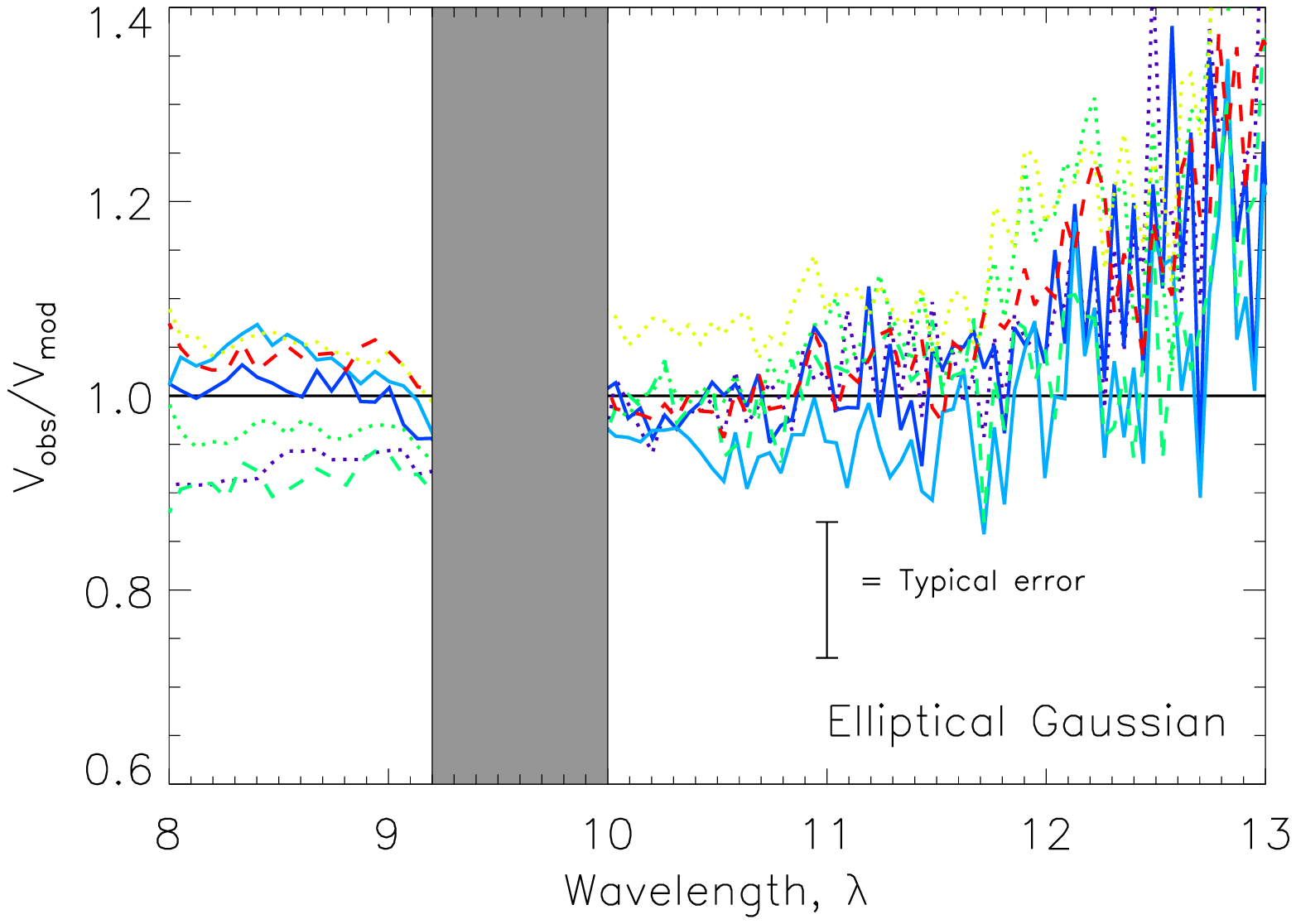}
\includegraphics[width=7cm]{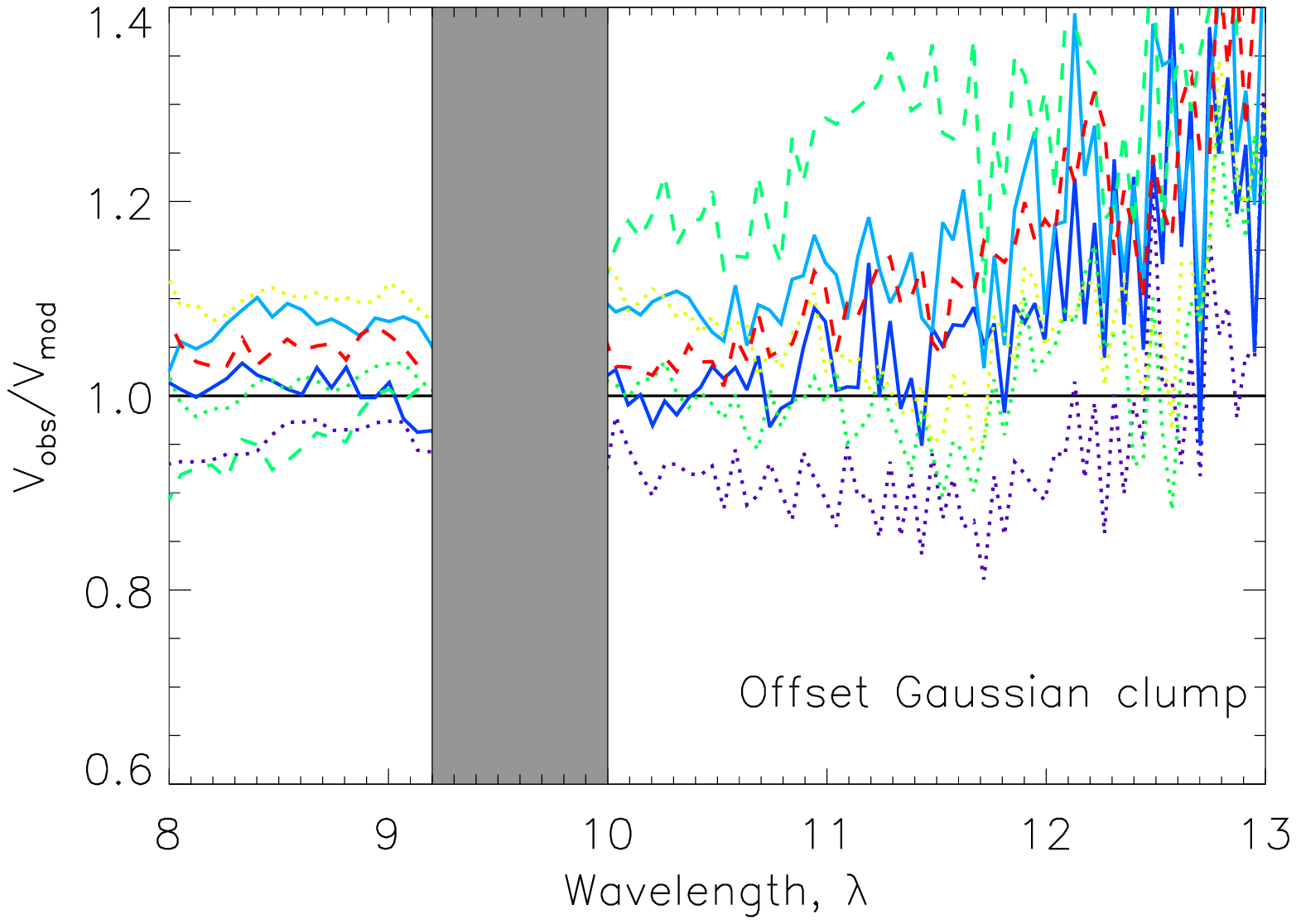}
\caption{\label{fig:172555_bestcomp} A comparison of the observed
  and best fitting model visibilities for HD172555.  A value of 1
  (solid black line) would represent a perfect fit of the model to the
  data.  Line colours and styles are as described in Figure
  \ref{fig:172555midi}.  A Gaussian of FWHM =30mas provides
  reasonable fit to the observed data according to the $\chi^2$
  analysis (errors on $V_{\rm{mod}}/V_{\rm{obs}}$
  are typically around 15\% across all baselines and the full
  wavelength range and are  calculated as outlined in Section 4.2),
  however a better fit can be found by allowing elliptical or inclined
  models (best fit elliptical Gaussian with $\Theta = 48$mas,
  $I$=75$^\circ$ and $\theta=120^\circ$EoN shown). An offset Gaussian
  clump offers no improvement over the circularly symmetric Gaussian
  model centered on the star.  See text for details.}
\end{figure} 

\begin{figure*}
\begin{minipage}{7cm}
\includegraphics[width=7cm]{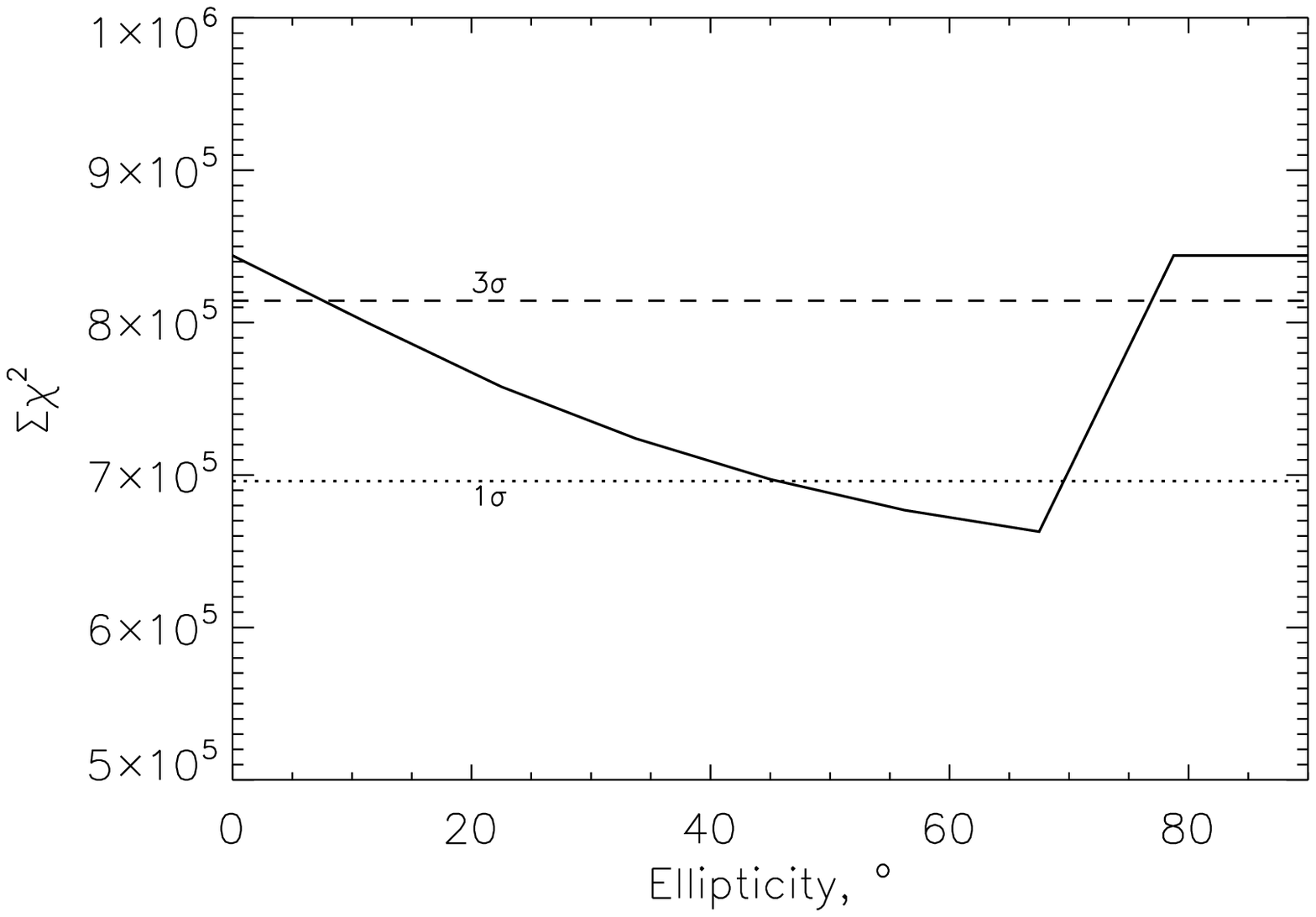}
\end{minipage}
\begin{minipage}{7cm}
\includegraphics[width=7cm]{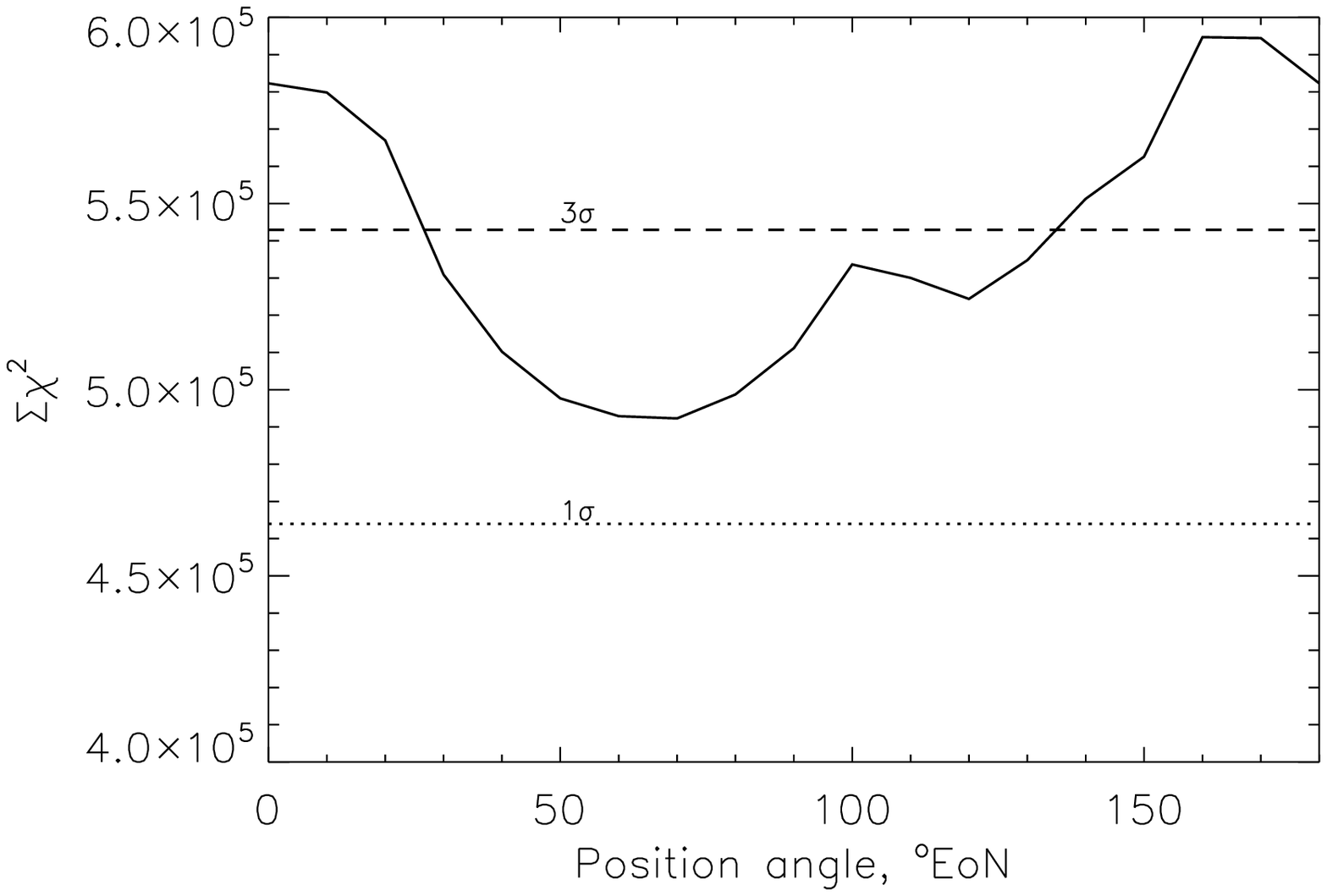}
\end{minipage} \\
\begin{minipage}{7cm}
\includegraphics[width=7cm]{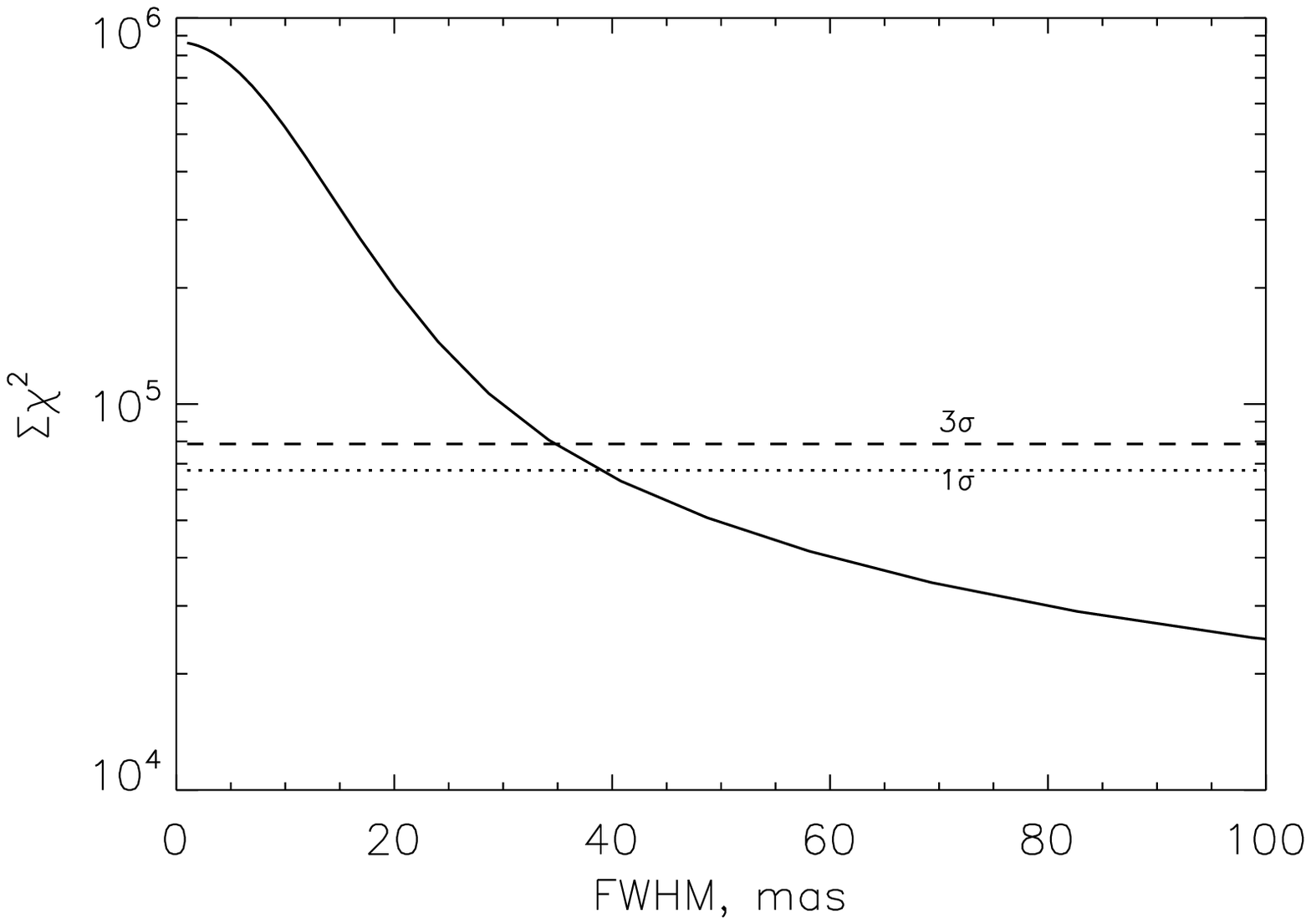}
\end{minipage}
\begin{minipage}{7cm}
\caption{\label{fig:172555_modelparams} The constraints we can place on
  the best fitting model parameters for an elliptical Gaussian model
  fitted to the observed visibilities of HD172555.  We tested the
  goodness-of-fit for models with varying ellipticity ($I$, top left),
  position angle ($\theta$, top right) and FWHM ($\Theta$ bottom left)
  using a Bayesian approach.  That is, for each parameter to be
  fitted, we computed an overall $\chi^2$ value by summing the
  individual $\chi^2$ values over all possible values of the other two
  parameters.   In the bottom
  left we can see that any size greater than 35mas is favoured, which
  provides a lower limit to the size of the disc. See text for
  detailed discussion.}  
\end{minipage}
\end{figure*}

To determine if a more extended offset distribution provides a good fit to
the observed visibilities, we also try to fit the data with a Gaussian
clump offset from the central star.  The clump is modelled in the same
way as a central Gaussian clump, with FWHM $2<\Theta<40$mas.  The
clump is offset from the central source (the star is again modelled as
a point source, so with uniform visibility of 1 everywhere) by a
distance $d$, with the tested range $ 5 < d < 300 $mas, and a position
angle $0 < \theta < 180^\circ$EoN. Note that we do not test the range
180--360$^\circ$EoN as the symmetry of the visibility function means
that the visibilities are the same for clumps at $\theta^\circ$EoN and
$\theta+180^\circ$EoN.   

The best fitting clump model is for a clump of FWHM $\Theta = 22$mas,
offset from the star by a distance $d = 10$mas at a position angle of
$100^\circ$EoN.  The size and offset of this best fitting clump model
are comparable to the best fitting circularly symmetric Gaussian
centered on the star (FWHM = 30mas).  The offset of $100^\circ$EoN
would result in higher visibilities on baselines around
$\sim10^\circ$EoN, which is what is seen in the observations on
baseline 80-D2 (see Figure \ref{fig:172555bybase}).  The reduced
$\chi^2$ for this best fitting model is 1.70.  
Although this seems at first sight like a reasonable fit, as
explained for HD113766 the absolute values of reduced $\chi^2$ are likely
to overstate the goodness-of-fit of the models tested here due to the
over-estimation of the degrees of freedom in the model.   
In addition to the poorer fit, the best-fitting Gaussian
clump model is a clump which encircles the star, and therefore is not
a geometry which we could expect either from a circumplanetary disc or
from the recent massive collision scenario suggested in
\citet{lisse09}.   

\subsubsection{Preferred model for HD172555} 

As with HD113766, the best fitting models have reduced $\chi^2<1$,
which is a reflection of the over-estimation of the number of
degrees-of-freedom.  In spite of this limitation to our interpretation
of the absolute values of reduced $\chi^2$, we can see from the
\emph{relative} values of reduced $\chi^2$ that the elliptical models and
inclined rings provide a better fit to the data than a circularly
symmetric Gaussian or clump model. This is made clearer by
consideration of the function $V_{\rm{obs}}/V_{\rm{mod}}$ for the best
fitting models.  These plots are shown in Figure
\ref{fig:172555_bestcomp}.  It is clear from this plot that the
elliptical Gaussian model offers a better fit than the circularly
symmetric or offset clump models.  We find there is no significant
difference in the goodness-of-fit offered by elliptical Gaussian and
inclined ring-like models.   The circularly symmetric Gaussian
and offset clump model offer similarly poor fits to the data from this
analysis.  We would expect a model with increased parameters to offer
an improved fit to the data, and the fact that the offset clump model
does not is reflected in the higher value of reduced $\chi^2$.  

As we have found that an elliptical or inclined disc structure offers
a better fit to the data than a circularly symmetric model, we want
to test the limits we can place on the inclination/ellipticity and
position angle of the disc with the data.  We use the elliptical
Gaussian model as the simpler model to test the parameter ranges.
To determine the limits on the inclination of the disc model,
  we sum the $\chi^2$ values over all values of the FWHM and position
  angle for each value of the inclination $I$ in turn.  The resulting
  probability distribution is shown in Figure
  \ref{fig:172555_modelparams}.  Models with inclinations of $I<8^\circ$ and
  $I>77^\circ$ are ruled out at the 3$\sigma$ level according to the
  percentage points of the $\chi^2$ distribution by this
  calculation, suggesting that a moderately inclined or elliptical
  disc is needed to fit the visibility data.  We use the same
  technique to determine the limits on the other model parameters
  (results shown in Figure \ref{fig:172555_modelparams}). These tests
  show that discs with FWHM $\Theta<35$mas and lying at position
  angles $\theta<27^\circ$EoN or $\theta>135^\circ$EoN are also ruled
  out at the 3$\sigma$ level.  This lower limit on the disc size
  translates to $\sim$1AU at 29.9pc.

\section{Discussion}

The results presented in this paper provide the most accurate direct
measurement of the dust at $\sim$1--8AU around any main-sequence star.
The MIDI data on HD113766 reveal a partially resolved disc.  The best
fitting models suggest a disc diameter of 9--12mas (1.2--1.6AU). A
circularly symmetric Gaussian provides a good fit to the data which
suggests we do not have significant evidence for asymmetries in the
disc structure which might be expected from a recent massive collision
for example. We also find no improvement in the fit if we consider
elliptical Gaussians, and thus there is no evidence that the disc is
inclined or that the binary is forcing an ellipticity on the disc.
The binary has a spectral type of F5 and lies at a projected distance
of 157AU.  At this distance the binary would not be expected to have a
strong influence on dust at $\sim$1.6AU (unless the binary has an
extremely elliptical orbit, $e_b>0.9$, \citealt{holman}).  There is no
evidence that this system has additional cold dust components.  The Q band
imaging data (which is more sensitive to cooler dust populations than
the N band) puts a limit of $<$500mJy on cooler discs
around HD113766 (see Figure \ref{fig:ext}).   This limit applies to
discs with radii $>$0\farcs15 (20AU) and assume a favourable disc
geometry (inclined discs or if face-on, narrow disc geometries offer
the best chance of detection).

For HD172555 we find visibility functions that are compatible with
completely resolved excess emission.   Modelling the observed data
with ring-like distributions of dust suggests that the emitting
material lies at a radial offset from the star of $>$35mas ($>$1AU
at a parallax distance of 29.2pc).   We also find 
tentative evidence for an inclined or elliptical disc structure, with
major axis orientated at $\sim120^\circ$EoN.  The TReCS imaging
data on this target places an upper limit of 7.9AU on the extent of the
disc in the N band, giving a combined constraint of 1--7.9AU on the
dust's location.  The TReCS Q-band imaging also indicates the
  presence of a more extended component, located at $\sim$8AU from
  the star.  This component is inclined or elliptical in structure,
  lying at a position angle of $\sim$110$^\circ$EoN.  The N band
  imaging shows that the 10$\mu$m emission must come from a region
  inside this, and from the MIDI results we infer a similar postion
  angle to that seen in the Q band image, but at a low significance.
  This suggests that the two may be linked in an extended disc, with
  the N and Q band data probing the inner and outer regions of the
  distribution respectively.  However, HD 172555 may also have two
  spatially distinct components, similar to other multi-component
  discs (e.g. $\eta$ Tel,
  \citealt{smitheta}; $\beta$ Leo, \citealt{stock}; $\eta$ Corvi,
  \citealt{smithmidi}).  HD 172555 could be viewed as another young
  solar system (see $\eta$ Tel ; \citealt{smitheta}), and so could
  alternatively be interpreted as having two debris belts somewhat
  analogous to the asteroid and Kuiper belts in the Solar System.  
  Neither configuration rules out the recent massive
  collision scenario suggested by \citet{lisse09} for the origin of
  the emission.  An alternative extended disc distribution could also
  explain the results presented in this paper, with the N and Q band data
  probing the inner and outer regions of the distribution
  respectively.  Again, such a distribution does not rule out the
  possibility of a recent massive collision, although if the majority
  of the emission is believed to have arisen from a collision,
  dynamical models will be required to confirm that the resulting
  ejecta can be spread over such a wide spatial range.

The young ages of the systems studied here (16Myr and 12Myr), combined
with the lack of observed cold dust in these systems and the location
from which the observed emission is seen to arise all suggest we are
witnessing ongoing terrestrial planet formation.  \citet{lisse09}
concluded that the spectrum of HD113766 is consistent with emission
from a disrupted S-type asteroid (of $\ge320$km in radius) or an
asteroid belt made up of a large number of small S-type asteroids.
Such a belt is a natural consequence of terrestrial planet formation
(see e.g., \citealt{kenyon04II}). For HD172555 \citet{lisse09} suggest
that the excess emission arises from a recent massive collision
similar to that believed to have formed the Earth-Moon system.  If
this is a true model of the source of the excess then we might expect
a clumpy dust distribution, however we do not see any evidence for
this in the visibility functions. \citet{wyattdent} presented a model
for the evolution of a clump following a massive collision in a debris
disc, which they found to be an unlikely source for the asymmetric
dust distribution in Fomalhaut.  Assuming a radius of 5.8AU, a parent
planetesimal radius of 1000km with density 2.5 g cm$^{-3}$ (from the 
fitting of \citealt{lisse09}) and a specific incident energy required
for catastrophic destruction of a planetesimal $Q_D^\ast$=200 J
kg$^{-1}$,  we find that the time taken for a dust cloud produced in a
massive collision to occupy half the disc (and thus no longer be
observable as a clump) is $\sim$200 years.  As the constraints from
the modelling of the IRS spectrum on the lifetime of the dust arising
from a massive collision suggest that the collision should have
occurred in the last 0.1Myr, clump-like emission would only be
expected to have lasted for the first 0.2\% of the lifetime of the
emission (this is consistent with a clump lifetime of 20--30 orbital
periods seen in the asteroid belt, \citealt{michel}).  It is not
surprising therefore that the emission we observe is not seen to be
clump-like if the origin of the emission is a massive collision. 
The poor fit to the observed visibilities provided by clump
models suggests that we can rule out the possibility of the emission
arising from circumplanetary regions (around an as yet undetected
planet).  

The constraints we can place on the radial location of the emitting
dust around HD113766 and HD172555 with these observations are of
particular interest, as when combined with spectral analysis they
allow the possibility of constraining models for the dust grains
themselves.  This is important because predictions of the location of
the debris based on spectral analyses alone can be highly dependent on
assumptions made about the grain properties such as size.  For
HD113766, the location of the dust inferred from spectral analysis is
13mas (1.7AU \citealt{lisse08}), slightly larger than our MIDI
observations imply.  For HD172555 dust has been predicted to arise
from a region around 5.8AU \citep{lisse09}, at the larger end of the 
1--7.9AU scales that the MIDI and TReCS observations presented here
imply.  We believe these discrepancies are due to the modelling
methods used, scaling observations made during the Deep Impact Temple
1 experiment to account for the luminosity of each target star.  
As our new interferometric data constrains the dust in both systems to
be located in regions smaller than predicted from the \citet{lisse08}
and \citet{lisse09} models, the implication is that the dust
composition is different from the dust released in the Deep Impact
event such that it would be hotter if placed at the same distance from
any given star.  These constraints
will allow the next-generation models to treat the dust composition
and location in a self-consistent way.

In this paper we have presented new VISIR spectroscopic and MIDI
interferometric observations of the youngest debris disc hosts with
emission on $\ll$10AU scales, HD113766 and HD172555.  We have
additionally presented new VISIR imaging observations of HD113766,
which rule out large-scale high surface brightness emission.  We find
that the visibility functions measured for HD113766 are consistent
with symmetric emission in the region between 0.6--0.8AU (or a
diameter of 1.2--1.6AU). For HD172555
the visibility functions observed suggest that the emission is
completely resolved with the VLTI, placing a lower limit of 1AU on
the spatial extent of the emitting region.  A new analysis of TReCS
data originally presented in \citet{moerchenastar} has revealed
extended emission around HD 172555 in the Q band, consistent with an
inclined disc at a median radius of 7.9AU from the star.  This means
that the HD172555  system could contain multiple disc components, or
that we are observing different regions of a broad dust distribution
around this stars in the different bands.  These observations are the
first that resolve the mid-infrared excess emission around these
targets.  In both cases the
observational data are compatible with current models that suggest we
are observing two systems in the midst of terrestrial planet forming
processes.

\section*{Acknowledgments}

Based on observations made with ESO Telescopes at the Paranal
Observatories under programme IDs 078.D-0808, 079.C-0259, 080.C-0737,
and 083.C-0775.  The authors wish to thank Christine Chen for
providing the Spitzer IRS spectra of HD113776 and HD172555.

\bibliographystyle{mn2e}  
\bibliography{/home/rachel/Work/PAPERs/thesis} 

\label{lastpage}

\end{document}